\documentclass[11pt,a4paper]{article} 
\usepackage{jheppub}

\usepackage{amsmath,amssymb,amscd,amsfonts,graphicx,bbm}
\usepackage{multirow,fancybox,latexsym,float,psfrag,bm,psfrag}

\newcommand{\llangle}{\langle\hspace*{-0.5mm}\langle}
\newcommand{\rrangle}{\rangle\hspace*{-0.5mm}\rangle}

\title{
One-loop operator matching in the static heavy and
domain-wall light quark system with $O(a)$ improvement
}

\author[a, b]{Tomomi Ishikawa,}
\author[b]{Yasumichi Aoki,
           \footnote{Present address:
	             Kobayashi-Maskawa Institute for the Origin of
                     Particles and the Universe (KMI), Nagoya University,
       		     Nagoya 464-8602, Japan.}}
\author[c]{Jonathan M. Flynn,}
\author[b, d]{Taku Izubuchi}
\author[e]{and Oleg Loktik}

\affiliation[a]{Physics Department, University of Connecticut, \\
                Storrs, CT 06269-3046, USA}
\affiliation[b]{RIKEN BNL Research Center, Brookhaven National Laboratory,\\
                Upton, NY 11973, USA}
\affiliation[c]{School of Physics and Astronomy, University of Southampton,\\
                Highfield, Southampton SO17 1BJ, UK}
\affiliation[d]{Physics Department, Brookhaven National Laboratory,\\
                Upton, NY 11973, USA}
\affiliation[e]{Physics Department, Columbia University,\\
                New York, NY 10027, USA}

\emailAdd{tomomi@phys.uconn.edu}
\emailAdd{yaoki@quark.phy.bnl.gov}
\emailAdd{j.m.flynn@soton.ac.uk}
\emailAdd{izubuchi@quark.phy.bnl.gov}
\emailAdd{oleg.loktik@gmail.com}

\abstract{
We discuss perturbative $O(g^2a)$ matching with static heavy quarks
and domain-wall light quarks for lattice operators relevant to
$B$-meson decays and $B^0$--$\bar B^0$ mixing.
The chiral symmetry of the light domain-wall quarks does not prohibit
operator mixing at $O(a)$ for these operators.
The $O(a)$ corrections to physical quantities are non-negligible and
must be included to obtain high-precision simulation results for CKM physics.
We provide results using plaquette, Symanzik, Iwasaki and DBW2 gluon actions
and applying APE, HYP1 and HYP2 link-smearing for the static quark action.
}


\begin{document}

\rightline{\sffamily RBRC-863}
\rightline{\sffamily SHEP-1043}\vspace*{-10mm}

\maketitle

\section{Introduction}

The Cabibbo-Kobayashi-Maskawa (CKM) quark mixing matrix
elements~\cite{Cabibbo:1963yz, Kobayashi:1973fv} play a key role in
elementary particle physics. Constraints on $V_{ts}$ and $V_{td}$ can
be obtained from $B^0-\bar{B}^0$ mixing, where the $SU(3)$-breaking
ratio $\xi=f_{B_s}\sqrt{B_{B_s}}/f_{B_d}\sqrt{B_{B_d}}$~\cite{Bernard:1998dg}
plays an important role.
Lattice QCD simulations for $B$-meson physics are challenging, however,
because of the difference in energy scales
between the light $u$ and $d$ quarks and the heavy $b$ quark.
One approach to this issue is to use Heavy Quark Effective Theory
(HQET)~\cite{Eichten:1989zv}.
Lattice calculations with HQET have several known difficulties (and solutions).
(i) The static propagator is noisy because the static self-energy
contains a $1/a$ power divergence.
The ALPHA collaboration introduced a modified static action which
improves the signal to noise ratio (S/N)~\cite{Della Morte:2005yc}.
The modification is to replace the gluon link variable in the static
action with a smeared one obtained by $3$-step hypercubic blocking.
Using this action the power-divergent contributions in the static
self-energy are significantly reduced.
(ii) Non-perturbative matching to the continuum is needed.
If we include $O(1/m_b)$ corrections in HQET, the continuum limit cannot be
reached using perturbative matching because of a power
divergence~\cite{Heitger:2003nj}.
Non-perturbative matching schemes include the Schr\"odinger functional
with step scaling or RI/MOM, but implementing them in practice is
not easy for HQET.

The static approximation (lowest order of HQET) is theoretically simple
and its implementation in lattice calculations is relatively easy.
Because of this it is often used in simulations as a first step.
This leads to errors of $O(\Lambda_{\rm QCD}/m_b)\sim 10\%$.
Although $O(1/m_b)$ corrections should be included for more
precise calculations, the static approximation works well in the
determination of the $SU(3)$-breaking ratio $\xi$, in which the
theoretical uncertainty is suppressed by $(m_s-m_d)/\Lambda_{\rm QCD}$
and is estimated to be about $2\%$.
In addition, perturbative matching is more justified for the static
approximation.
Hence the aim of this paper is to calculate the one-loop perturbative matching
factor for the static approximation to enable the determination of $\xi$.

Discretization errors in lattice simulations can limit the precision
with which physical quantities are determined.
Therefore the $O(a)$ improvement program on the lattice is important.
The $O(a)$ improvement of heavy-light currents with clover Wilson light quarks
was investigated using one-loop perturbation theory in
non-relativistic QCD~\cite{Morningstar:1997ep} and the static
approximation~\cite{Ishikawa:1998rv}.
The $O(a)$ effects give large corrections to $B$ meson quantities.
In this work we consider the case where the light quarks are simulated
using a lattice action with good chiral symmetry.
In the light-light quark system, the chiral symmetry guarantees the absence
of $O(a)$ errors in the operator.
For the static heavy-light quark system, however, there are
$O(a)$ effects even if we use chiral fermions for the light quarks.
This result was already found for the clover Wilson light quark with
Wilson parameter $r=0$
(which is chirally symmetric, but has doublers)~\cite{Ishikawa:1998rv}.
In this study we use the domain-wall (DW) fermion
action~\cite{Kaplan:1992bt, Narayanan:1992wx, Shamir:1993zy}
to realize chiral symmetry for the light quarks.
This paper is an extension of several previous works:
the perturbative matching at $O(g^2)$ in the static heavy and
domain-wall light quark system~\cite{Loktik:2006kz} and its version with
a link-smeared static action~\cite{Albertus:2007zz, Christ:2007cn}.
The main results are the matching between continuum HQET and lattice HQET
and $O(a)$ improvement of the heavy-light quark bilinear
operator~(Eq.~(\ref{EQ:matching_bilinear})) and of the $\Delta B=2$
four-quark operator~(Eq.~(\ref{EQ:matching_four-quark_L})).
The calculations are performed using mean-field (MF) improved one-loop
lattice perturbation theory and include the link smearing of the
static action.

The paper is organized as follows.
In Section~\ref{SEC:general_discussion} we give our field notation and
discuss the matching procedure.
We give symmetry-based restrictions on the forms of the HQET operators;
these apply for non-perturbative treatments as well.
In Section~\ref{SEC:one-loop_matching} we give the form of the one-loop
matching expressions while in Section~\ref{SEC:lattice_PT} we provide
more discussion of the lattice perturbation theory calculations.
Expressions for the continuum to HQET to lattice HQET matching with
mean-field improvement are given in Eqs.~(\ref{EQ:matching_bilinear})
to (\ref{EQ:matching_four-quark_S}).
In Section \ref{SEC:example} we compute numerical values
of the one-loop matching factors and estimate the $O(a)$ effect for
physical quantities including the $SU(3)$ breaking ratio $\xi$.
Concluding remarks are given in Section \ref{SEC:conclusion}.
The lattice Feynman rules are listed in Appendix~\ref{SEC:feynman_rules}
and details of the lattice perturbation theory calculation are shown
in Appendix~\ref{SEC:lattice_perturbation}.
Numerical values of several integrals are listed in tables in
Appendices~\ref{SEC:tables_light} and \ref{SEC:tables}.


\section{General discussion of matching in HQET}
\label{SEC:general_discussion}

\subsection{Definition of the heavy quark field and state normalization}

We regard the $b$ quark as heavy and give it an on-shell velocity
$v=(1,0,0,0)$;
then the on-shell momentum is $p=m_bv=(m_b,0,0,0)$, where $m_b$ is
the b quark mass.
We build a heavy quark field $h$ as a sum of quark, $h_+$,
and anti-quark, $h_-$:
\begin{eqnarray}
h(x)\equiv h_+(x)+h_-(x),\;\;\;
\overline{h}(x)\equiv \overline{h}_+(x)+\overline{h}_-(x)=
h_+^{\dag}(x)-h_-^{\dag}(x),
\end{eqnarray}
where
\begin{eqnarray}
h_{\pm}(x)\equiv
e^{\mp im_bv\cdot x}\frac{1\pm{\not\!v}}{2}b(x)
=e^{\mp m_bt}\frac{1\pm\gamma_0}{2}b(x).\label{EQ:static_field}
\end{eqnarray}
In the static limit, $h_+$ and $h_-$ decouple.
The definition of the creation and annihilation operators for them is as
follows:
\begin{list}{}{\leftmargin=37mm}
\item[$h_+^{\dag}$ :] creates the outgoing heavy quark,\\
                      cannot annihilate the incoming anti-heavy quark.
\item[$h_+$ :] annihilates the incoming heavy quark,\\
	       cannot create the outgoing anti-heavy quark.
\item[$h_-$ :] creates the outgoing anti-heavy quark,\\
	       cannot annihilate the incoming heavy quark.
\item[$h_-^{\dag}$ :] annihilates the incoming anti-heavy quark,\\
	              cannot create the outgoing heavy quark.
\end{list}

The HQET state normalization differs from that normally used in QCD.
The relation of the states is
\begin{eqnarray}
|B\rangle_{\rm QCD}=\sqrt{m_B}\left[|B\rangle_{\rm HQET}+O(1/m_b)\right],
\label{EQ:state_normalization}
\end{eqnarray}
where $m_B$ is the mass of the state $|B\rangle_{\rm QCD}$.

We use the PDG notation in which the quark content of the $B$
meson is $B=(\overline{b}q)$ and $\overline{B}=(b\overline{q})$, where
$q$ denotes light $d$ or $s$ quarks~\cite{Nakamura:2010zzi}.

\subsection{Operators relevant for $B$ meson decays and $B^0-\overline{B}^0$
            mixing}

The QCD operators considered in this paper are the heavy-light quark bilinear 
\begin{eqnarray}
J_{\Gamma}^{\rm QCD}=\overline{b}\Gamma q,
\end{eqnarray}
where
$\Gamma=\{1, \gamma_5, \gamma_{\mu}, \gamma_{\mu}\gamma_5, \sigma_{\mu\nu}\}$
with $\sigma_{\mu\nu}=\frac{i}{2}[\gamma_{\mu}, \gamma_{\nu}]$,
and the $\Delta B=2$ four-quark operator
\begin{eqnarray}
O_L^{\rm QCD}=
[\overline{b}\gamma_{\mu}^Lq][\overline{b}\gamma_{\mu}^Lq],
\end{eqnarray}
where $\gamma_{\mu}^R=\gamma_{\mu}P_R=\gamma_{\mu}(1+\gamma_5)/2$ and
$\gamma_{\mu}^L=\gamma_{\mu}P_L=\gamma_{\mu}(1-\gamma_5)/2$.
Quark colors are contracted within the square brackets.
The static effective theory operators needed for matching to
the QCD operators are
\begin{eqnarray}
J_{\Gamma}^{\rm HQET}&=&\overline{h}\Gamma q,\\
O_L^{\rm HQET}&=&
[\overline{h}\gamma_{\mu}^Lq][\overline{h}\gamma_{\mu}^Lq],\label{EQ:O_L0}\\
O_S^{\rm HQET}&=&
[\overline{h}P_Lq][\overline{h}P_Lq].\label{EQ:O_S0}
\end{eqnarray}
Omitting parts of these operators which do not contribute in the matrix element
for $B^0-\overline{B}^0$ mixing, Eqs.~(\ref{EQ:O_L0}) and (\ref{EQ:O_S0}) become
\begin{eqnarray}
O_L^{\rm HQET}&=&
[\![\overline{h}\gamma_{\mu}^Lq]\!][\![\overline{h}\gamma_{\mu}^Lq]\!]
\nonumber\\
&=&[\overline{h}_+\gamma_{\mu}^Lq][\overline{h}_-\gamma_{\mu}^Lq]
+[\overline{h}_-\gamma_{\mu}^Lq][\overline{h}_+\gamma_{\mu}^Lq]
=
2[\overline{h}_+\gamma_{\mu}^Lq][\overline{h}_-\gamma_{\mu}^Lq],\label{EQ:O_L}\\
O_S^{\rm HQET}&=&[\![\overline{h}P_Lq]\!][\![\overline{h}P_Lq]\!]
\nonumber\\
&=&[\overline{h}_+P_Lq][\overline{h}_-P_Lq]
+[\overline{h}_-P_Lq][\overline{h}_+P_Lq]
=2[\overline{h}_+P_Lq][\overline{h}_-P_Lq],\label{EQ:O_S}
\end{eqnarray}
where we define the doubled square bracket by
$[\![\overline{h}X]\!][\![\overline{h}Y]\!]
=[\overline{h}_+X][\overline{h}_-Y]+[\overline{h}_-X][\overline{h}_+Y]$.

\subsection{Matching procedure}
\label{SEC:matching_procedure}

We adopt a two step matching procedure:
\begin{eqnarray}
 \parbox[b][37mm][c]{130mm}{
\begin{center}
\psfrag{cqcd}[c][c]{continuum QCD}
\psfrag{CQCD}[c][c]{(CQCD)}
\psfrag{chqet}[c][c]{continuum HQET}
\psfrag{CHQET}[c][c]{(CHQET)}
\psfrag{lhqet}[c][c]{lattice HQET}
\psfrag{LHQET}[c][c]{(LHQET)}
\psfrag{mb}[c][c]{$\mu=m_b$}
\psfrag{ainv}[c][c]{$\mu=a^{-1}$}
 \psfrag{run}[l][c]{RG-evolution}
 \includegraphics[scale=0.50, viewport = 0 0 730 210, clip]
{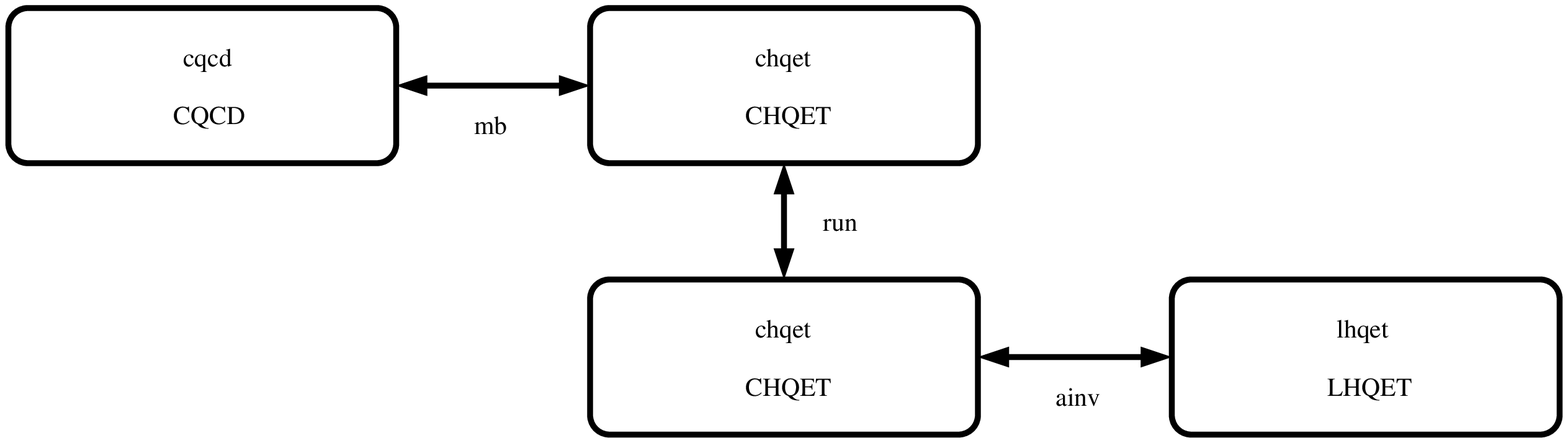}
\end{center}
}\;,\nonumber
\end{eqnarray}
in which we first perform the matching between continuum QCD (CQCD) and
continuum HQET (CHQET) and subsequently match CHQET to lattice HQET (LHQET).
Some comments on this matching:
\begin{list}{}{}
\item[(1)]
The continuum QCD (CQCD) operators are renormalized
in $\overline{\rm MS}$(NDR) at scale $\mu_b$ which is usually
chosen to be the $b$ quark mass $m_b$.
Fierz transformations in arbitrary dimensions are specified in the
NDR scheme introduced by Buras and Weisz~\cite{Buras:1989xd}.
The introduction of evanescent operators gives vanishing finite terms
at one-loop but is needed to obtain the correct anomalous dimensions
at two-loop.
\item[(2)]
The CHQET operators are also renormalized in $\overline{\rm MS}$(NDR)
at some scale $\mu$.
Matching between the continuum theories is performed in perturbation theory by
calculating and comparing matrix elements of the operators between an initial
state $|{\rm i}\rangle$ and final state $|{\rm f}\rangle$ for each theory.
The calculation has been done for quark bilinears at
one-loop~\cite{Eichten:1989zv} and two-loop~\cite{Broadhurst:1994se} levels,
and for the $\Delta B=2$ four-quark operator at one-loop~\cite{Flynn:1990qz}.
\item[(3)]
The continuum matching between QCD and HQET is done at scale $\mu=m_b$ to
avoid a large logarithm of $\mu/m_b$.
We use renormalization group (RG) running in CHQET to move to a lower
scale at which the HQET matching between continuum and lattice is done.
We employ the two-loop anomalous dimension calculations in
Refs.~\cite{Ji:1991pr,Broadhurst:1991fz} for the bilinear and in
Refs.~\cite{Gimenez:1992is,Ciuchini:1996sr,Buchalla:1996ys}
for the four-quark operator.
\item[(4)]
Matching between CHQET and LHQET is performed at scale ${\mu=a^{-1}}$,
where $a$ denotes the lattice spacing.
The calculation is performed in one-loop perturbation theory taking into
account $O(a)$ discretization errors on the lattice.
For this we introduce external momenta, e.g.,
\begin{eqnarray}
\langle{\rm f}|O|{\rm i}\rangle&=&
\langle h(p')|O|q(p)\rangle \hspace*{+28.5mm}
\mbox{for bilinear operators},\nonumber\\
\langle{\rm f}|O|{\rm i}\rangle&=&
\langle h(p'_2), q(p_2)|O|h(p'_1), q(p_1)\rangle \;\;\;\;\;
\mbox{for four-quark operators},
\end{eqnarray}
where $O$ denotes the bilinear and four-quark operators.
On-shell improvement is used,
in which we impose the equations of motion on the external quarks:
\begin{eqnarray}
D_0h=0,\;\;\;
\left({\not\!\!D}+m_q\right)q=0,
\end{eqnarray}
where $m_q$ denotes the light quark mass.
The transition amplitudes are expanded around zero in the external momenta
up to first order, giving rise to $O(pa)$ effects.
The $O(p'a)$ effects always vanish due to the on-shell condition $p'=0$.
The amplitudes are also expanded to first order in the light quark mass
to reveal $O(ma)$ effects.
\end{list}

\subsection{HQET operators and symmetries}

In this section we introduce symmetry transformations that appear to be
powerful tool to study the mixing structure of the HQET operators under
consideration.
The arguments in this section are used in Secs.~\ref{EQ:continuum_matching}
and \ref{EQ:HQET_matching}.
Similar discussions were presented in
Refs.~\cite{Becirevic:2003hd,Blossier:2007hg}.

\subsubsection{Symmetries in the static heavy and domain-wall
               light quark system}
\label{SEC:symmetries_HL-system}

We use following symmetries to restrict the operator mixing.

\subsubsection*{Chiral symmetry}

We assume an $SU(N_F)\otimes SU(N_F)$ chiral symmetry for the
light quark sector ($N_F$ denotes number of light quark flavors):
\begin{eqnarray}
q_R\longrightarrow U_R q_R,\;\;\; q_L\longrightarrow U_L q_L,
\end{eqnarray}
where $q_R=P_Rq$, $q_L=P_Lq$, and $U_R$ and $U_L$ represent the
$SU(N_F)$ chiral transformation matrices.
To account for explicit symmetry-breaking by the light quark masses
we assume that the light quark mass matrix
$M={\rm diag}(m_1, \cdots, m_{N_F})$ transforms as follows:
\begin{eqnarray}
M\longrightarrow U_LMU_R^{\dag},\;\;\;
M^{\dag}\longrightarrow U_RM^{\dag}U_L^{\dag},
\label{EQ:mass_transformation}
\end{eqnarray}
This restricts the $O(ma)$ operator mixing.
We use the domain-wall fermion formulation for light quarks on the lattice
in this paper and treat these quarks as exactly chiral.
In practice the domain-wall quark action breaks chiral symmetry
by an amount proportional to powers of the residual quark mass which is
of $O(10^{-3})$ and hence small.

\subsubsection*{Heavy quark spin symmetry (HQS)}

The static heavy quark obeys the heavy quark
symmetry~\cite{Isgur:1989vq,Isgur:1989ed}
\begin{eqnarray}
h_{\pm}\longrightarrow{\cal V}_ih_{\pm},\;\;\;
\overline{h}_{\pm}\longrightarrow\overline{h}_{\pm}{\cal V}_i^{\dag},\;\;\;
{\rm with}\;\; {\cal V}_i
=\exp\Biggl[-\frac{i}{2}\phi_i\sum_{j,k}\epsilon_{ijk}\sigma_{jk}\Biggr],
\end{eqnarray}
where $\{i,j,k\} \in \{1,2,3\}$ and $\phi_i$ is a rotation parameter.
This symmetry comes from the spin-independence of the static quark interaction.

\subsubsection*{Spatial rotational symmetry}

In the static limit the 4-dimensional Lorentz symmetry
(hypercubic symmetry on the lattice) breaks to 3-dimensional rotational
symmetry.
For this reason we treat the temporal and spatial directions separately.

\subsubsection*{Discrete symmetries}

The static and domain-wall actions respect the discrete symmetries:
parity ${\cal P}$, time reversal ${\cal T}$ and charge conjugation ${\cal C}$.
The transformations are given by (we follow the definitions in the pedagogical
reference~\cite{Gupta:1997nd}.):
\begin{eqnarray}
{\cal P}:\;\;\;&&
q(\vec{x}, t)\longrightarrow\gamma_0q(-\vec{x}, t),\;\;\;
\overline{q}(\vec{x}, t)\longrightarrow\overline{q}(-\vec{x}, t)\gamma_0,
\nonumber\\
&&h_{\pm}(\vec{x}, t)\longrightarrow\pm h_{\pm}(-\vec{x}, t),\;\;\;
\overline{h}_{\pm}(\vec{x}, t)\longrightarrow
\pm\overline{h}_{\pm}(-\vec{x}, t),\\
&&D_0\longrightarrow D_0,\;\;\;
D_i\longrightarrow -D_i,\nonumber\\
{\cal T}:\;\;\;&&
q(\vec{x}, t)\longrightarrow\gamma_0\gamma_5q(\vec{x}, -t),\;\;\;
\overline{q}(\vec{x}, t)\longrightarrow
\overline{q}(\vec{x}, -t)\gamma_5\gamma_0,
\nonumber\\
&&h_{\pm}(\vec{x}, t)\longrightarrow\pm\gamma_5 h_{\mp}(\vec{x}, -t),\;\;\;
\overline{h}_{\pm}(\vec{x}, t)\longrightarrow
\pm\overline{h}_{\mp}(\vec{x}, -t)\gamma_5,\\
&&D_0\longrightarrow -D_0,\;\;\;
D_i\longrightarrow D_i,\nonumber
\end{eqnarray}
\begin{eqnarray}
{\cal C}:\;\;\;&&
q(\vec{x}, t)\longrightarrow{\cal C}\overline{q}(\vec{x}, t)^T,\;\;\;
\overline{q}(\vec{x}, t)\longrightarrow -q(\vec{x}, t)^T{\cal C}^{-1},
\nonumber\\
&&h_{\pm}(\vec{x}, t)\longrightarrow
{\cal C}\overline{h}_{\mp}(\vec{x}, t)^T,\;\;\;
\overline{h}_{\pm}(\vec{x}, t)\longrightarrow
-h_{\mp}(\vec{x}, t)^T{\cal C}^{-1},\\
&&D_{\mu}\longrightarrow D_{\mu}^{\ast},\nonumber\\
&&{\cal C}=\gamma_0\gamma_2,\;\;\;
{\cal C}\gamma_{\mu}{\cal C}^{-1}=-\gamma_{\mu}^T.\nonumber
\end{eqnarray}
Their combinations are also useful:
\begin{eqnarray}
{\cal P}\cdot{\cal T}:\;\;\;&&
q(\vec{x}, t)\longrightarrow \gamma_5q(-\vec{x}, -t),\;\;\;
\overline{h}_{\pm}(\vec{x}, t)\longrightarrow
\overline{h}_{\mp}(-\vec{x}, -t)\gamma_5,\\
&&D_{\mu}\longrightarrow -D_{\mu},\nonumber\\
{\cal T}\cdot{\cal C}:\;\;\;&&
q(\vec{x}, t)\longrightarrow
\gamma_0\gamma_5{\cal C}\overline{q}(\vec{x}, -t)^T,\;\;\;
\overline{h}_{\pm}(\vec{x}, t)\longrightarrow
\mp h_{\pm}(\vec{x}, -t)^T{\cal C}^{-1}\gamma_5,\\
&&D_0\longrightarrow-D_0^{\ast},\;\;\;D_i\longrightarrow D_i^{\ast}.\nonumber
\end{eqnarray}

\subsubsection{Quark bilinear operators}\label{SEC:symmetry_bilinear}

We introduce higher dimensional operators to improve $J_{\pm\Gamma}$
at $O(pa)$ and $O(ma)$:
\begin{eqnarray}
J_{\pm\Gamma D}=\overline{h}_{\pm}\Gamma
(\mbox{\boldmath$\gamma$}\cdot\overrightarrow{\mbox{\boldmath$D$}})q,
\;\;\;
J_{\pm\Gamma M}=m_q\overline{h}_{\pm}\Gamma q=m_qJ_{\pm\Gamma}^{(0)},
\label{EQ:bilinear_Oa_operators}
\end{eqnarray}
where we use spatial rotational symmetry and equations of motion
(following the on-shell improvement program) to reduce the number of operators.
In the following we see how the HQET operators transform under
the symmetries listed above in Sec.~\ref{SEC:symmetries_HL-system}.
\begin{list}{}{\leftmargin=+4mm}
\item[--] $\cal P$

The bilinear operators transform under parity as
\begin{eqnarray}
J_{\pm\Gamma}\xrightarrow{\cal P}GJ_{\pm\Gamma},\;\;\;
J_{\pm\Gamma D}\xrightarrow{\cal P}GJ_{\pm\Gamma D},\;\;\;
J_{\pm\Gamma M}\xrightarrow{\cal P}GJ_{\pm\Gamma M},
\end{eqnarray}
where $G$ is defined by $\gamma_0\Gamma\gamma_0=G\Gamma$ and
takes the values $+1$ or $-1$.

\item[--] HQS

Under the HQS transformation the bilinear operators form multiplets
classified by $G$.

\item[--] Chiral symmetry

Chiral transformations interchange multiplets classified by $G=\pm 1$.
The rotational direction in this transformation of
$J_{\pm\Gamma D}$ and $J_{\pm\Gamma M}$ is opposite to $J_{\pm\Gamma}$.

\item[--] ${\cal P}\cdot{\cal T}$

Under the ${\cal P}\cdot{\cal T}$ transformation,
\begin{eqnarray}
J_{\pm\Gamma}\xrightarrow{{\cal P}\cdot{\cal T}}J_{\mp\Gamma},\;\;\;
J_{\pm\Gamma D}\xrightarrow{{\cal P}\cdot{\cal T}}J_{\mp\Gamma D},\;\;\;
J_{\pm\Gamma M}\xrightarrow{{\cal P}\cdot{\cal T}}J_{\mp\Gamma M}.
\end{eqnarray}

\item[--] ${\cal T}\cdot{\cal C}$

Under the ${\cal T}\cdot{\cal C}$ transformation,
\begin{eqnarray}
J_{\pm\Gamma}\xrightarrow{{\cal T}\cdot{\cal C}}
-H_5K\left(J_{\pm\Gamma}\right)^{\dag},\;\;\;
J_{\pm\Gamma D}\xrightarrow{{\cal T}\cdot{\cal C}}
-H_5K\left(J_{\pm\Gamma D}\right)^{\dag},\;\;\;
J_{\pm\Gamma M}\xrightarrow{{\cal T}\cdot{\cal C}}
-H_5K\left(J_{\pm\Gamma M}\right)^{\dag},\nonumber\\
\end{eqnarray}
where $H_5$ and $K$ are defined by $\gamma_5\Gamma\gamma_5=H_5\Gamma$
and ${\cal C}^{-1}\Gamma{\cal C}=K\Gamma^{\ast}$ respectively.
\end{list}

\subsubsection{$\Delta B=2$ four-quark operator}

We introduce dimension seven operators to improve the $\Delta B=2$ operators.
Using the equations of motion, spatial rotational symmetry and
chiral symmetry, the independent operators can be reduced to
\begin{eqnarray}
Q_{\pm ND}=2[\overline{h}_{\pm}\gamma_{\mu}^R
(\mbox{\boldmath$\gamma$}\cdot\overrightarrow{\mbox{\boldmath$D$}})q]
[\overline{h}_{\mp}\gamma_{\mu}^Lq],\;
&&Q_{\pm ND}'=2[\overline{h}_{\pm}P_R
(\mbox{\boldmath$\gamma$}\cdot\overrightarrow{\mbox{\boldmath$D$}})q]
[\overline{h}_{\mp}P_Lq],\label{EQ:O_LD_O_SD}\\
Q_{\pm NM}=2m_q[\overline{h}_{\pm}\gamma_{\mu}^Rq]
[\overline{h}_{\mp}\gamma_{\mu}^Lq],\;
&&Q_{\pm NM}'=2m_q[\overline{h}_{\pm}P_Rq]
[\overline{h}_{\mp}P_Lq],\label{EQ:O_LM_O_SM}
\end{eqnarray}
where $Q_{\pm ND}$ and $Q_{\pm ND}'$ are $O(pa)$ operators,
$Q_{\pm NM}$ and $Q_{\pm NM}'$ are $O(ma)$ operators.
The factor $2$ in Eqs.~(\ref{EQ:O_LD_O_SD}) and (\ref{EQ:O_LM_O_SM})
is chosen for convenience.

\begin{list}{}{\leftmargin=+4mm}
\item[--] ${\cal P}\cdot{\cal T}$

Under the ${\cal P}\cdot{\cal T}$ transformation,
\begin{eqnarray}
&&O_L\xrightarrow{{\cal P}\cdot{\cal T}}O_L,\\
&&Q_{\pm ND}\xrightarrow{{\cal P}\cdot{\cal T}}Q_{\mp ND},\;
Q_{\pm ND}'\xrightarrow{{\cal P}\cdot{\cal T}}Q_{\mp ND}',\;
Q_{\pm NM}\xrightarrow{{\cal P}\cdot{\cal T}}Q_{\mp NM},\;
Q_{\pm NM}'\xrightarrow{{\cal P}\cdot{\cal T}}Q_{\mp NM}'.\nonumber
\end{eqnarray}
Thus the $O(pa)$ and $O(ma)$ operators are written as
\begin{eqnarray}
Q_{ND}=Q_{+ND}+Q_{-ND},\;
&&Q_{ND}'=Q_{+ND}'+Q_{-ND}',\label{EQ:four-quark_Oa_operators}\\
Q_{NM}=Q_{+NM}+Q_{-NM},\;
&&Q_{NM}'=Q_{+NM}'+Q_{-NM}'.\nonumber
\end{eqnarray}

\item[--] ${\cal T}\cdot{\cal C}$

Under the ${\cal T}\cdot{\cal C}$ transformation,
\begin{eqnarray}
O_L\xrightarrow{{\cal T}\cdot{\cal C}}\left(O_L\right)^{\dag},
\;\;\;
Q_{ND}\xrightarrow{{\cal T}\cdot{\cal C}}\left(Q_{ND}\right)^{\dag},
\;\;\;
Q_{NM}\xrightarrow{{\cal T}\cdot{\cal C}}\left(Q_{NM}\right)^{\dag},\\
O_S\xrightarrow{{\cal T}\cdot{\cal C}}\left(O_S\right)^{\dag},
\;\;\;
Q_{ND}'\xrightarrow{{\cal T}\cdot{\cal C}}\left(Q_{ND}'\right)^{\dag},
\;\;\;
Q_{NM}'\xrightarrow{{\cal T}\cdot{\cal C}}\left(Q_{NM}'\right)^{\dag}.
\nonumber
\end{eqnarray}

\item[--] HQS

We define the following bilinear operators:
\begin{eqnarray}
J_{\pm\mu}=\overline{h}_{\pm}\gamma_{\mu}^Lq,\;\;\;
K_{\pm\mu}=\overline{h}_{\pm}\gamma_{\mu}^Rq,
\end{eqnarray}
for convenience.
Under an HQS transformation along the $i$-th direction ($i$-HQS),
these operators transform as
\begin{eqnarray}
&&J_{\pm 0}\xrightarrow{i\rm -HQS}-J_{\pm i},\;\;\;
J_{\pm i}\xrightarrow{i\rm -HQS}J_{\pm 0},\;\;\;
J_{\pm j(\not=i)}\xrightarrow{i\rm -HQS}\epsilon_{ijk}J_{\pm k}.\nonumber\\
&&K_{\pm 0}\xrightarrow{i\rm -HQS}K_{\pm i},\;\;\;
K_{\pm i}\xrightarrow{i\rm -HQS}-K_{\pm 0},\;\;\;
K_{\pm j(\not=i)}\xrightarrow{i\rm -HQS}\epsilon_{ijk}K_{\pm k},\nonumber
\end{eqnarray}
where $X\xrightarrow{i\rm -HQS}Y$ means ${\cal V}_i(X)=1+\phi_iY+O(\phi_i^2)$.
Next, we consider the four-quark operators:
\begin{eqnarray}
L_{\alpha}\equiv
[\![\overline{h}\gamma_{\alpha}^Lq]\!][\![\overline{h}\gamma_{\alpha}^Lq]\!]
&=&J_{+\alpha}J_{-\alpha}+J_{-\alpha}J_{+\alpha},\nonumber\\
N_{\alpha}\equiv
[\![\overline{h}\gamma_{\alpha}^Rq]\!][\![\overline{h}\gamma_{\alpha}^Lq]\!]
&=&K_{+\alpha}J_{-\alpha}+K_{-\alpha}J_{+\alpha},\\
R_{\alpha}\equiv
[\![\overline{h}\gamma_{\alpha}^Rq]\!][\![\overline{h}\gamma_{\alpha}^Rq]\!]
&=&K_{+\alpha}K_{-\alpha}+K_{-\alpha}K_{+\alpha},\nonumber
\end{eqnarray}
where the free index $\alpha$ is not summed.
Under the $i$-HQS transformation, these operators transform as
\begin{eqnarray}
L_0\xrightarrow{i\rm -HQS}L_i,\;\;\;
&L_i\xrightarrow{i\rm -HQS}L_0,\;\;\;&
L_{j(\not=i)}\xrightarrow{i\rm -HQS}L_{k(\not=i,j)},\nonumber\\
N_0\xrightarrow{i\rm -HQS}-N_i,\;\;\;
&N_i\xrightarrow{i\rm -HQS}-N_0,\;\;\;&
N_{j(\not=i)}\xrightarrow{i\rm -HQS}N_{k(\not=i,j)},\\
R_0\xrightarrow{i\rm -HQS}R_i,\;\;\;
&R_i\xrightarrow{i\rm -HQS}R_0,\;\;\;&
R_{j(\not=i)}\xrightarrow{i\rm -HQS}R_{k(\not=i,j)}.\nonumber
\end{eqnarray}
Using this, we can define the $\sum$-HQS transformation as
\begin{eqnarray}
h_{\pm}\xrightarrow{\sum\rm -HQS}{\cal V}h_{\pm},\;\;\;
\overline{h}_{\pm}\xrightarrow{\sum\rm -HQS}\overline{h}_{\pm}{\cal V}^{\dag},
\;\;\;
{\rm with}\;\; {\cal V}=\prod_{i=1}^3{\cal V}_i,
\end{eqnarray}
such that
\begin{eqnarray}
L_0\xrightarrow{\sum\rm -HQS}\sum_iL_i,\;\;\;&&
L_i\xrightarrow{\sum\rm -HQS}L_0+\sum_{k\not=i}L_k,\nonumber\\
N_0\xrightarrow{\sum\rm -HQS}-\sum_iN_i,\;\;\;&&
N_i\xrightarrow{\sum\rm -HQS}-N_0+\sum_{k\not=i}N_k,\\
R_0\xrightarrow{\sum\rm -HQS}\sum_iR_i,\;\;\;&&
R_i\xrightarrow{\sum\rm -HQS}R_0+\sum_{k\not=i}R_k.
\nonumber
\end{eqnarray}
Under the $\sum$-HQS transformation,
\begin{eqnarray}
{\cal V}_L=
\begin{bmatrix}
L_0+\sum_i L_i \\ -3L_0+\sum_iL_i
\end{bmatrix},\;\;\;
{\cal V}_N=
\begin{bmatrix}
-N_0+\sum_iN_i \\ 3N_0+\sum_iN_i
\end{bmatrix},\;\;\;
{\cal V}_R=
\begin{bmatrix}
R_0+\sum_i R_i \\ -3R_0+\sum_iR_i
\end{bmatrix},
\end{eqnarray}
behave as eigenvectors:
\begin{eqnarray}
{\cal V}_{L,N,R}\xrightarrow{\sum\rm -HQS}
\begin{bmatrix}
3 & 0 \\ 0 & -1
\end{bmatrix}
{\cal V}_{L,N,R}.
\label{EQ:sigma_eigen}
\end{eqnarray}
Applying the transformation behavior (\ref{EQ:sigma_eigen}) and
imposing the chiral symmetry, we find $O(1)$, $O(pa)$ and $O(ma)$
eigenvectors which have the same symmetry transformation property as $O_L$:
\begin{eqnarray}
{\cal V}^{(1)}\equiv
\begin{bmatrix}
O_L \\ O_L+4O_S
\end{bmatrix},\;\;\;
{\cal V}^{(pa)}\equiv
\begin{bmatrix}
O_{ND} \\ O_{\overline{ND}}
\end{bmatrix},\;\;\;
{\cal V}^{(ma)}\equiv
\begin{bmatrix}
O_{NM} \\ O_{\overline{NM}}
\end{bmatrix},
\label{EQ:four-quark_operators_chiral}
\end{eqnarray}
where we define:
\begin{eqnarray}
O_{ND}&\equiv&Q_{ND}+2Q_{ND}'
=2[\overline{h}\gamma_{\mu}^R(\bm{\gamma}\cdot\overrightarrow{\bm{D}})q]
[\overline{h}\gamma_{\mu}^Lq]
+4[\overline{h}P_R(\bm{\gamma}\cdot\overrightarrow{\bm{D}})q]
[\overline{h}P_Lq],\\
O_{\overline{ND}}&\equiv&Q_{ND}-2Q_{ND}'
=2[\overline{h}\gamma_{\mu}^R(\bm{\gamma}\cdot\overrightarrow{\bm{D}})q]
[\overline{h}\gamma_{\mu}^Lq]
-4[\overline{h}P_R(\bm{\gamma}\cdot\overrightarrow{\bm{D}})q]
[\overline{h}P_Lq],\\
O_{NM}&\equiv&Q_{NM}+2Q_{NM}'
=m_q\left(2[\overline{h}\gamma_{\mu}^Rq][\overline{h}\gamma_{\mu}^Lq]
+4[\overline{h}P_Rq][\overline{h}P_Lq]\right),\\
O_{\overline{NM}}&\equiv&Q_{NM}-2Q_{NM}'
=m_q\left(2[\overline{h}\gamma_{\mu}^Rq][\overline{h}\gamma_{\mu}^Lq]
-4[\overline{h}P_Rq][\overline{h}P_Lq]\right).
\end{eqnarray}
When the chiral symmetry is not imposed, more operators arise besides
those in Eq.~(\ref{EQ:four-quark_operators_chiral}).
This is the case both for the Wilson-type quarks, also for the
domain-wall fermion with finite extent of the fifth dimension.
We will discuss this issue in Appendix~\ref{SEC:Wilson_quark}.
\end{list}

\subsection{Continuum matching: CQCD $\longleftrightarrow$ CHQET}
\label{EQ:continuum_matching}

The operator matching between QCD and HQET in the continuum is expressed by
\begin{eqnarray}
J_{\Gamma}^{\rm CQCD}(\mu_b)&=&
C_{\Gamma}(\mu_b,\mu)J_{\Gamma}^{\rm CHQET}(\mu)+O(\Lambda_{\rm QCD}/m_b),\\
O_L^{\rm CQCD}(\mu_b)&=&
Z_1(\mu_b,\mu)O_L^{\rm CHQET}(\mu)+Z_2(\mu_b,\mu)O_S^{\rm CHQET}(\mu)
+O(\Lambda_{\rm QCD}/m_b).
\end{eqnarray}
We employ one-loop perturbative matching.
For the bilinear operators, we quote Ref.~\cite{Eichten:1989zv}:
\begin{eqnarray}
C_{\Gamma}(\mu_b,\mu)
=1+\left(\frac{g}{4\pi}\right)^2\left[
C_F\left(\frac{H^2}{4}-\frac{5}{2}\right)\ln\left(\frac{\mu^2}{\mu_b^2}\right)
+A_{\Gamma}\right],
\label{EQ:bilinear_CQCD-CHQET}
\end{eqnarray}
with
\begin{eqnarray}
A_{\Gamma}=C_F
\left(D_{\rm QCD}-D_{\rm HQET}+\frac{E_{\rm QCD}-E_{\rm HQET}}{2}\right),\\
D_{\rm QCD}=-\frac{HG}{2}+\frac{3}{4}H^2-HH'-1,\;\;\;
E_{\rm QCD}=-4,\\
D_{\rm HQET}=1,\;\;\;\;\;
E_{\rm HQET}=0,\label{EQ:D_E_HQET}
\end{eqnarray}
where $C_F=(N_c^2-1)/(2N_c)$ ($N_c=3$ is the number of colors)
is a second Casimir, and we introduced $H$ defined by
$H\Gamma=\gamma_{\mu}\Gamma\gamma_{\mu}$ and $H'$ which is the
derivative of $H$ with respect to $d$ in $d$ dimensions.
For the four-quark operator, we quote Ref.~\cite{Flynn:1990qz, Buchalla:1996ys}:
\begin{eqnarray}
Z_1(\mu_b,\mu)
=1+\left(\frac{g}{4\pi}\right)^2\left[
-6\ln\left(\frac{\mu^2}{\mu_b^2}\right)+B_L\right],\;\;\;
Z_2(\mu_b,\mu)
=\left(\frac{g}{4\pi}\right)^2B_S,
\label{EQ:four-quark_CQCD-CHQET}
\end{eqnarray}
where
\begin{eqnarray}
B_L=-\frac{8N_c^2+9N_c-15}{2N_c},\;\;\;
B_S=-2(N_c+1).
\end{eqnarray}

To avoid large logarithms of $\mu/\mu_b$ in
Eqs.~(\ref{EQ:bilinear_CQCD-CHQET}) and (\ref{EQ:four-quark_CQCD-CHQET})
we match at scale $\mu=\mu_b=m_b$ and use renormalization group
running in the effective theory to reach a lower scale $\mu<m_b$:
\begin{eqnarray}
\mu^2\frac{d}{d\mu^2}C_{\Gamma}(\mu_b, \mu)&=&
\frac{1}{2}C_{\Gamma}(\mu_b, \mu)\gamma_{\Gamma},\\
\mu^2\frac{d}{d\mu^2}
\begin{bmatrix}
Z_1(\mu_b, \mu) & Z_2(\mu_b, \mu)
\end{bmatrix}
&=&
\frac{1}{2}
\begin{bmatrix}
Z_1(\mu_b, \mu) & Z_2(\mu_b, \mu)
\end{bmatrix}
\begin{bmatrix}
\gamma_{11} & \gamma_{12} \\
\gamma_{21} & \gamma_{22}
\end{bmatrix},
\label{EQ:RG_eq_Z12}
\end{eqnarray}
where $\gamma$'s denote the anomalous dimension of each operator.
Here HQS (\ref{EQ:sigma_eigen}) ensures that
\begin{eqnarray}
\gamma_{12}=0,\;\;\;
\gamma_{22}=\gamma_{11}+4\gamma_{21}.
\label{EQ:gamma_L}
\end{eqnarray}
Solving the RG-equations, we obtain the RG-evolution of the
renormalization factors:
\begin{eqnarray}
C_{\Gamma}(m_b, \mu)&=&
C_{\Gamma}(m_b, m_b)U_{\Gamma}(m_b, \mu),
\label{EQ:C_Gamma_using_RG}\\
\begin{bmatrix}
Z_1(m_b, \mu) & Z_2(m_b, \mu)
\end{bmatrix}
&=&
\begin{bmatrix}
Z_1(m_b, m_b) & Z_2(m_b, m_b)
\end{bmatrix}
\begin{bmatrix}
U_L^{(11)}(m_b, \mu) & U_L^{(12)}(m_b, \mu) \\
U_L^{(21)}(m_b, \mu) & U_L^{(22)}(m_b, \mu)
\end{bmatrix},\nonumber\\
\label{EQ:Z_12_using_RG}
\end{eqnarray}
where the factors $U(m_b, \mu)$'s represent the RG-evolution on the HQET side,
and the symmetry constraints (\ref{EQ:gamma_L}) give
\begin{eqnarray}
U_L^{(12)}(m_b, \mu)=0,\;\;\;
U_L^{(22)}(m_b, \mu)=U_L^{(11)}(m_b, \mu)+4U_L^{(21)}(m_b, \mu).
\end{eqnarray}
The one-loop matching requires a two-loop calculation of the anomalous dimension
\begin{eqnarray}
\gamma(\alpha_s)=\frac{\alpha_s}{4\pi}\gamma^{(0)}
+\left(\frac{\alpha_s}{4\pi}\right)^2\gamma^{(1)}+O(\alpha_s^3),
\end{eqnarray}
where $\alpha_s=g^2/(4\pi)$, and of the beta-function
\begin{eqnarray}
\beta(\alpha_s)&=&-\alpha_s\left\{\frac{\alpha_s}{4\pi}\beta_0
+\left(\frac{\alpha_s}{4\pi}\right)^2\beta_1+O(\alpha_s^3)\right\},
\end{eqnarray}
with
\begin{eqnarray}
\beta_0=\frac{11}{3}C_A-\frac{4}{3}T_FN_f,\;\;\;
\beta_1=\frac{34}{3}C_A^2-4\left(\frac{5}{3}C_A+C_F\right)T_FN_f,
\end{eqnarray}
where $C_A=N_c$ and $T_F=1/2$ are color factors for the fundamental
representation of $SU(N_c)$ and $N_f$ denotes number of flavors of
quarks with masses below $\mu$.
The two-loop anomalous dimension was calculated.
in Refs.~\cite{Ji:1991pr,Broadhurst:1991fz} for the bilinear and
in Refs.~\cite{Gimenez:1992is,Ciuchini:1996sr,Buchalla:1996ys} for the
four-quark operator.
Here we quote the results:
\begin{eqnarray}
U_{\Gamma}(\mu_b, \mu)&=&
\left(1+\frac{\alpha_s(\mu)-\alpha_s(\mu_b)}{4\pi}J_{\Gamma}\right)
\left[\frac{\alpha_s(\mu_b)}{\alpha_s(\mu)}\right]
^{d_{\Gamma}}+O(\alpha_s^2),\\
U_L^{(11)}(\mu_b, \mu)&=&
\left(1+\frac{\alpha_s(\mu)-\alpha_s(\mu_b)}{4\pi}J_1\right)
\left[\frac{\alpha_s(\mu_b)}{\alpha_s(\mu)}\right]
^{d_1}+O(\alpha_s^2),\\
U_L^{(21)}(\mu_b, \mu)&=&
-\frac{1}{4}
\left(\left[\frac{\alpha_s(\mu_b)}{\alpha_s(\mu)}\right]
^{d_1}
-\left[\frac{\alpha_s(\mu_b)}{\alpha_s(\mu)}\right]
^{d_2}\right)+O(\alpha_s),\\
U_L^{(22)}(\mu_b, \mu)&=&
\left[\frac{\alpha_s(\mu_b)}{\alpha_s(\mu)}\right]
^{d_2}+O(\alpha_s),
\end{eqnarray}
where
\begin{eqnarray}
d_{\Gamma}=\frac{\gamma_{\Gamma}^{(0)}}{2\beta_0},\;\;\;
d_1=\frac{\gamma_{11}^{(0)}}{2\beta_0},\;\;\;
d_2=\frac{\gamma_{22}^{(0)}}{2\beta_0},
\end{eqnarray}
\begin{eqnarray}
J_{\Gamma}=\frac{\gamma_{\Gamma}^{(0)}}{2\beta_0}
\left(
\frac{\beta_1}{\beta_0}-\frac{\gamma_{\Gamma}^{(1)}}{\gamma_{\Gamma}^{(0)}}
\right),\;\;\;
J_1=\frac{\gamma_{11}^{(0)}}{2\beta_0}
\left(\frac{\beta_1}{\beta_0}
-\frac{\gamma_{11}^{(1)}}{\gamma_{11}^{(0)}}\right),
\label{EQ:J_Gamma_J_1}
\end{eqnarray}
with
\begin{eqnarray}
\gamma_{\Gamma}^{(0)}&=&-3C_F,\\
\gamma_{\Gamma}^{(1)}&=&-C_F\left\{
\frac{49}{6}C_A-\frac{5}{2}C_F-\frac{10}{3}T_FN_f-4(C_A-4C_F)\zeta(2)
\right\},\\
\gamma_{11}^{(0)}&=&-6C_F,\;\;\;
\gamma_{22}^{(0)}=-6C_F+4\frac{N_c+1}{N_c},\\
\gamma_{11}^{(1)}&=&\frac{1}{6}\Biggl[
C_F\left\{126C_F-190C_A+56T_FN_f-48C_A\zeta(2)\right\}\\
&&+\frac{N_c-1}{2N_c}
\left\{-240C_F+104C_A+(24-16T_F)N_f+48(8C_F-C_A)\zeta(2)\right\}
\Biggr],\nonumber
\end{eqnarray}
where $\zeta(2)=\pi^2/6$.

We assume the lattice cutoff is higher than the charm quark mass $m_c$.
Since dynamical lattice QCD simulations are usually performed with up,
down and strange sea quarks, not including charm,
we employ a two step running~\cite{Albertus:2010nm}:
\begin{eqnarray}
U_{\Gamma}(m_b, a^{-1})&=&
U_{\Gamma}^{N_f=4}(m_b, m_c)U_{\Gamma}^{N_f=3}(m_c, a^{-1}),\\
\begin{bmatrix}
U_L^{(11)}(m_b, a^{-1}) & U_L^{(12)}(m_b, a^{-1}) \\
U_L^{(21)}(m_b, a^{-1}) & U_L^{(22)}(m_b, a^{-1})
\end{bmatrix}
&=&
\begin{bmatrix}
U_L^{(11)}(m_b, m_c) & U_L^{(12)}(m_b, m_c) \\
U_L^{(21)}(m_b, m_c) & U_L^{(22)}(m_b, m_c)
\end{bmatrix}^{N_f=4}
\nonumber\\
&&\times
\begin{bmatrix}
U_L^{(11)}(m_c, a^{-1}) & U_L^{(12)}(m_c, a^{-1}) \\
U_L^{(21)}(m_c, a^{-1}) & U_L^{(22)}(m_c, a^{-1})
\end{bmatrix}^{N_f=3},
\end{eqnarray}
in which numerical values of Eq.~(\ref{EQ:J_Gamma_J_1}) are given by
\begin{eqnarray}
J_{\Gamma}^{N_f=4}=0.910, && J_1^{N_f=4}=1.864,\\
J_{\Gamma}^{N_f=3}=0.755, && J_1^{N_f=3}=1.698.
\end{eqnarray}

\subsection{Static effective theory matching:
CHQET $\longleftrightarrow$ LHQET}
\label{EQ:HQET_matching}

In the matching between continuum and the lattice HQET we include $O(pa)$
and $O(ma)$ contributions, where $p$ and $m$ represent the light quark
external momenta and light quark mass respectively.
There is no $O(a)$ correction for the light-light quark system with
exact chiral symmetry, but this correction is not prohibited for
the static heavy-light case.

\subsubsection*{Quark bilinear}

To obtain the matching factors, we calculate the on-shell transition
amplitude of the operator $J_{\pm\Gamma}$ with initial state
$|{\rm i}\rangle$ and final state $|{\rm f}\rangle$ with external momenta,
as described in Sec.~\ref{SEC:matching_procedure}.
Taking into account the $O(a)$ error,
we define a column vector of the HQET operators
\begin{eqnarray}
{\cal J}_{\pm\Gamma}=
\begin{bmatrix}J_{\pm\Gamma} & aJ_{\pm\Gamma D} & aJ_{\pm\Gamma M}
\end{bmatrix}^T.
\end{eqnarray}
Using them, the transition amplitudes in the CHQET through the first order
in $O(pa,ma)$ are given by
\begin{eqnarray}
\langle{\rm f}|J_{\pm\Gamma}|{\rm i}\rangle_{\rm CHQET}
&=&
\begin{bmatrix}
{\cal A}^{(1)} &
G{\cal A}^{(pa)} &
G{\cal A}^{(ma)}
\end{bmatrix}_{\rm cont}
\llangle{\rm f}|{\cal J}_{\pm\Gamma}|{\rm i}\rrangle_{\rm CHQET}\nonumber\\
&\equiv&{\cal A}_{G,\;\rm cont}
\llangle{\rm f}|{\cal J}_{\pm\Gamma}|{\rm i}\rrangle_{\rm CHQET},
\label{EQ:amplitude_CHQET}
\end{eqnarray}
where $\llangle\cdots\rrangle$ denotes a tree-level amplitude and we
use the abbreviation $[A_X ~B_X ~\cdots]=[A ~B ~\cdots]_X$.
In this expression, ${\cal A}^{(1)}$, ${\cal A}^{(pa)}$
and ${\cal A}^{(ma)}$ are all real and independent of $\pm\Gamma$.
That fact and the existence of $G$ in Eq.~(\ref{EQ:amplitude_CHQET})
are read from the transformation properties in Sec.~\ref{SEC:symmetry_bilinear}.
The transition amplitudes for the LHQET are generically written as
\begin{eqnarray}
\langle{\rm f}|{\cal J}_{\pm\Gamma}|{\rm i}\rangle_{\rm LHQET}&=&
\begin{bmatrix}
 {\cal A}^{( 1,1)} & G{\cal A}^{( 1,pa)} & G{\cal A}^{( 1,ma)} \\
G{\cal A}^{(pa,1)} &  {\cal A}^{(pa,pa)} &  {\cal A}^{(pa,ma)}\\
G{\cal A}^{(ma,1)} &  {\cal A}^{(ma,pa)} &  {\cal A}^{(ma,ma)}
\end{bmatrix}_{\rm latt}
\llangle{\rm f}|{\cal J}_{\pm\Gamma}|{\rm i}\rrangle_{\rm LHQET}\nonumber\\
&\equiv&{\cal A}_{G,\;\rm latt}
\llangle{\rm f}|{\cal J}_{\pm\Gamma}|{\rm i}\rrangle_{\rm LHQET}.
\label{EQ:amplitude_LHQET}
\end{eqnarray}
Here we have introduced the operator mixing matrix elements
${\cal A}^{(pa,1)}$ and ${\cal A}^{(ma,1)}$ which represent $1/a$
power-divergent mixing with lower dimensional operators.
This possibility needs to be taken into account because the reduced
symmetry in lattice theories may not be sufficient to prohibit power
divergences~\cite{Sachrajda:1988ak}.
We will discuss this more in Sec.~\ref{SEC:power_divergence}.
Combining Eqs.~(\ref{EQ:amplitude_CHQET}) and (\ref{EQ:amplitude_LHQET}),
the matching between the continuum and the lattice HQET can be written as
\begin{eqnarray}
\langle{\rm f}|J_{\pm\Gamma}|{\rm i}\rangle_{\rm CHQET}
&=&
{\cal A}_{G,\;\rm cont}
{\rm diag}[{\cal S}(J_{\pm\Gamma}), {\cal S}(aJ_{\pm\Gamma D}),
{\cal S}(aJ_{\pm\Gamma M})]
\left({\cal A}_{G,\;\rm latt}\right)^{-1}
\langle{\rm f}|{\cal J}_{\pm\Gamma}|{\rm i}\rangle_{\rm LHQET}
\nonumber\\
&\equiv&
Z_{\Gamma}
\langle{\rm f}|J_{\pm\Gamma}|{\rm i}\rangle_{\rm LHQET}
+Z_{\Gamma D}G
\langle{\rm f}|aJ_{\pm\Gamma D}|{\rm i}\rangle_{\rm LHQET}
+Z_{\Gamma M}G
\langle{\rm f}|aJ_{\pm\Gamma M}|{\rm i}\rangle_{\rm LHQET},
\nonumber\\
\label{EQ:matching_bilinear_npr}
\end{eqnarray}
where ${\cal S}(X)=\llangle{\rm f}|X|{\rm i}\rrangle_{\rm CHQET}
\llangle{\rm f}|X|{\rm i}\rrangle_{\rm LHQET}^{-1}$,
which absorbs the difference in the tree-level amplitude between CHQET
and LHQET.

${\cal A}_G$ and ${\cal A}_G^{-1}$ can be expanded in
perturbation theory both in the continuum and on the lattice as
\begin{eqnarray}
{\cal A}_G=I+\left(\frac{g}{4\pi}\right)^2C_F\hat{\cal A}_G+O(g^4),\;\;\;
{\cal A}_G^{-1}=I-\left(\frac{g}{4\pi}\right)^2C_F\hat{\cal A}_G+O(g^4).
\end{eqnarray}
At one-loop it is sufficient to calculate 
\begin{eqnarray}
\langle{\rm f}|J_{\pm\Gamma}|{\rm i}\rangle_{\rm CHQET}&=&
\left\{1+\left(\frac{g}{4\pi}\right)^2C_F
\left(\hat{\cal A}_{\rm cont}^{(1)}-\hat{\cal A}_{\rm latt}^{(1,1)}\right)
\right\}
{\cal S}(J_{\pm\Gamma})
\langle{\rm f}|J_{\pm\Gamma}|{\rm i}\rangle_{\rm LHQET}\nonumber\\
&&\hspace*{-30mm}+\left(\frac{g}{4\pi}\right)^2C_FG
\left(\hat{\cal A}^{(pa)}_{\rm cont}-\hat{\cal A}^{(1,pa)}_{\rm latt}\right)
{\cal S}(aJ_{\pm\Gamma D})
\langle{\rm f}|aJ_{\pm\Gamma D}|{\rm i}\rangle_{\rm LHQET}\nonumber\\
&&\hspace*{-30mm}+\left(\frac{g}{4\pi}\right)^2C_FG
\left(\hat{\cal A}^{(ma)}_{\rm cont}-\hat{\cal A}^{(1,ma)}_{\rm latt}\right)
{\cal S}(aJ_{\pm\Gamma M})
\langle{\rm f}|aJ_{\pm\Gamma M}|{\rm i}\rangle_{\rm LHQET}+O(g^4).
\label{EQ:one-loop_trans_amp_bilinear}
\end{eqnarray}

\subsubsection*{Four-quark operator}

From Eq.~(\ref{EQ:sigma_eigen}) there are two distinct multiplets
associated with the $\Sigma$-HQS transformation.
Defining column vectors of the HQET operators:
\begin{eqnarray}
{\cal O}_1&=&
\begin{bmatrix}
\;\;\;\;\;\;\;\;\;\;\;
O_L & ~~aO_{ND} & ~~aO_{NM}
\end{bmatrix}^T,\\
{\cal O}_2&=&
\begin{bmatrix}
O_L+4O_S & ~~aO_{\overline{ND}} & ~~aO_{\overline{NM}}
\end{bmatrix}^T,
\end{eqnarray}
the transition amplitudes in the CHQET can be written as
\begin{eqnarray}
\langle{\rm f}|O_L|{\rm i}\rangle_{\rm CHQET}
&=&
\begin{bmatrix}
{\cal B}_1^{(1)} & {\cal B}_1^{(pa)} & {\cal B}_1^{(ma)}
\end{bmatrix}_{\rm cont}
\llangle{\rm f}|{\cal O}_1|{\rm i}\rrangle_{\rm CHQET}\nonumber\\
&\equiv&{\cal B}_{1,\;\rm cont}
\llangle{\rm f}|{\cal O}_1|{\rm i}\rrangle_{\rm CHQET},
\label{EQ:amplitude_four-quark1_CHQET}\\
\langle{\rm f}|O_L+4O_S|{\rm i}\rangle_{\rm CHQET}
&=&
\begin{bmatrix}
{\cal B}_2^{(1)} & {\cal B}_2^{(pa)} & {\cal B}_2^{(ma)}
\end{bmatrix}_{\rm cont}
\llangle{\rm f}|{\cal O}_2|{\rm i}\rrangle_{\rm CHQET}\nonumber\\
&\equiv&{\cal B}_{2,\;\rm cont}
\llangle{\rm f}|{\cal O}_2|{\rm i}\rrangle_{\rm CHQET}.
\label{EQ:amplitude_four-quark2_CHQET}
\end{eqnarray}
The transition amplitudes for the LHQET are generically
\begin{eqnarray}
\langle{\rm f}|{\cal O}_{\alpha}|{\rm i}\rangle_{\rm LHQET}&=&
\begin{bmatrix}
{\cal B}_{\alpha}^{( 1, 1)} & {\cal B}_{\alpha}^{( 1,pa)} &
{\cal B}_{\alpha}^{( 1,ma)} \\
{\cal B}_{\alpha}^{(pa, 1)} & {\cal B}_{\alpha}^{(pa,pa)} &
{\cal B}_{\alpha}^{(pa, ma)} \\
{\cal B}_{\alpha}^{(ma, 1)} & {\cal B}_{\alpha}^{(ma,pa)} &
{\cal B}_{\alpha}^{(ma,ma)}
\end{bmatrix}_{\rm latt}
\llangle{\rm f}|{\cal O}_{\alpha}|{\rm i}\rrangle_{\rm LHQET}\nonumber\\
&\equiv&{\cal B}_{\alpha,\;\rm latt}
\llangle{\rm f}|{\cal O}_{\alpha}|{\rm i}\rrangle_{\rm LHQET},
\label{EQ:amplitude_four-quark_LHQET}
\end{eqnarray}
where $\alpha$ takes on the values $1$ or $2$.
Using these relations, the matching between the continuum and the lattice
HQET can be written as
\begin{eqnarray}
&&\langle{\rm f}|O_L|{\rm i}\rangle_{\rm CHQET}
\label{EQ:matching_four-quark1_npr}\\
&=&
{\cal B}_{1,\;\rm cont}
{\rm diag}[{\cal S}(O_L), {\cal S}(aO_{ND}), {\cal S}(aO_{NM})]
\left({\cal B}_{1,\;\rm latt}\right)^{-1}
\langle{\rm f}|{\cal O}_1|{\rm i}\rangle_{\rm LHQET}\nonumber\\
&\equiv&
Z_L\langle{\rm f}|O_L|{\rm i}\rangle_{\rm LHQET}
+Z_{ND}\langle{\rm f}|aO_{ND}|{\rm i}\rangle_{\rm LHQET}
+Z_{NM}\langle{\rm f}|aO_{NM}|{\rm i}\rangle_{\rm LHQET},
\nonumber\\
&&\langle{\rm f}|\left(O_L+4O_S\right)|{\rm i}\rangle_{\rm CHQET}
\label{EQ:matching_four-quark2_npr}\\
&=&
{\cal B}_{2,\;\rm cont}
{\rm diag}[{\cal S}(O_L+4O_S), {\cal S}(aO_{\overline{ND}}),
{\cal S}(aO_{\overline{NM}})]
\left({\cal B}_{2,\;\rm latt}\right)^{-1}
\langle{\rm f}|{\cal O}_2|{\rm i}\rangle_{\rm LHQET}\nonumber\\
&\equiv&
Z_{L+4S}\langle{\rm f}|(O_L+4O_S)|{\rm i}\rangle_{\rm LHQET}
+Z_{\overline{ND}}\langle{\rm f}|aO_{\overline{ND}}|{\rm i}\rangle_{\rm LHQET}
+Z_{\overline{NM}}\langle{\rm f}|aO_{\overline{NM}}|{\rm i}\rangle_{\rm LHQET}.
\nonumber
\end{eqnarray}
In perturbation theory both in the continuum and on the lattice,
${\cal B}_{\alpha}$ and ${\cal B}_{\alpha}^{-1}$ are expressed as
\begin{eqnarray}
{\cal B}_{\alpha}
=I+\left(\frac{g}{4\pi}\right)^2\hat{\cal B}_{\alpha}+O(g^4),\;\;\;
{\cal B}_{\alpha}^{-1}
=I-\left(\frac{g}{4\pi}\right)^2\hat{\cal B}_{\alpha}+O(g^4),
\end{eqnarray}
and such that expression for the matching between CHQET and LHQET is:
\begin{eqnarray}
\langle{\rm f}|
\begin{bmatrix}
O_L \\ O_S
\end{bmatrix}
|{\rm i}\rangle_{\rm CHQET}
&=&\left\{I+\left(\frac{g}{4\pi}\right)^2\left(
\begin{bmatrix}
\hat{\cal B}_1 & 0 \\
\frac{\hat{\cal B}_2-\hat{\cal B}_1}{4} & \hat{\cal B}_2 \\ 
\end{bmatrix}_{\rm cont}^{(1)}
-
\begin{bmatrix}
\hat{\cal B}_1 & 0 \\
\frac{\hat{\cal B}_2-\hat{\cal B}_1}{4} & \hat{\cal B}_2 \\ 
\end{bmatrix}_{\rm latt}^{(1,1)}
\right)\right\}
\nonumber\\
&&\;\;\;\;\times
{\rm diag}[{\cal S}(O_L), {\cal S}(O_S)]
\langle{\rm f}|
\begin{bmatrix}
O_L \\ O_S
\end{bmatrix}
|{\rm i}\rangle_{\rm LHQET}\nonumber\\
&&+\left(\frac{g}{4\pi}\right)^2\left(
\begin{bmatrix}
{\cal B}_1 & 0 \\
-\frac{{\cal B}_1}{4} & \frac{{\cal B}_2}{4}
\end{bmatrix}_{\rm cont}^{(pa)}
-
\begin{bmatrix}
{\cal B}_1 & 0 \\
-\frac{{\cal B}_1}{4} & \frac{{\cal B}_2}{4}
\end{bmatrix}_{\rm latt}^{(1, pa)}
\right)\nonumber\\
&&\;\;\;\;\times
{\rm diag}[{\cal S}(aO_{ND}), {\cal S}(aO_{\overline{ND}})]
\langle{\rm f}|
\begin{bmatrix}
aO_{ND} \\ aO_{\overline{ND}}
\end{bmatrix}
|{\rm i}\rangle_{\rm LHQET}
\nonumber\\
&&+\left(\frac{g}{4\pi}\right)^2\left(
\begin{bmatrix}
{\cal B}_1 & 0 \\
-\frac{{\cal B}_1}{4} & \frac{{\cal B}_2}{4}
\end{bmatrix}_{\rm cont}^{(ma)}
-
\begin{bmatrix}
{\cal B}_1 & 0 \\
-\frac{{\cal B}_1}{4} & \frac{{\cal B}_2}{4}
\end{bmatrix}_{\rm latt}^{(1, ma)}
\right)
\nonumber\\
&&\;\;\;\;\times
{\rm diag}[{\cal S}(aO_{NM}), {\cal S}(aO_{\overline{NM}})]
\langle{\rm f}|
\begin{bmatrix}
aO_{NM} \\ aO_{\overline{NM}}
\end{bmatrix}
|{\rm i}\rangle_{\rm LHQET}
+O(g^4).\;\;\;\;\;\;\;
\label{EQ:one-loop_trans_amp_four-quark}
\end{eqnarray}


\section{One-loop perturbative lattice to continuum matching in HQET}
\label{SEC:one-loop_matching}

In the continuum perturbative calculations 
we use the $\overline{\rm MS}$ renormalization scheme.
Feynman gauge is employed and IR divergences are regulated
using a gluon mass $\lambda$ in the continuum and lattice calculations.
In this section we only collect the results.
The actual calculations are shown in Sec.~\ref{SEC:lattice_PT} and
Appendices~\ref{SEC:feynman_rules} and \ref{SEC:lattice_perturbation}.
For simplicity, the lattice spacing $a$ is set to be $1$ in the calculations.

\subsection{Quark bilinear operators}

We consider one-loop perturbative matching of HQET operators
between the continuum and the lattice.
The one-loop transition amplitudes (\ref{EQ:amplitude_CHQET}) and
(\ref{EQ:amplitude_LHQET}) for the HQET can be written as
\begin{eqnarray}
\langle{\rm f}|J_{\pm\Gamma}|{\rm i}\rangle_{\rm HQET}
&=&\left\{1+\left(\frac{g}{4\pi}\right)^2C_F\hat{\cal A}^{(1)}\right\}
\llangle{\rm f}|J_{\pm\Gamma}|{\rm i}\rrangle_{\rm HQET}\\
&&\hspace*{-10mm}
+\left(\frac{g}{4\pi}\right)^2C_FG\hat{\cal A}^{(pa)}
\llangle{\rm f}|aJ_{\pm\Gamma D}|{\rm i}\rrangle_{\rm HQET}
+\left(\frac{g}{4\pi}\right)^2C_FG\hat{\cal A}^{(ma)}
\llangle{\rm f}|aJ_{\pm\Gamma M}|{\rm i}\rrangle_{\rm HQET}.\nonumber
\end{eqnarray}
In this expression the continuum coefficients $\hat{\cal A}_{\rm HQET}^{(1)}$,
$\hat{\cal A}_{\rm HQET}^{(pa)}$ and $\hat{\cal A}_{\rm HQET}^{(ma)}$ are
\begin{eqnarray}
\hat{\cal A}_{\rm CHQET}^{(1)}&=&
\left(-\ln\frac{\lambda^2}{\mu^2}+D_{\rm HQET}\right)
+\frac{1}{2}\left(-2\ln\frac{\lambda^2}{\mu^2}+E_{\rm HQET}\right)
+\frac{1}{2}\left(\ln\frac{\lambda^2}{\mu^2}+F\right),\\
\hat{\cal A}_{\rm CHQET}^{(pa)}&=&
-\frac{8\pi}{3a\lambda},\\
\hat{\cal A}_{\rm CHQET}^{(ma)}&=&
-\frac{4\pi}{a\lambda},
\end{eqnarray}
where $D_{\rm HQET}$ and $E_{\rm HQET}$ are shown in Eq.~(\ref{EQ:D_E_HQET})
and
\begin{eqnarray}
F=\frac{1}{2}.\label{EQ:F}
\end{eqnarray}
The lattice coefficients are
\begin{eqnarray}
\hat{\cal A}_{\rm LHQET}^{(1)}&=&
\left(-\ln\left(a^2\lambda^2\right)+d^{(1)}\right)
+\frac{1}{2}\left(-2\ln\left(a^2\lambda^2\right)+e\right)
+\frac{1}{2}\left(\ln\left(a^2\lambda^2\right)+f\right),\\
\hat{\cal A}_{\rm LHQET}^{(pa)}&=&
-\frac{8\pi}{3a\lambda}+d^{(pa)},\\
\hat{\cal A}_{\rm LHQET}^{(ma)}&=&
-\frac{4\pi}{a\lambda}+d^{(ma)}.
\end{eqnarray}
Using them, the matching factors that appear in
Eq.~(\ref{EQ:one-loop_trans_amp_bilinear}) are
\begin{eqnarray}
\hat{\cal A}_{\rm CHQET}^{(1)}
-\hat{\cal A}_{\rm LHQET}^{(1)}
&=&\frac{3}{2}\ln\left(a^2\mu^2\right)
+D_{\rm HQET}-d^{(1)}+\frac{E_{\rm HQET}-e}{2}+\frac{F-f}{2},\\
\hat{\cal A}_{\rm CHQET}^{(pa)}
-\hat{\cal A}_{\rm LHQET}^{(pa)}
&=&-d^{(pa)},\\
\hat{\cal A}_{\rm CHQET}^{(ma)}
-\hat{\cal A}_{\rm LHQET}^{(ma)}
&=&-d^{(ma)}.
\end{eqnarray}
Infrared divergences cancel between CHQET and LHQET.
We also note that the tree-level amplitudes of CHQET and LHQET are not
generally the same.
The finite values $f$, $e$, $d^{(1)}$, $d^{(pa)}$ and $d^{(ma)}$ are
obtained from the following one-loop calculations.

\subsection*{Light quark propagator correction}

The light quark self-energy terms from the rising-sun (RS)
and tadpole (TP) diagrams are
\begin{eqnarray}
\Sigma_q^{\rm RS}(p)=
\parbox[b][20mm][b]{40mm}{
\includegraphics[scale=0.28, viewport = 0 0 380 160, clip]
                {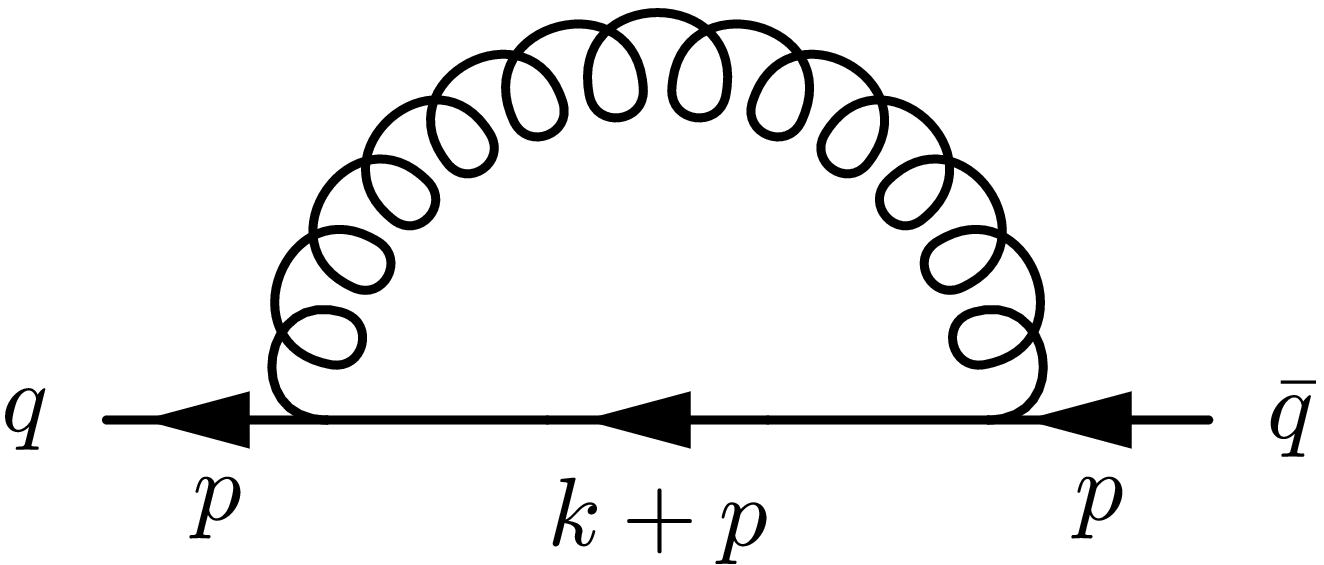}
		\vspace*{-3mm}},\;\;\;
\Sigma_q^{\rm TP}(p)=
\parbox[b][20mm][b]{40mm}{
\includegraphics[scale=0.28, viewport = 0 0 380 250, clip]
                {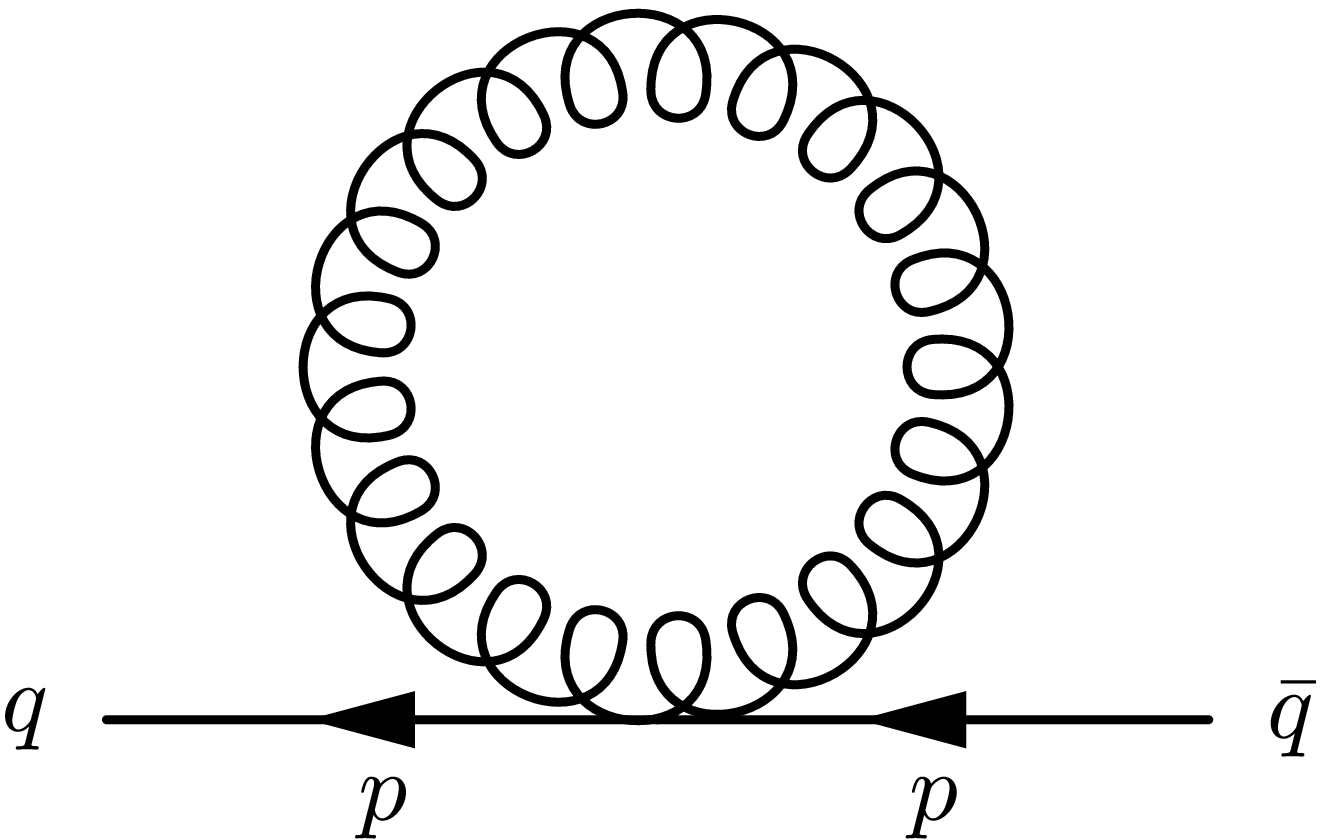}
		\vspace*{-3mm}}.
\label{EQ:static_self-energy}
\end{eqnarray}
The wave function renormalization is
\begin{eqnarray}
Z_q=1-\left.
\frac{\partial\left(\Sigma_q^{\rm RS}(p)+\Sigma_q^{\rm TP}(p)\right)}
{\partial (i{\not\!p})}\right|_{{\not p}=0}
=1+\left(\frac{g}{4\pi}\right)^2C_F\left[\ln\frac{\lambda^2}{\mu^2}+f\right].
\end{eqnarray}
For the chiral light quarks employed in this paper there are
no $O(a)$ contributions to this part.
The numerical value of $f$ for the domain-wall quarks is given in
Refs.~\cite{Aoki:1998vv,Aoki:2002iq}.
We cite their results and quote them in Tab.~\ref{TAB:f_v}.

\subsection*{Static heavy quark propagator correction}

The static self-energies from RS and TP contribution are represented by
\begin{eqnarray}
\Sigma_{h_{\pm}}^{\rm RS}(p_0)=
\parbox[b][20mm][b]{43mm}{
\includegraphics[scale=0.28, viewport = 0 0 420 160, clip]
                {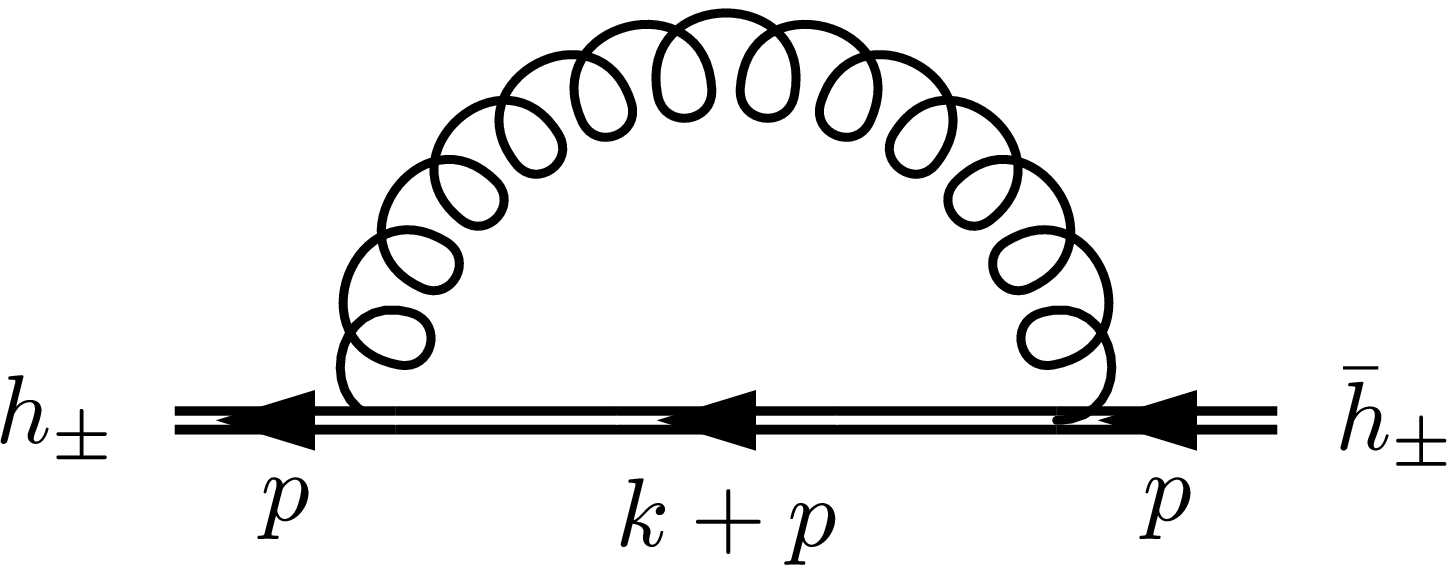}
		\vspace*{-3mm}},\;\;\;
\Sigma_{h_{\pm}}^{\rm TP}(p_0)=
\parbox[b][20mm][b]{43mm}{
\includegraphics[scale=0.28, viewport = 0 0 420 250, clip]
                {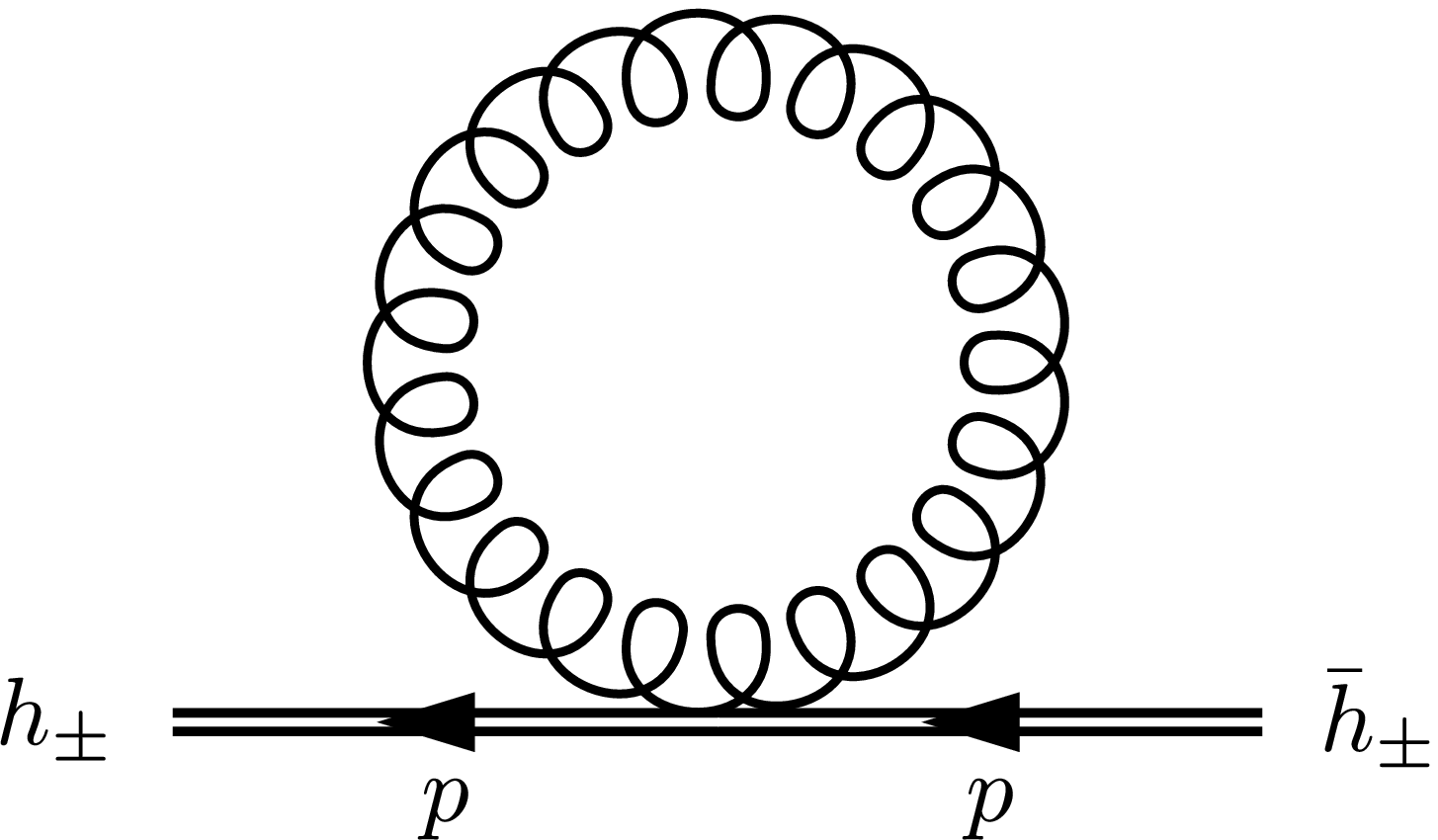}
		\vspace*{-3mm}},
\label{EQ:static_RS_TP}
\end{eqnarray}
and the radiative correction to the static mass is
\begin{eqnarray}
\delta M=
-\Sigma_{h_{\pm}}^{\rm RS}(p_0=0)-\Sigma_{h_{\pm}}^{\rm TP}(p_0=0)
\equiv\left(\frac{g}{4\pi}\right)^2C_F\delta\hat{M}.
\label{EQ:delta_M}
\end{eqnarray}
The wave function renormalization is
\begin{eqnarray}
Z_h=1-\left.
\frac{\partial
\left(\Sigma_{h_{\pm}}^{\rm RS}(p_0)+\Sigma_{h_{\pm}}^{\rm TP}(p_0)\right)}
{\partial e^{\mp ip_0}}\right|_{p_0=0}
=1+\left(\frac{g}{4\pi}\right)^2C_F\left[-2\ln(a^2\lambda^2)+e\right].
\label{EQ:Z_h}
\end{eqnarray}
As mentioned in Ref.~\cite{Borrelli:1992fy}, the static heavy quark propagator
has no $O(a)$ error after imposing the on-shell condition $p_0=0$.
The finite part $e$ can be decomposed as
\begin{eqnarray}
e={\cal R}+\delta\hat{M},\label{EQ:definition_e}
\end{eqnarray}
where $\cal R$ is defined in Eq.~(\ref{EQ:definition_cal-R}).

\subsection*{Heavy-light vertex correction}

The heavy-light vertex correction has an $O(a)$ part which is
determined by expanding in powers of the external momentum and the
light quark masses.
In this expansion, the equations of motion are used.
The expansion with respect to static quark external momentum always vanishes
owing to the on-shell condition, whereas that with respect to light quark
survives due to anisotropy between space and time in the system.
The expansion of the vertex correction on the lattice has the form:
\begin{eqnarray}
\delta\Gamma_{h_{\pm}q}(p;m_q)&=&
\parbox[b][15mm][b]{43mm}{
\includegraphics[scale=0.28, viewport = 0 0 400 190, clip]
                {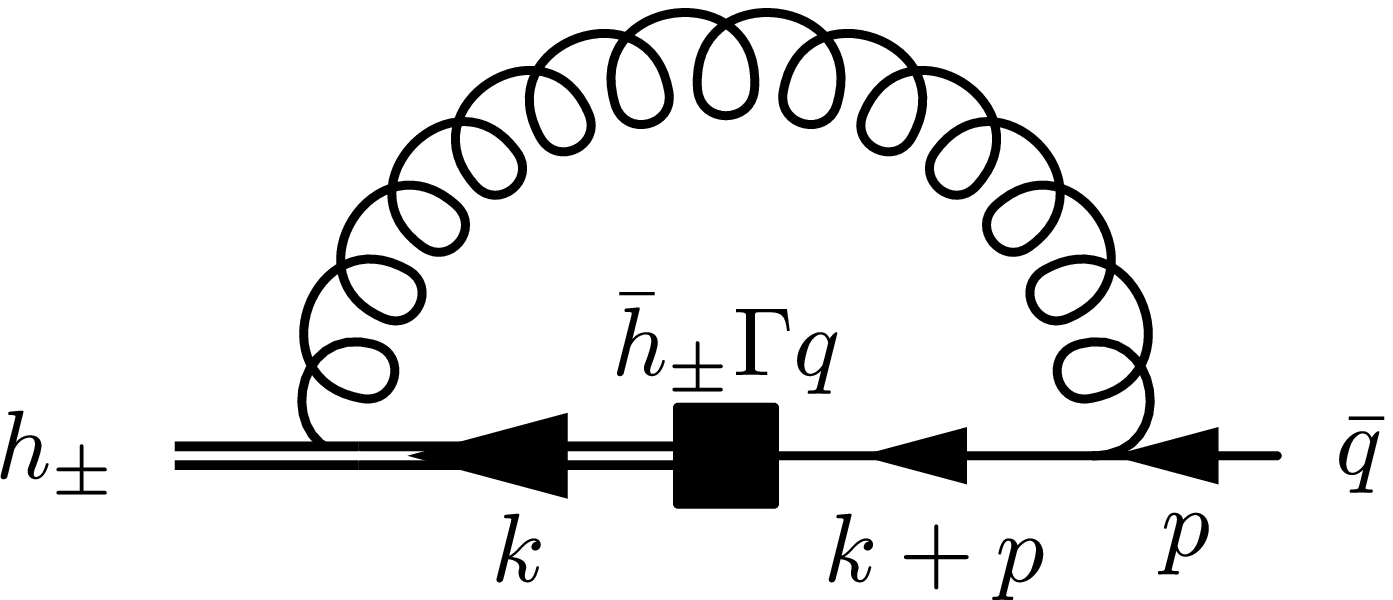}
		\vspace*{-3mm}}
\label{EQ:heavy-light_vertex_correction}\\
\nonumber\\
&=&\delta\Gamma_{hq}^{(1)}
+G\delta\Gamma_{hq}^{(pa)}\left(i\bm{\gamma}\cdot\bm{p}\right)
+G\delta\Gamma_{hq}^{(ma)}m_q+O(p^2, pm_q, m_q^2)\nonumber\\
&=&\left(\frac{g}{4\pi}\right)^2C_F\Gamma\biggl\{
\left[-\ln(a^2\lambda^2)+d^{(1)}\right]
+G\left[-\frac{8\pi}{3a\lambda}+d^{(pa)}\right]
\left(i\bm{\gamma}\cdot\bm{p}\right)\nonumber\\
&&\hspace*{+25mm}
+G\left[-\frac{4\pi}{a\lambda}+d^{(ma)}\right]m_q
+O(p^2, pm_q, m_q^2)\biggr\}+O(g^4).\nonumber
\end{eqnarray}

\subsection{$\Delta B=2$ four-quark operator}

The one-loop transition amplitudes (\ref{EQ:amplitude_four-quark1_CHQET}),
(\ref{EQ:amplitude_four-quark2_CHQET}) and
(\ref{EQ:amplitude_four-quark_LHQET}),
which are the relevant parts for the one-loop matching, are
\begin{eqnarray}
\langle{\rm f}|O_L|{\rm i}\rangle_{\rm HQET}&=&
\left\{1+\left(\frac{g}{4\pi}\right)^2\hat{{\cal B}}^{(1)}\right\}
\llangle{\rm f}|O_L|{\rm i}\rrangle_{\rm HQET}\\
&&+\left(\frac{g}{4\pi}\right)^2\hat{{\cal B}}^{(pa)}
\llangle{\rm f}|aO_{ND}|{\rm i}\rrangle_{\rm HQET}
+\left(\frac{g}{4\pi}\right)^2\hat{{\cal B}}^{(ma)}
\llangle{\rm f}|aO_{NM}|{\rm i}\rrangle_{\rm HQET}
+O(g^4),\nonumber\\
\langle{\rm f}|O_S|{\rm i}\rangle_{\rm HQET}&=&
\llangle{\rm f}|O_S|{\rm i}\rrangle_{\rm HQET}+O(g^2),
\end{eqnarray}
where $\hat{{\cal B}}^{(1)}=\hat{{\cal B}}_1^{(1)}$,
$\hat{{\cal B}}^{(pa)}=\hat{{\cal B}}_1^{(pa)}$ and
$\hat{{\cal B}}^{(ma)}=\hat{{\cal B}}_1^{(ma)}$, respectively.
Note that only the tree-level amplitude is needed for the amplitude of $O_S$,
since the coefficient $Z_2(\mu_b, \mu)$
in Eq.~(\ref{EQ:four-quark_CQCD-CHQET}) begins at $O(g^2)$.
In the expression above, the continuum values
$\hat{{\cal B}}^{(1)}$, $\hat{{\cal B}}^{(pa)}$ and $\hat{{\cal B}}^{(ma)}$ are
\begin{eqnarray}
\hat{{\cal B}}_{\rm CHQET}^{(1)}&=&2\left(C_F+\frac{N_c-1}{2N_c}\right)
\left(-\ln\frac{\lambda^2}{\mu^2}+D_{\rm HQET}\right)\\
&&+\frac{N_c-1}{2N_c}
\left(-2\ln\frac{\lambda^2}{\mu^2}+C_{\rm HQET}\right)
+\frac{N_c-1}{2N_c}
\left(4\ln\frac{\lambda^2}{\mu^2}+V\right)\nonumber\\
&&+C_F\left(-2\ln\frac{\lambda^2}{\mu^2}+E_{\rm HQET}\right)
+C_F\left(\ln\frac{\lambda^2}{\mu^2}+F\right),\nonumber\\
\hat{{\cal B}}_{\rm CHQET}^{(pa)}&=&
-\left(C_F+\frac{N_c+1}{2N_c}\right)\left(-\frac{8\pi}{3a\lambda}\right),\\
\hat{{\cal B}}_{\rm CHQET}^{(ma)}&=&
-\left(C_F+\frac{N_c+1}{2N_c}\right)\left(-\frac{4\pi}{a\lambda}\right),
\end{eqnarray}
where the values of $D_{\rm HQET}$ and $E_{\rm HQET}$
are shown in Eq.~(\ref{EQ:D_E_HQET}), $F$ in Eq.~(\ref{EQ:F}) and also
\begin{eqnarray}
C_{\rm HQET}=0,\;\;\;
V=-5.
\end{eqnarray}
The lattice counterparts are
\begin{eqnarray}
\hat{{\cal B}}_{\rm LHQET}^{(1)}&=&2\left(C_F+\frac{N_c-1}{2N_c}\right)
\left(-\ln\left(a^2\lambda^2\right)+d^{(1)}\right)\nonumber\\
&&+\frac{N_c-1}{2N_c}
\left(-2\ln\left(a^2\lambda^2\right)+c\right)
+\frac{N_c-1}{2N_c}
\left(4\ln\left(a^2\lambda^2\right)+v\right)\nonumber\\
&&+C_F\left(-2\ln\left(a^2\lambda^2\right)+e\right)
+C_F\left(\ln\left(a^2\lambda^2\right)+f\right),
\label{EQ:4quark_lattice_O_0}\\
\hat{{\cal B}}_{\rm LHQET}^{(pa)}&=&
-\left(C_F+\frac{N_c+1}{2N_c}\right)
\left(-\frac{8\pi}{3a\lambda}+d^{(pa)}\right),\\
\hat{{\cal B}}_{\rm LHQET}^{(ma)}&=&
-\left(C_F+\frac{N_c+1}{2N_c}\right)
\left(-\frac{4\pi}{a\lambda}+d^{(ma)}\right).
\end{eqnarray}
Using them, the matching factors that appear in
Eq.~(\ref{EQ:one-loop_trans_amp_four-quark}) are
\begin{eqnarray}
\hat{\cal B}_{\rm CHQET}^{(1)}-\hat{\cal B}_{\rm LHQET}^{(1)}
&=&
3C_F\ln\left(a^2\mu^2\right)
+2\left(C_F+\frac{N_c-1}{2N_c}\right)\left(D_{\rm HQET}-d^{(1)}\right)
\label{EQ:4quark_matching_O_0}\\
&&
+\frac{N_c-1}{2N_c}\left(C_{\rm HQET}-c+V_{\rm HQET}-v\right)
+C_F\left(E_{\rm HQET}-e+F-f\right),\nonumber
\end{eqnarray}
\begin{eqnarray}
\hat{\cal B}_{\rm CHQET}^{(pa)}-\hat{\cal B}_{\rm LHQET}^{(pa)}
&=&
\left(C_F+\frac{N_c+1}{2N_c}\right)d^{(pa)},\\
\hat{\cal B}_{\rm CHQET}^{(ma)}-\hat{\cal B}_{\rm LHQET}^{(ma)}
&=&
\left(C_F+\frac{N_c+1}{2N_c}\right)d^{(ma)}.
\end{eqnarray}
To obtain $c$ and $v$ in the expressions (\ref{EQ:4quark_lattice_O_0}) and
(\ref{EQ:4quark_matching_O_0}), we need the following loop calculations.

\subsection*{Light-light vertex correction}

For the $\Delta B=2$ four-quark operator, we need the light-light vertex
correction:
\begin{eqnarray}
\delta\Gamma_{qq}&=&
\parbox[b][15mm][b]{40mm}{
\includegraphics[scale=0.28, viewport = 0 0 380 180, clip]
                {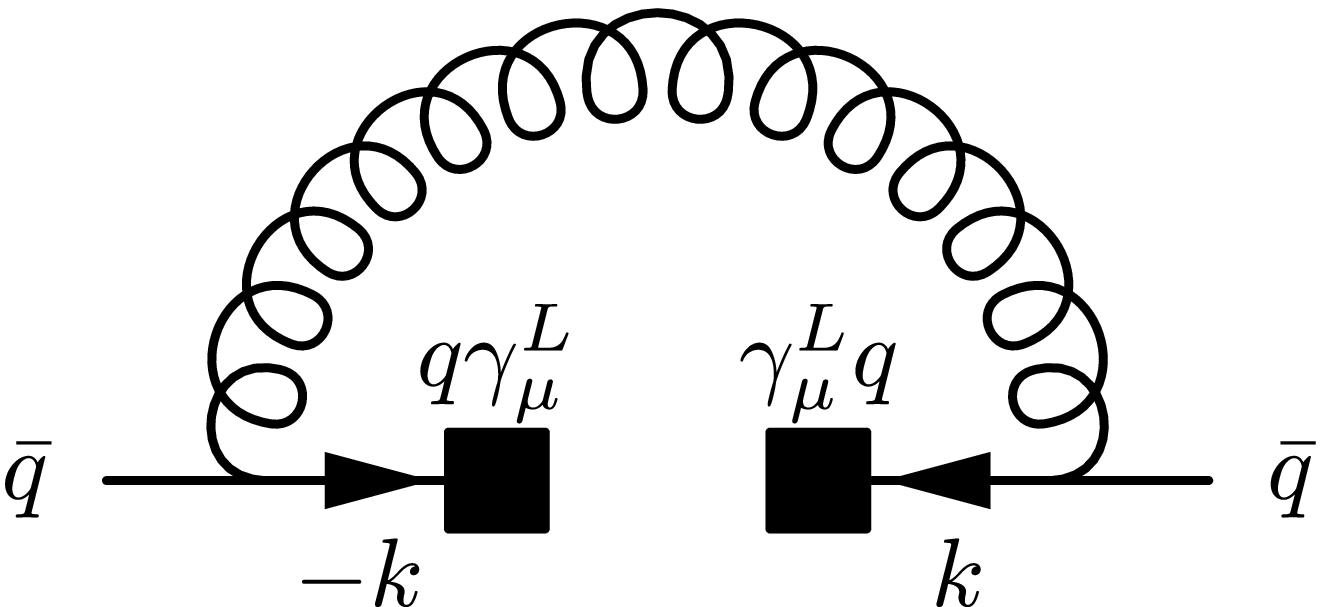}
		\vspace*{-3mm}}
=(T^A\gamma_{\mu}^L\otimes T^A\gamma_{\mu}^L)\left(\frac{g}{4\pi}\right)^2
\left[4\ln (a^2\lambda^2)+v\right].
\label{EQ:light-light_vertex}
\end{eqnarray}
For domain-wall quarks, $v$ in Eq.~(\ref{EQ:light-light_vertex}) was
calculated in Ref.~\cite{Aoki:2002iq,Aoki:1999ky}.
We quote those results in Tab.~\ref{TAB:f_v}.
For domain-wall quarks there is no $O(a)$ error in this part even
when the quarks are off-shell.

\subsection*{Heavy-heavy vertex correction}

For the four-quark operator we also need the heavy-heavy vertex;
\begin{eqnarray}
\delta\Gamma_{hh}&=&
\parbox[b][15mm][b]{43mm}{
\includegraphics[scale=0.28, viewport = 0 0 420 180, clip]
                {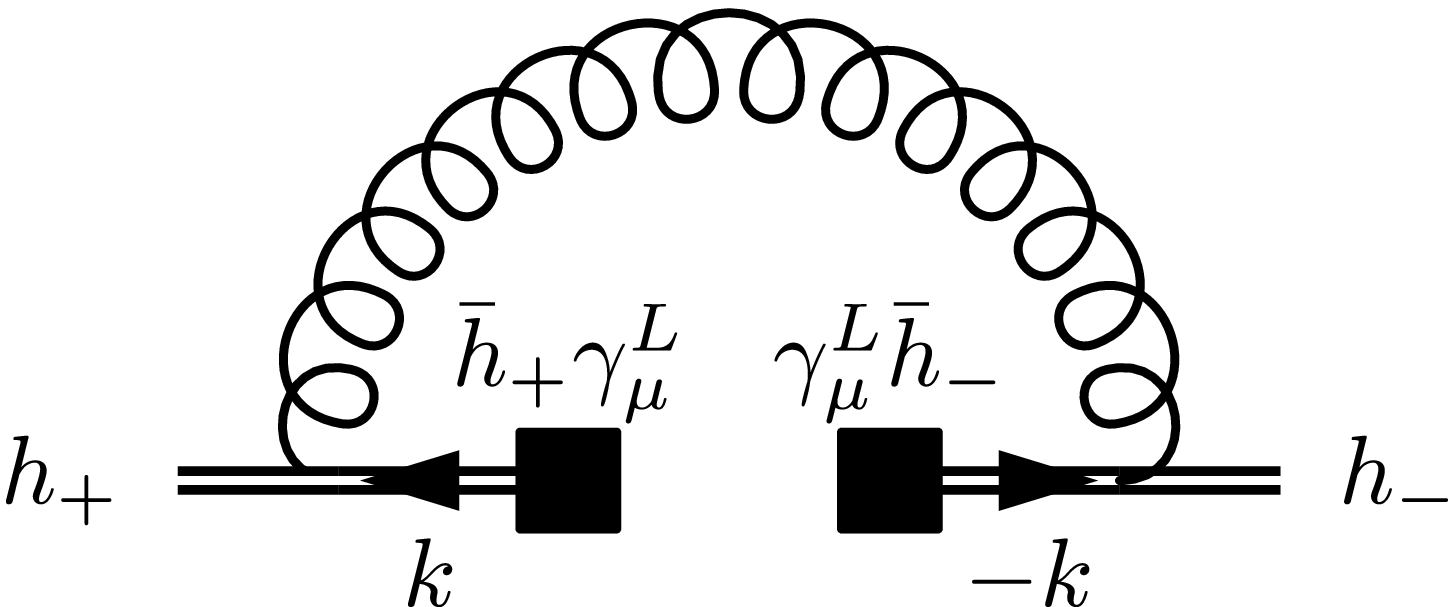}
		\vspace*{-3mm}}
=(T^A\gamma_{\mu}^L\otimes
T^A\gamma_{\mu}^L)\left(\frac{g}{4\pi}\right)^2
\left[-2\ln (a^2\lambda^2)+c\right].\;\;\;
\end{eqnarray}
The finite part $c$ is exactly
\begin{eqnarray}
c=e-\delta\hat{M}.
\end{eqnarray}
Note also that if we impose the on-shell condition, this part has
no $O(a)$ error.


\section{Lattice perturbation theory}
\label{SEC:lattice_PT}

The full one-loop lattice perturbation theory calculation is
presented in Appendices~\ref{SEC:feynman_rules}
and~\ref{SEC:lattice_perturbation}.
Here we make some general comments on the calculation.

\subsection{Lattice action}

The lattice action comprises three pieces:
\begin{eqnarray}
S=S_{\rm static}+S_{\rm DW}+S_{\rm gluon},
\end{eqnarray}
where $S_{\rm static}$ is the static quark action describing the heavy
($b$) quark, $S_{\rm DW}$ is the domain-wall fermion action describing
the light ($u$, $d$, $s$) quarks and $S_{\rm gluon}$ is the gluon action.
This action satisfies the symmetries presented
in Sec.~\ref{SEC:symmetries_HL-system}.

\subsubsection{Standard static action with link smearing}

The standard static quark action~\cite{Eichten:1989kb} is given by
\begin{eqnarray}
S_{\rm static}
&=&\sum_x\left\{\overline{h}_+(x)
\Bigl[h_+(x)-U_0^{\dag}(x-\hat{0})h_+(x-\hat{0})\Bigr]\right.\nonumber\\
&&\left.-\overline{h}_-(x)
\Bigl[U_0(x)h_-(x+\hat{0})-h_-(x)\Bigr]\right\}.
\label{EQ:static_action}
\end{eqnarray}
The lattice derivatives used here are not symmetric in order to avoid
fermion doublers.
We introduce link smearing to improve the signal to noise
ratio~\cite{Della Morte:2005yc}.
The modification is just to replace the link variable $U_0(x)$ with
a 3-step hypercubic blocked one $V_0(x)$, defined by
\begin{eqnarray}
V_{\mu}(x)&=&{\rm Proj}_{\rm SU(3)}\biggl[
(1-\alpha_1)U_{\mu}(x)
+\frac{\alpha_1}{6}\sum_{\pm\nu\not=\mu}
\widetilde{V}_{\nu;\mu}(x)\widetilde{V}_{\mu;\nu}(x+\hat{\nu})
\widetilde{V}_{\nu;\mu}^{\dag}(x+\hat{\mu})\Biggr],\\
\widetilde{V}_{\mu;\nu}(x)&=&{\rm Proj}_{\rm SU(3)}\biggl[
(1-\alpha_2)U_{\mu}(x)
+\frac{\alpha_2}{4}\sum_{\pm\rho\not=\nu,\mu}
\overline{V}_{\rho;\nu\mu}(x)\overline{V}_{\mu;\rho\nu}(x+\hat{\rho})
\overline{V}_{\rho;\nu\mu}^{\dag}(x+\hat{\mu})\Biggr],\;\;\;\;\;\;\\
\overline{V}_{\mu;\nu\rho}(x)&=&{\rm Proj}_{\rm SU(3)}\biggl[
(1-\alpha_3)U_{\mu}(x)
+\frac{\alpha_3}{2}\sum_{\pm\eta\not=\rho,\nu,\mu}
U_{\eta}(x)U_{\mu}(x+\hat{\eta})U_{\eta}^{\dag}(x+\hat{\mu})\Biggr],
\end{eqnarray}
where ${\rm Proj}_{\rm SU(3)}$ denotes a $SU(3)$ projection and
$(\alpha_1, \alpha_2, \alpha_3)$ are the hypercubic blocking
parameters ~\cite{Hasenfratz:2001hp}.
We use the parameter choices:
\begin{eqnarray}
(\alpha_1, \alpha_2, \alpha_3)=\begin{cases}
(0.0 , 0.0, 0.0) & \hspace*{5mm} :{\rm unsmeared}~(V_{\mu}=U_{\mu}) \\
(1.0 , 0.0, 0.0) & \hspace*{5mm} :{\rm APE~with}~\alpha=1
~\mbox{\cite{Albanese:1987ds}}   \\
(0.75, 0.6, 0.3) & \hspace*{5mm} :{\rm HYP1}
~\mbox{\cite{Hasenfratz:2001hp}} \\
(1.0 , 1.0, 0.5) & \hspace*{5mm} :{\rm HYP2}
~\mbox{\cite{Della Morte:2005yc}}.
                              \end{cases}
\end{eqnarray}

\subsubsection{Domain-wall fermion action}
\label{SEC:dwf_action}

The domain-wall fermion action with fermion mass parameter $m_f$ is
\footnote{The expression of the action we employ here is different from
that used in many simulations. See Appendix \ref{SEC:different_DWF_definition}.}
\begin{eqnarray}
S_{\rm DW}&=&\sum_{s, s'=1}^{L_s}\sum_{x, y}\overline{\psi}_{s}(x)
D_{ss'}^{\rm DW}(x, y)\psi_{s'}(y) + \sum_{x}m_f\overline{q}(x)q(x),\\
D_{ss'}^{\rm DW}(x, y)&=&D^4(x, y)\delta_{ss'}+D^5(s, s')\delta_{xy}
+(M_5-5)\delta_{ss'}\delta_{xy},\\
D^4(x, y)&=&\sum_{\mu}\frac{1}{2}
\left[(1+\gamma_{\mu})U_{\mu}(x)\delta_{x+\hat{\mu}, y}
+(1-\gamma_{\mu})U_{\mu}^{\dagger}(y)\delta_{x-\hat{\mu}, y}\right],\\
D^5(s,s')&=&\left\{
\begin{array}{ll}
 P_R\delta_{2, s'} & (s=1) \\
 P_R\delta_{s+1, s'} + P_L\delta_{s-1, s'} & (1<s<L_s) \\
 P_L\delta_{L_s-1, s'} & (s=L_s) \\
\end{array}
\right.,
\end{eqnarray}
where $\psi_s(x)$ is a $4{+}1$-dimensional fermion field.
The fifth dimension extends from $1$ to $L_s$ and is labeled by the
indices $s$ and $s'$.
The domain-wall height (fifth dimensional mass) $M_5$ is a parameter of
the theory which can be set between $0 < M_5 < 2$.
The physical four-dimensional quark field $q(x)$ is constructed
from the $4{+}1$-dimensional field $\psi_s(x)$ at $s=1$ and $L_s$:
\begin{eqnarray}
q(x)&=&P_R\psi_1(x)+P_L\psi_{L_s}(x),\\
\overline{q}(x)&=&\overline{\psi}_1(x)P_L+\overline{\psi}_{L_s}(x)P_R.
\end{eqnarray}
In our perturbative calculations we assume that $L_s$ is
infinity such that the right and left-handed modes are decoupled and
chiral symmetry is restored.
We follow the formalism developed in Refs.~\cite{Aoki:1998vv,Aoki:1997xg}.
The physical quark propagator is written as
\begin{eqnarray}
S_q(p)=\langle q(-p)\overline{q}(p)\rangle=
\frac{(1-w_0^2)}{i{\not\!p}+(1-w_0^2)m_f}
\left(1+O(p^2, pm_f, m_f^2)\right),
\label{EQ:quark_propagator}
\end{eqnarray}
where $w_0=1-M_5$.
From Eq.~(\ref{EQ:quark_propagator}) we see that the quark wave function
has a domain-wall specific factor $(1-w_0^2)^{1/2}$ and that the relation
between the quark mass $m_q$ and the domain-wall fermion mass parameter
$m_f$ is $m_q=(1-w_0^2)m_f$.
Hence the relationships between the lattice and continuum tree-level
amplitudes in Eqs.~(\ref{EQ:matching_bilinear_npr}),
(\ref{EQ:one-loop_trans_amp_bilinear}),
(\ref{EQ:matching_four-quark1_npr}), (\ref{EQ:matching_four-quark2_npr})
and (\ref{EQ:one-loop_trans_amp_four-quark}) are:
\begin{eqnarray}
&{\cal S}(J_{\pm\Gamma})={\cal S}(J_{\pm\Gamma D})
={\cal S}(J_{\pm\Gamma M})=(1-w_0^2)^{-1/2},&\\
&{\cal S}(O_L)={\cal S}(O_{ND})={\cal S}(O_{NM})={\cal S}(O_S)
={\cal S}(O_{\overline{ND}})={\cal S}(O_{\overline{NM}})
=(1-w_0^2)^{-1}.&\nonumber
\end{eqnarray}

As is discussed in Refs.~\cite{Aoki:1998vv, Blum:1999xi},
the domain-wall height $M_5$ receives an additive quantum correction.
This leads to a renormalization of the factor $(1-w_0^2)$,
which we write as $Z_w$.
The one-loop expression for $Z_w$ is
\begin{eqnarray}
Z_w=1+\frac{g^2}{(4\pi)^2}C_F\frac{2w_0}{1-w_0^2}\Sigma_w
\equiv 1+\frac{g^2}{(4\pi)^2}C_F\hat{z}_w,
\label{EQ:Z_w}
\end{eqnarray}
where values for $\Sigma_w$ are listed in Ref.~\cite{Aoki:2002iq}
and quoted in Tab.~\ref{TAB:Sigma_w}.
This domain-wall specific factor can be treated as a part of the wave
function renormalization.
Note that the domain-wall fermion is automatically off-shell $O(a)$
improved~\cite{Noaki:1999ij}.

\subsubsection{Gluon action}
\label{subsec:gluon} 

We consider a class of RG-improved gluon actions:
\begin{equation}
S_{\rm gluon}=-\frac{2}{g^2_0}\left( (1-8c_1)\sum_P {\rm ReTr}[U_{P}]
+c_1\sum_R {\rm ReTr}[U_R]\right) , 
\end{equation}
where $g_{0}$ denotes the bare lattice coupling, $U_P$ is the path-ordered
product of links around the $1\times 1$ plaquette $P$ and $U_R$ is the
path-ordered product of links around the $1\times 2$ rectangle $R$.
We will present numerical results for the following choices of $c_1$:
\begin{eqnarray}
c_1=
\begin{cases}
0        & \mbox{(standard plaquette action)}\\
-1/12    & \mbox{(Symanzik action)~\cite{Luscher:1984xn,Weisz:1983bn}}\\
-0.331   & \mbox{(Iwasaki action)~\cite{Iwasaki:1985we,Iwasaki:1983ck}}\\
-1.40686 & \mbox{(DBW2 action)~\cite{deForcrand:1999bi}}
\end{cases}.
\end{eqnarray}

\subsection{Mean field improvement}\label{SEC:MF_impr}

Mean field (MF) improvement is important for precise estimation in
lattice perturbation theory~\cite{Lepage:1992xa}.
The lattice action is improved by replacing the gluon links $U_{\mu}$
with $U_{\mu}/u_0$, where $u_0$ represents the MF value of $U_{\mu}$.
For the value of $u_0$ we choose the fourth root of the expectation value of
the plaquette $P$ taken from the simulation and to accomplish
MF improvement we also need the one-loop perturbative expansion of $u_0$:
\begin{eqnarray}
u_0=P^{1/4}=1-g^2C_F\frac{T_{\rm MF}}{2},
\label{EQ:PT_u0}
\end{eqnarray}
where the MF factor $T_{\rm MF}$ takes the values~\cite{Aoki:2002iq}
\begin{eqnarray}
T_{\rm MF}=
\begin{cases}
1/8       & \mbox{(standard plaquette action)} \\
0.0915657 & \mbox{(Symanzik action)} \\
0.0525664 & \mbox{(Iwasaki action)} \\
0.0191580 & \mbox{(DBW2 action)}
\end{cases}.
\end{eqnarray}

First we present MF improvement for the light quark sector.
For the domain-wall height $M_5$ we have the following
replacements~\cite{Aoki:2002iq}:
\begin{eqnarray}
M_5
&\longrightarrow&M_5^{\rm MF}=M_5-4(1-u_0),\label{EQ:MFimp-M5}\\
w_0(M_5)
&\longrightarrow& w_0^{\rm MF}=w_0(M_5^{\rm MF})=w_0(M_5)+4(1-u_0),\\
\hat{z}_w(M_5)
&\longrightarrow& \hat{z}_w^{\rm MF}(M_5)
=\hat{z}_w(M_5^{\rm MF})
+\frac{4w_0^{\rm MF}}{1-(w_0^{\rm MF})^2}(4\pi)^2T_{\rm MF}.
\label{EQ:MFimp-hat_zw}
\end{eqnarray}
Replacements are also required as in the usual (Wilson quark) treatment:
\begin{eqnarray}
f(M_5)
&\longrightarrow& f^{\rm MF}(M_5)=f(M_5^{\rm MF})-(4\pi)^2\frac{T_{\rm MF}}{2},
\\
v(M_5)
&\longrightarrow& v^{\rm MF}(M_5)=v(M_5^{\rm MF}),\\
q
&\longrightarrow& u_0^{1/2}q,\\
m_q=(1-w_0^2)m_f
&\longrightarrow&
m_q^{\rm MF}=\frac{1}{u_0}(1-(w_0^{\rm MF})^2)m_f,\\
U_{\mu}
&\longrightarrow& \frac{U_{\mu}}{u_0}.
\end{eqnarray}

MF improvement for static quarks was first discussed
in Ref.~\cite{Hernandez:1994bx} and is summarized here:
\begin{list}{}{}
\item[(1)]
As mentioned in Ref.~\cite{Eichten:1989kb}, the power divergent piece in
the static heavy quark self-energy causes a change in
the fitting function of the $B$ meson correlator.
Here, let us illustrate this issue using the static quark propagator.
The propagator is given by Eq.~(\ref{EQ:static_propagator}) and its inverse is
\begin{eqnarray}
(S_h(p))^{-1}=1-e^{-ip_0},
\end{eqnarray}
where only the $h_+$ part is taken.
The renormalization of the static propagator is calculated from
the static self-energy $\Sigma_h(p_0)$ (\ref{EQ:static_self-energy})
and the inverse bare propagator is written as
\begin{eqnarray}
(S_h(p))^{-1}\sim Z_h^{-1}(1-e^{-ip_0}+\delta M)\;\;\;\;\;\;
(p_0\ll 1),
\end{eqnarray}
where $Z_h$ and $\delta M$ at one-loop are given by Eqs.~(\ref{EQ:Z_h})
and (\ref{EQ:delta_M}), respectively.
Then, the coordinate space static quark propagator is
\begin{eqnarray}
S_h(t)\sim Z_he^{-\ln(1+\delta M)(t+1)}\;\;\;\;\;\;
(t\gg 1).
\label{EQ:static-quark_propagator_real-space}
\end{eqnarray}
Here we consider two types of fit function for the propagator
(\ref{EQ:static-quark_propagator_real-space}):
\begin{eqnarray}
F_1(t)&=&Ae^{-B(t+1)},\label{EQ:fit_t+1}
\\
F_2(t)&=&Ae^{-Bt}.\label{EQ:fit_t}
\end{eqnarray}
Although the difference between them naively seems to be an
irrelevant $O(a)$ effect in the continuum limit, it requires
special attention owing to the $1/a$ divergence in $\delta M$.
When we choose the fit function (\ref{EQ:fit_t}),
Eq.~(\ref{EQ:static-quark_propagator_real-space}) is rewritten as
\begin{eqnarray}
S_h(t)=\frac{Z_h}{1+\delta M}e^{-\ln(1+\delta M)t}
=Z_h'e^{-\ln(1+\delta M)t},
\end{eqnarray}
where $Z_h'=Z_h/(1+\delta M)$, which leads to a ``reduced value''
for $e$~(\ref{EQ:definition_e}):
\begin{eqnarray}
e_R=e-\delta\hat{M}. \label{EQ:reduced_value}
\end{eqnarray}
We use the fit function (\ref{EQ:fit_t}) with the reduced value $e_R$
rather than (\ref{EQ:fit_t+1}) to avoid an unwanted $O(a)$ error in the
light quark propagator which obeys the function (\ref{EQ:fit_t}).

\item[(2)]
The link variables in the static quark action are smeared.
One way to implement MF improvement for them is to replace the smeared
gauge links $V_0$ with $V_0/v_0$, where $v_0$ represents
an MF value of $V_0$~\cite{Horsley:2008ap}.
Associated with this, the one-loop perturbative expansion of $v_0$ is
introduced using the expectation value of the plaquette with the link
smearing $P_{\rm smeared}$:
\begin{eqnarray}
v_0=(P_{\rm smeared})^{1/4}
=1-g^2C_F\frac{T_{\rm MF}^{\rm smeared}}{2}.
\end{eqnarray}

\item[(3)]
In coordinate space, the static quark propagator $S_h(t)$
is related to the MF-improved propagator $\widetilde{S}_h(t)$ by
\begin{eqnarray}
S_h(t)=v_0^t\widetilde{S}_h(t),
\label{EQ:static_prop_relation}
\end{eqnarray}
since each link variable in the propagator is multiplied by
$1/v_0$ in MF-improved case.
The relation~(\ref{EQ:static_prop_relation}) is satisfied nonperturbatively
but we show the perturbative derivation for pedagogical reasons.
Starting from the static quark action without MF improvement,
the action with MF improvement is obtained by rescaling the static quark
field by $v_0^{1/2}$ and adding a static quark mass $M_0=(1/v_0-1)$.
The interesting feature of the static quark is that the mass can be
absorbed into a shift of the external momentum $p$ in the propagator,
whose momentum is $k+p$ ($k$ is the loop momentum),
\begin{eqnarray}
\left(S_{\pm h}\left(k+p; M_0\right)\right)^{-1}
&=&1-e^{\mp i(k_0+p_0)}+M_0\nonumber\\
&=&(1+M_0)\left(S_{\pm h}\left(k+p'; M_0=0\right)\right)^{-1}
\end{eqnarray}
where $p_0'=p_0\mp i\ln (1+M_0)$, a shifted external momentum.
Perturbative calculations are not affected by this mass shift but the
position of the pole in the propagator is changed.
The value of $e$ is not altered but the static quark mass correction changes:
\begin{eqnarray}
\delta M(M_0)=(1+M_0)\delta M(M_0=0).
\end{eqnarray}
Thus we find the same result as Eq.~(\ref{EQ:static_prop_relation}):
\begin{eqnarray}
S_h(t)\sim
\frac{1}{v_0}\widetilde{Z}_he^{-\ln\left(1+\frac{1}{v_0}\delta\widetilde{M}
+\left(\frac{1}{v_0}-1\right)\right)(t+1)}
=v_0^t\widetilde{S}_h(t).
\end{eqnarray} 
\end{list}
Combining the results above, we relate the coordinate space quark
propagator to the MF-improved one by
\begin{eqnarray}
Z_h'e^{-\ln\left(1+\delta M\right)t}
=\widetilde{Z}_h'
e^{-\ln\frac{1}{v_0}\left(1+\delta\widetilde{M}\right)t},
\end{eqnarray}
in which $\widetilde{Z}_h$ and $\delta\widetilde{M}$ are estimated using
the MF improved coupling $g_{\rm MF}$.
The replacements needed for MF improvement are
\begin{eqnarray}
c
&\longrightarrow& c^{\rm MF}=c,
\label{EQ:MFimp-c}\\
d^{(1,pa,ma)}(M_5)
&\longrightarrow& d^{(1,pa,ma)\rm MF}(M_5)=d^{(1,pa,ma)}(M_5^{\rm MF}),\\
e_R
&\longrightarrow& e_R^{\rm MF}=e_R,\\
h
&\longrightarrow& h,\label{EQ:MFimp-h}\\
\delta M
&\longrightarrow& \delta M^{\rm MF}=
\frac{1}{v_0}\left\{1-v_0+\frac{g^2}{(4\pi)^2}C_F
\left(\delta\hat{M}-(4\pi)^2\frac{T_{\rm MF}^{\rm smeared}}{2}\right)\right\}.
\end{eqnarray}

At one-loop in MF-improved perturbation theory,
the renormalized coupling in the continuum $\overline{\rm MS}$ scheme
$g_{\overline{\rm MS}}(\mu)$ at scale $\mu$ and the bare
lattice coupling $g_0$ are related by
\begin{eqnarray}
\frac{1}{g_{\overline{\rm MS}}^2(\mu)}=
\frac{P}{g_0^2}+d_g+c_p+\frac{22}{16\pi^2}\ln(\mu a)
+N_f\left(d_f-\frac{4}{48\pi^2}\ln(\mu a)\right),
\label{EQ:g_MSbar}
\end{eqnarray}
where $d_g$ and $c_p$ are dependent only on the gluon action and
$d_f$ is dependent only on the fermion action.
$c_p$ is obtained from
\begin{eqnarray}
c_p=2C_FT_{\rm MF},
\end{eqnarray}
and the value of $d_g$ is given by~\cite{Hasenfratz:1980kn, Hasenfratz:1981tw,
Weisz:1982zw, Weisz:1983bn, Iwasaki:1983zm, Iwasaki:1984cj, Sakai:2000jm}:
\begin{eqnarray}
d_g=
\begin{cases}
-0.4682      & \mbox{(standard plaquette action)} \\
-0.2361      & \mbox{(Symanzik action)} \\
\;\;\:0.1053 & \mbox{(Iwasaki action)} \\
\;\;\:0.5317 & \mbox{(DBW2 action)}.
\end{cases}
\end{eqnarray}
Values of $d_f$ for domain-wall fermions are given in
Ref.~\cite{Aoki:2003uf} and quoted in Tab.~\ref{TAB:df}.

Collecting the pieces needed for the MF improvement,
we now have final expressions:
\begin{eqnarray}
J_{\pm\Gamma}^{\rm CHQET}
&=&(1-(w_0^{\rm MF})^2)^{-1/2}(Z_w^{\rm MF})^{-1/2}u_0^{1/2}
\label{EQ:matching_bilinear}\\
&&
\times\Biggl[Z_{\Gamma (R)}^{(1)\rm MF}
J_{\pm\Gamma}^{\rm LHQET}
+\frac{Z_{\Gamma}^{(pa)\rm MF}}{u_0}
GaJ_{\pm\Gamma D}^{\rm LHQET}
+\frac{Z_{\Gamma}^{(ma)\rm MF}}{u_0}
GaJ_{\pm\Gamma M}^{\rm LHQET}
\Biggr]\nonumber\\
&\equiv&
{\cal Z}_{\Gamma}\Biggl[
J_{\pm\Gamma}^{\rm LHQET}
+c_{\Gamma}^{(pa)}aJ_{\pm\Gamma D}^{\rm LHQET}
+c_{\Gamma}^{(ma)}a\widetilde{J}_{\pm\Gamma M}^{\rm LHQET}
\Biggr],\nonumber\\
O_L^{\rm CHQET}
&=&(1-(w_0^{\rm MF})^2)^{-1}(Z_w^{\rm MF})^{-1}u_0
\label{EQ:matching_four-quark_L}\\
&&\times\Biggl[Z_{L (R)}^{(1)\rm MF}O_L^{\rm LHQET}
+\frac{Z_L^{(pa)\rm MF}}{u_0}aO_{ND}^{\rm LHQET}
+\frac{Z_L^{(ma)\rm MF}}{u_0}aO_{NM}^{\rm LHQET}
\Biggr]\nonumber\\
&\equiv&
{\cal Z}_L\Biggl[O_L^{\rm LHQET}
+c_L^{(pa)}aO_{ND}^{\rm LHQET}
+c_L^{(ma)}a\widetilde{O}_{NM}^{\rm LHQET}
\Biggr],\nonumber\\
O_S^{\rm CHQET}
&=&(1-(w_0^{\rm MF})^2)^{-1}(Z_w^{\rm MF})^{-1}u_0
O_S^{\rm LHQET}\label{EQ:matching_four-quark_S}\\
&\equiv&
{\cal Z}_SO_S^{\rm LHQET},\nonumber
\end{eqnarray}
where $\widetilde{J}_{\pm\Gamma M}^{\rm LHQET}$ and
$\widetilde{O}_{NM}^{\rm LHQET}$ are
introduced by
\begin{eqnarray}
\frac{J_{\pm\Gamma M}^{\rm LHQET}}{\widetilde{J}_{\pm\Gamma M}^{\rm LHQET}}=
\frac{O_{NM}^{\rm LHQET}}{\widetilde{O}_{NM}^{\rm LHQET}}=
1-(w_0^{\rm MF})^2.
\end{eqnarray}
In the above equations the matching coefficients are written as
\begin{eqnarray}
Z_{\Gamma (R)}^{(1)\rm MF}&=&1+\left(\frac{g}{4\pi}\right)^2C_F
\left[\hat{\cal A}_{\rm CHQET}-\hat{\cal A}_{\rm LHQET}\right]_R^{(1)\rm MF}
\equiv 1+\left(\frac{g}{4\pi}\right)^2C_F\hat{z}_{\Gamma (R)}^{(1)\rm MF},
\label{EQ:hat_z_gamma_O1_R}\\
Z_{\Gamma}^{(pa)\rm MF}&=&\left(\frac{g}{4\pi}\right)^2C_F
\left[\hat{\cal A}_{\rm CHQET}-\hat{\cal A}_{\rm LHQET}\right]^{(pa)\rm MF}
\equiv\left(\frac{g}{4\pi}\right)^2C_F\hat{z}_{\Gamma}^{(pa)\rm MF},
\label{EQ:hat_z_gamma_Opa}\\
Z_{\Gamma}^{(ma)\rm MF}&=&\left(\frac{g}{4\pi}\right)^2C_F
\left[\hat{\cal A}_{\rm CHQET}-\hat{\cal A}_{\rm LHQET}\right]^{(ma)\rm MF}
\equiv\left(\frac{g}{4\pi}\right)^2C_F\hat{z}_{\Gamma}^{(ma)\rm MF},
\label{EQ:hat_z_gamma_Oma}\\
Z_{L (R)}^{(1)\rm MF}&=&1+\left(\frac{g}{4\pi}\right)^2
\left[\hat{\cal B}_{\rm CHQET}-\hat{\cal B}_{\rm LHQET}\right]_R^{(1)\rm MF}
\equiv 1+\left(\frac{g}{4\pi}\right)^2\hat{z}_{L (R)}^{(1)\rm MF},
\label{EQ:hat_z_L_O1_R}\\
Z_L^{(pa)\rm MF}&=&\left(\frac{g}{4\pi}\right)^2
\left[\hat{\cal B}_{\rm CHQET}-\hat{\cal B}_{\rm LHQET}\right]^{(pa)\rm MF}
\equiv \left(\frac{g}{4\pi}\right)^2\hat{z}_L^{(pa)\rm MF},
\label{EQ:hat_z_L_Opa}\\
Z_L^{(ma)\rm MF}&=&\left(\frac{g}{4\pi}\right)^2
\left[\hat{\cal B}_{\rm CHQET}-\hat{\cal B}_{\rm LHQET}\right]^{(ma)\rm MF}
\equiv \left(\frac{g}{4\pi}\right)^2\hat{z}_L^{(ma)\rm MF},
\label{EQ:hat_z_L_Oma}
\end{eqnarray}
where $g$ is replaced by $g_{\overline{\rm MS}}(\mu=a^{-1})$
in Eq.~(\ref{EQ:g_MSbar}), and $e_R$ is used instead of $e$ in
the matching factors~(\ref{EQ:hat_z_gamma_O1_R}) and (\ref{EQ:hat_z_L_O1_R}).
The numerical values of the one-loop coefficients
in Eqs.~(\ref{EQ:hat_z_gamma_O1_R})--(\ref{EQ:hat_z_L_Oma}) are shown in
Tabs.~\ref{TAB:hat_z_GR_1_hat_z_G_pa_plaquette} --
\ref{TAB:hat_z_L_pa_hat_z_L_ma_DBW2}.

\subsection{Alternative form for the $O(pa)$ terms}
\label{SEC:total_derivative}

We defined $O(pa)$ operators containing covariant derivatives
for the quark bilinears and for the $\Delta B=2$ four-quark operator
in Eq.~(\ref{EQ:bilinear_Oa_operators}) and
Eq.~(\ref{EQ:four-quark_Oa_operators}) respectively.
These operators can be conveniently rewritten at $O(g^0)$ using the
equations of motion as:
\begin{eqnarray}
\widetilde{J}_{\pm\Gamma D}
&=&\left(G(\mp u_0\partial_0+1-u_0\right)-m_q)J_{\pm\Gamma},
\label{EQ:J_O(pa)_2}\\
\widetilde{O}_{ND}&=&2u_0[\overline{h}_+\gamma_{\mu}^Lq]
\left(\overleftarrow{\partial}_0-\overrightarrow{\partial}_0\right)
[\overline{h}_-\gamma_{\mu}^Lq]-2(1-u_0)O_L
-O_{NM},\;\;\;\;\;\;\;\;
\label{EQ:OL_O(pa)_2}
\end{eqnarray}
where we used the tree-level MF-improved equations of motion for the light quark
\begin{eqnarray}
\left[\frac{{\not\!\!D}}{u_0}+\frac{m_q}{u_0}\right]q=0,
\end{eqnarray}
and for the static quark
\begin{eqnarray}
\overline{h}_{\pm}
\left[\pm\frac{\overleftarrow{D}_0^{\rm smeared}}{v_0}
-\left(\frac{1}{v_0}-1\right)\right]=0,
\end{eqnarray}
(The unimproved case is obtained by setting $u_0=1$ and $v_0=1$.)
In the derivation of Eqs.~(\ref{EQ:J_O(pa)_2}) and (\ref{EQ:OL_O(pa)_2}),
we replace $v_0$ with $u_0$ because the difference between the smeared
and unsmeared gauge field is $O(a^2)$~\cite{Lee:2002ui, Lee:2002fj} and
it is irrelevant for the $O(a)$ improvement program.
We also note that there are derivatives rather than covariant derivatives
in Eqs.~(\ref{EQ:J_O(pa)_2}) and~(\ref{EQ:OL_O(pa)_2}).
The cancellation of gluon fields occurs between the covariant derivative
of the static and that of the light quark:
\begin{eqnarray}
\overleftarrow{D}_0^{\rm smeared}+\overrightarrow{D}_0
&=&
\left(\overleftarrow{\partial}_0-igA_0^{\rm smeared}+O(g^2a)\right)
+\left(\overrightarrow{\partial}_0+igA_0+O(g^2a)\right)\nonumber\\
&=&
\left(\overleftarrow{\partial}_0-ig(1+O(a^2))A_0+O(g^2a)\right)
+\left(\overrightarrow{\partial}_0+igA_0+O(g^2a)\right)\nonumber\\
&=&
\overleftarrow{\partial}_0+\overrightarrow{\partial}_0+O(g^2a, ga^2).
\label{EQ:gauge_cancelle}
\end{eqnarray}
This cancellation is valid at full order in $g$ and the $O(a)$ deficit
in Eq.~(\ref{EQ:gauge_cancelle}) becomes an $O(a^2)$ error
in the $O(a)$ improvement program.

Now we apply the operator $\widetilde{J}_{\pm\Gamma D}$
(\ref{EQ:J_O(pa)_2}) to the $B$ meson correlator:
\begin{eqnarray}
\langle \widetilde{J}_{-\Gamma D}(t) (J_{-\Gamma}(0))^{\dag}\rangle
=\left\{G\left(-u_0E^{\rm bind}+1-u_0\right)-m_q\right\}
\langle J_{-\Gamma}(t) (J_{-\Gamma}(0))^{\dag}\rangle,
\label{EQ:correlator_total_derivative}
\end{eqnarray}
where $E^{\rm bind}$ denotes the $B$ meson binding energy, which comes
from fitting the correlator to a function proportional to $e^{-E^{\rm bind}t}$.
The total time derivative in $\widetilde{J}_{-\Gamma D}$
picks up $E_{\rm bind}$ which contains a $1/a$ divergence originating
from the static quark mass shift $\delta M$.
At first glance, this appears to conflict with using $J_{\pm\Gamma D}$,
the original $O(pa)$ operator, because the spatial derivative in
$J_{\pm\Gamma D}$ seems not to pick up the $1/a$ divergence
in the static quark momentum. 
We will discuss this point in more detail in next subsection.
While Eq.~(\ref{EQ:correlator_total_derivative}) is a part of $O(a)$
term, a combination of the $O(a)$ and the $1/a$ affects the $O(1)$ term.
This is, however, an effect at higher order in the one-loop perturbative
calculation.

\subsection{Power divergence from $O(pa)$ operator}
\label{SEC:power_divergence}

As already mentioned, we need to pay special attention to the $1/a$
power divergence when taking the continuum limit.
One issue is the fit function ambiguity of the correlator,
which we discussed in Sec.~\ref{SEC:MF_impr}.
Another issue is the power divergence coming from the $O(pa)$ operator,
mentioned above in Sec.~\ref{SEC:total_derivative}.
The loss of Lorentz invariance in the lattice discretization allows
the $O(pa)$ operators to give rise to a power divergence.
The divergence comes from quantum corrections, that is, $O(g^2)$ effects,
but because these operators appear at $O(g^2)$ in the $O(a)$ improvement terms,
the power divergent piece affects only higher orders in perturbation theory.
It is, however, useful to see how this divergence arises
from the $O(pa)$ operators both for the higher order
perturbative and for non-perturbative renormalization.

Consider first the $O(pa)$ operator for the quark bilinear
$J_{\pm\Gamma D}$ in Eq.~(\ref{EQ:bilinear_Oa_operators}).
To locate the power divergences, we calculate the transition amplitude
of this operator at $O(g^2)$.
When the on-shell condition is imposed, the external quark momenta
should be the renormalized ones.
The external light quark momentum does not cause the power divergence,
since the massless point is well defined for chiral fermions.
Note that even for the Wilson fermion action, which has intrinsic chiral
symmetry breaking, a power divergence can be subtracted by adjusting
the massless point using the critical hopping parameter $\kappa_c$.
The on-shell external static quark momentum $p'$, however, receives a static
quark mass renormalization, leading to $\pm ip_0'+\delta M=0$,
which means $p'$ includes a power divergence.
In the analysis up to one-loop we include tree and one-loop diagrams.
Some tree diagrams acquire a power divergence in the on-shell external
momentum $p'$ and these tree diagrams need to be taken into account.
Divergences from $p'$ are also contained in the one-loop diagrams, 
but are a higher order effect by $g^2$ and therefore can be omitted
in our analysis at the one-loop level.

For the operator $J_{\pm\Gamma D}$ the tree-level amplitude does not pick up
a power divergence since the operator contains a derivative
acting only on the light quark: 
\begin{eqnarray}
 \parbox[c][10mm][c]{43mm}{
\includegraphics[scale=0.28, viewport = 0 0 400 180, clip]
                {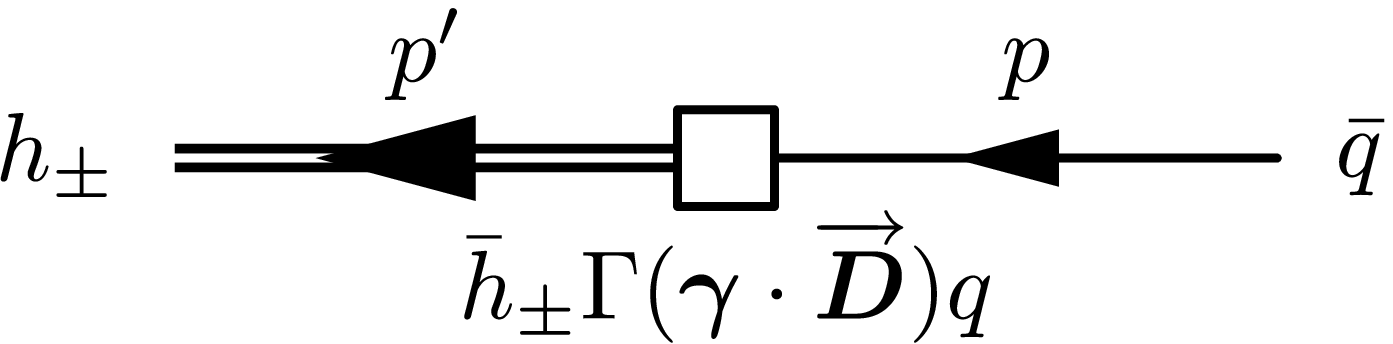}
		\vspace*{+6mm}}
&=&\frac{1}{a}\left[0+O(pa)\right].
\end{eqnarray}
The following one-loop diagrams need to be calculated:
\begin{eqnarray}
\parbox[b][15mm][b]{43mm}{
\includegraphics[scale=0.28, viewport = 0 0 400 190, clip]
                {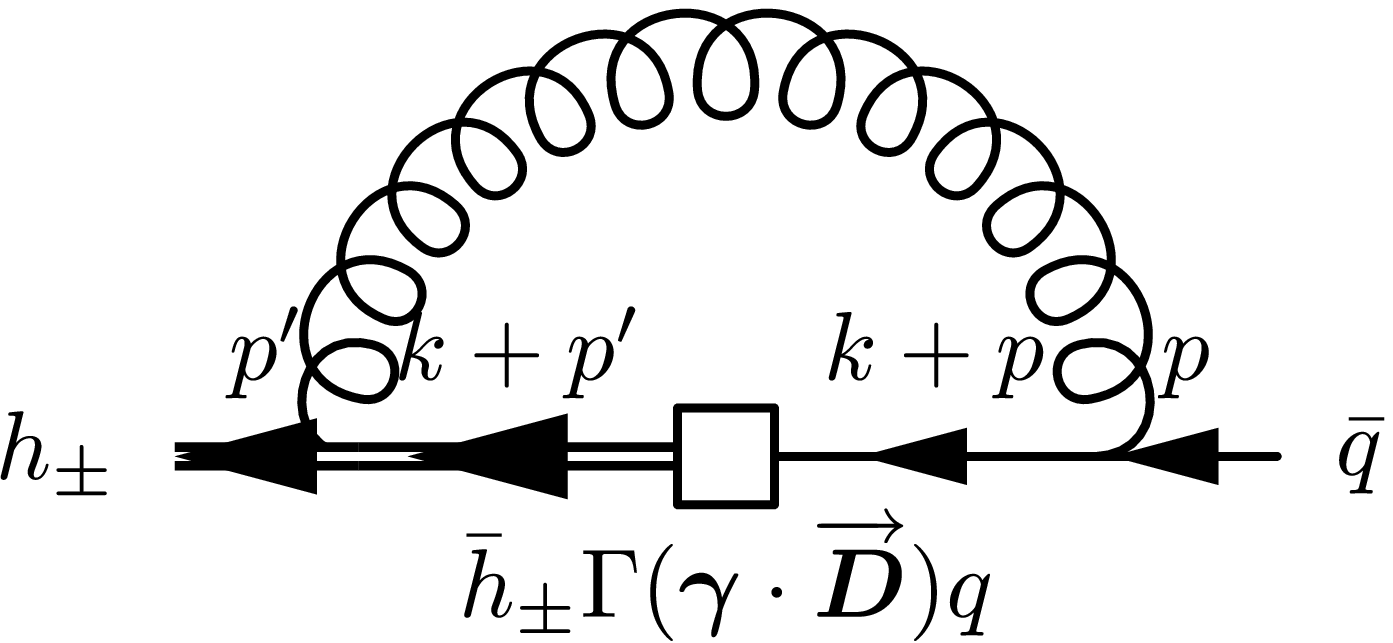}
		\vspace*{-5mm}}
&=&\frac{1}{a}\left[-G\delta M+O(g^4, pa)\right],
\label{EQ:heavy-light_Opa_type0_loop2}\\
\parbox[b][15mm][b]{43mm}{
\includegraphics[scale=0.28, viewport = 0 0 400 190, clip]
                {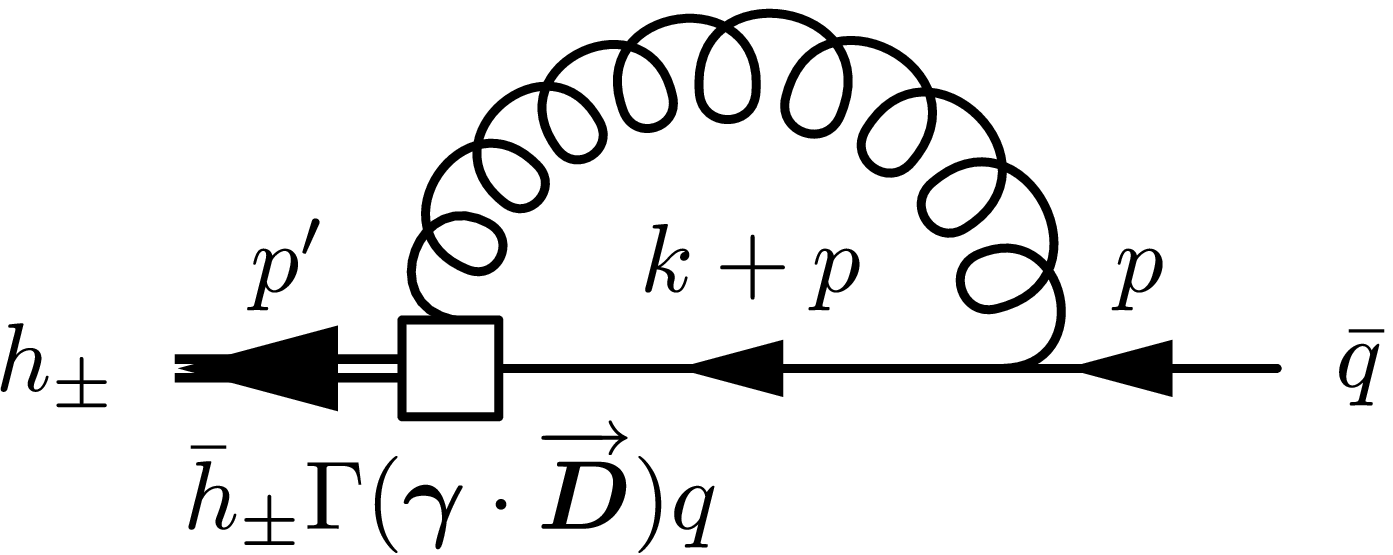}
		\vspace*{-5mm}}
&=&\frac{1}{a}\left[0+O(g^4, pa)\right],\\
\parbox[b][15mm][b]{43mm}{
\includegraphics[scale=0.28, viewport = 0 0 400 190, clip]
                {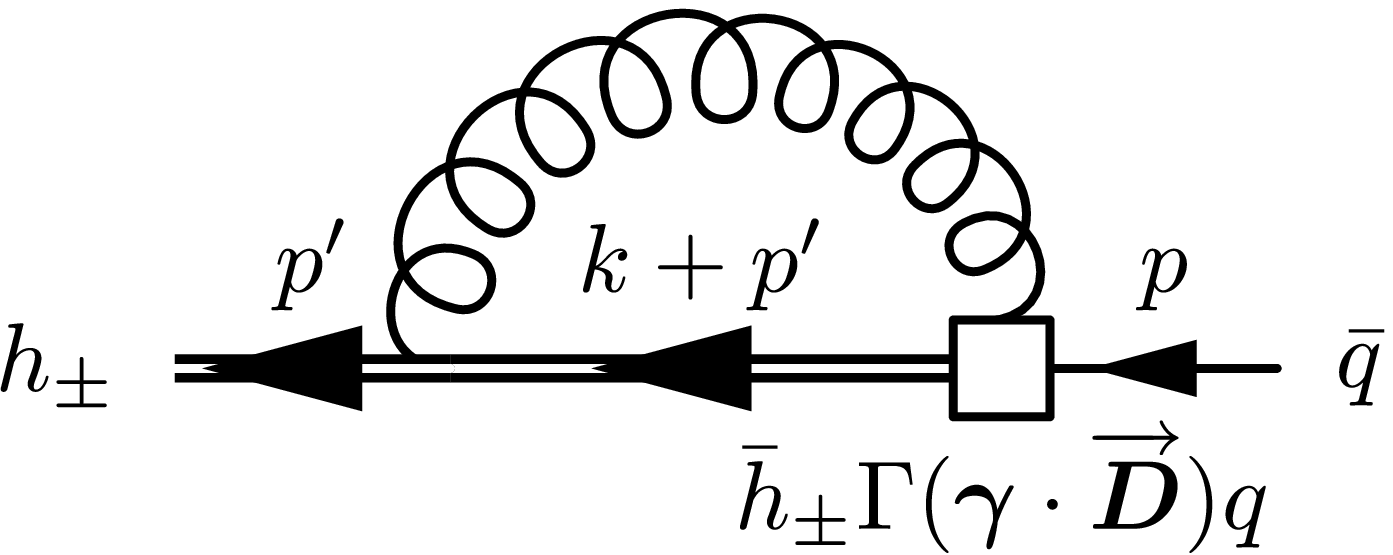}
		\vspace*{-5mm}}
&=&\frac{1}{a}\left[0+O(g^4, pa)\right],
\end{eqnarray}
\begin{eqnarray}
\parbox[b][25mm][b]{43mm}{
\includegraphics[scale=0.28, viewport = 0 0 400 300, clip]
                {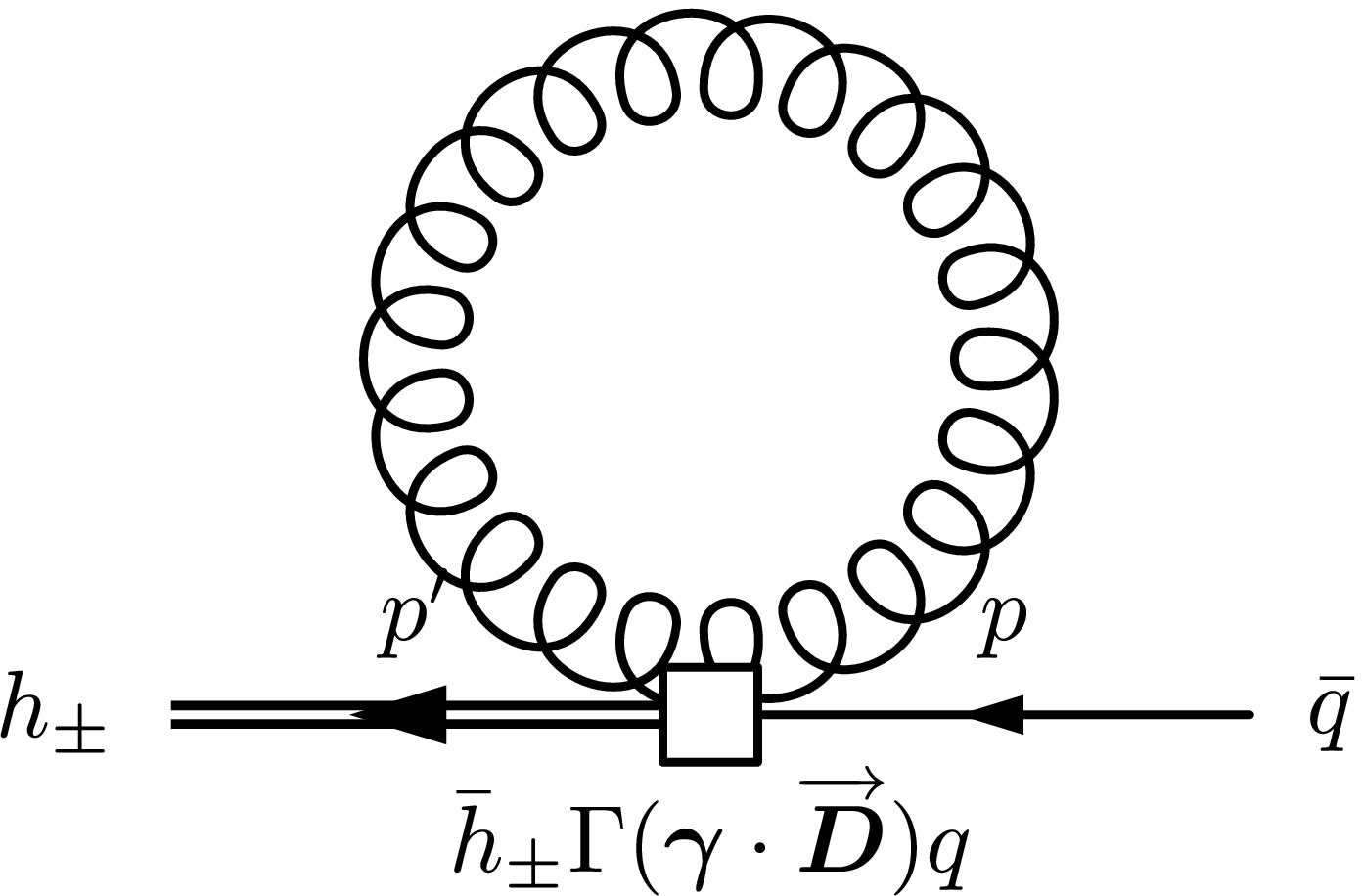}
		\vspace*{-5mm}}
&=&\frac{1}{a}\left[0+O(g^4, pa)\right].\\
\nonumber
\end{eqnarray}
The power divergent piece comes from the loop diagram of
Eq.~(\ref{EQ:heavy-light_Opa_type0_loop2}).
Again, this power divergence contributes at $O(g^4)$ which is higher
order than the one-loop calculation in this paper, since the $O(pa)$ term
is already $O(g^2)$.

In Sec.~\ref{SEC:total_derivative}, the transformation of the
original $O(pa)$ operator to a total time derivative operator was discussed.
Using the equations of motion the operator
$\mp\partial_0(\overline{h}_{\pm}\Gamma\gamma_0 q)$ is obtained.
This operator gives a power divergence from the tree diagrams
with the on-shell condition:
\begin{eqnarray}
\parbox[c][10mm][c]{43mm}{
\includegraphics[scale=0.28, viewport = 0 0 400 180, clip]
                {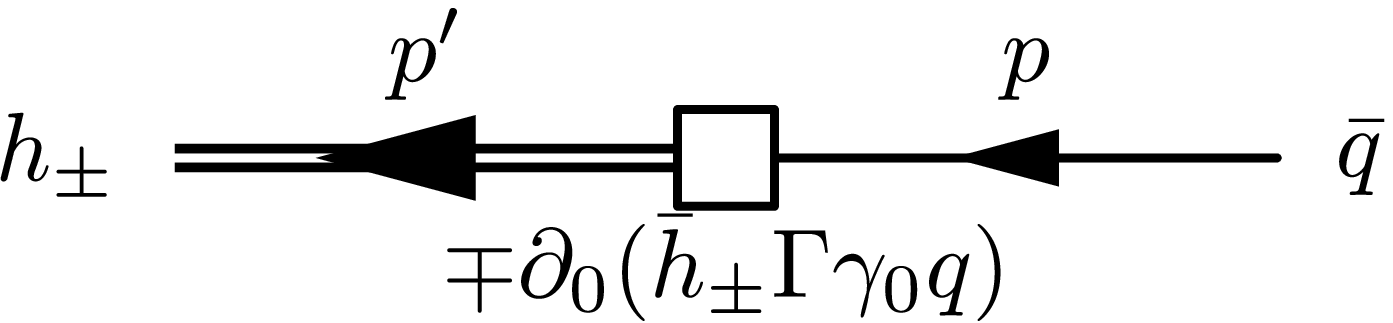}
		\vspace*{+7mm}}
&=&\frac{1}{a}\left[-G\delta M+O(g^4, pa)\right],\\
\parbox[b][18mm][b]{43mm}{
\includegraphics[scale=0.28, viewport = 0 0 400 190, clip]
                {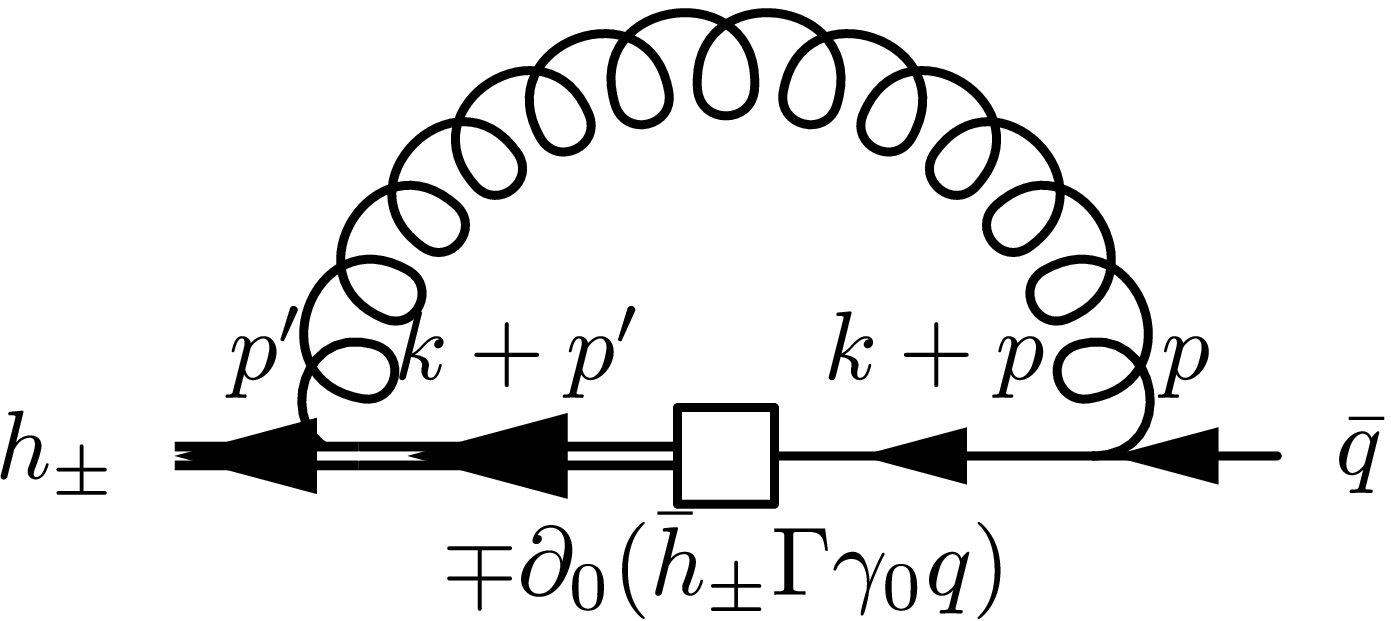}
		\vspace*{-4mm}}
&=&\frac{1}{a}\left[0+O(g^4, pa)\right].
\label{EQ:heavy-light_Opa_type4_loop1}
\end{eqnarray}
The loop diagram (\ref{EQ:heavy-light_Opa_type4_loop1}), however, does not
produce a divergence owing to the nature of total derivative at $O(g^2)$.
The power divergence is the same as that in the original operator
$J_{\pm\Gamma D}$ which is consistent with the fact that
the equations of motion do not contain any power divergence.
(The power divergent structure in the equations of motion is presented
in Appendix~\ref{SEC:power-div_eom}.)
Now we answer the puzzle introduced in the last paragraph in
Sec.~\ref{SEC:total_derivative}.
The power divergence in the operator
$\mp\partial_0(\overline{h}_{\pm}\Gamma\gamma_0 q)$ comes from the on-shell
static quark momentum, while the divergence for the original $O(pa)$ operator
$J_{\pm\Gamma D}$ arises through quantum correction to its vertex function.


\section{Numerical value of the matching factor: an example}
\label{SEC:example}

In this section we show an example how to obtain the numerical values of
the renormalization constants and $O(a)$ improvement coefficients using
the simulation settings from Ref.~\cite{Aoki:2010dy}.
The simulation parameters are presented in Tab.~\ref{TAB:simulation_parameters}.
\begin{table}[t]
\begin{center}
\begin{tabular}{c|cc}
\hline\hline
                        & Configurations (A)     & Configurations (B)\\ 
\hline
$\beta$                 & $2.13$                 & $2.25$\\
$L^3\times T\times L_s$ & $24^3\times64\times16$ & $32^3\times64\times16$\\
$a^{-1}$ (GeV)          & $1.73(3)$              & $2.28(3)$\\
$am_{\rm res}$          & $0.003152(43)$         & $0.0006664(76)$\\
$M_5$                   & $1.8$                  & $1.8$\\
$P$~(chiral limit)      & $0.5883$               & $0.6156$\\
$M_5^{\rm MF}$          & $1.3032$               & $1.3432$\\
\hline\hline
\end{tabular}
\caption{
Simulation data for $2+1$ flavor dynamical domain-wall fermions with
Iwasaki gluons from the RBC-UKQCD Collaborations~\cite{Aoki:2010dy}.}
\label{TAB:simulation_parameters}
\end{center}
\end{table}

\subsection{Continuum matching factor}

We fix the coupling constant at the $Z$-boson mass to the PDG value
$\alpha_s(m_Z=91.1876~\mbox{GeV})=0.1184$~\cite{Nakamura:2010zzi},
from which $\alpha_s$ at different scales are calculated using four-loop
RG running~\cite{Chetyrkin:1997dh, Vermaseren:1997fq}.
Because the simulation is performed not including dynamical charm quark,
we employ a two step running as mentioned in Sec.~\ref{EQ:continuum_matching}.
The obtained coupling constants are:
$\alpha_s(m_b=4.19~\mbox{GeV\cite{Nakamura:2010zzi}})=0.2260$,
$\alpha_s(m_c=1.27~\mbox{GeV\cite{Nakamura:2010zzi}})=0.3919$,
$\alpha_s(a^{-1}=1.73~\mbox{GeV})=0.3204$ and
$\alpha_s(a^{-1}=2.28~\mbox{GeV})=0.2773$.
Using these couplings, the matching factors between CQCD and CHQET,
(\ref{EQ:C_Gamma_using_RG}) and (\ref{EQ:Z_12_using_RG}), are:
\begin{eqnarray}
C_{\Gamma=\gamma_0\gamma_5}(m_b, a^{-1}=1.73)&=&
0.9520\times 1.1550\times 0.9521 = 1.0470,\\
C_{\Gamma=\gamma_0\gamma_5}(m_b, a^{-1}=2.28)&=&
0.9520\times 1.1550\times 0.9196 = 1.0112,
\end{eqnarray}
\begin{eqnarray}
\begin{bmatrix} Z_1(m_b, a^{-1}=1.73) \\ Z_2(m_b, a^{-1}=1.73) \end{bmatrix}^T
&=&
\begin{bmatrix} 0.7483 \\ -0.1439 \end{bmatrix}^T
\begin{bmatrix} 1.3345 & 0 \\ -0.0526 & 1.0921 \end{bmatrix}
\begin{bmatrix} 0.9055 & 0 \\  0.0141 & 0.9706 \end{bmatrix}
=
\begin{bmatrix} 0.9088 \\ -0.1525 \end{bmatrix}^T,\nonumber\\
\\
\begin{bmatrix} Z_1(m_b, a^{-1}=2.28) \\ Z_2(m_b, a^{-1}=2.28) \end{bmatrix}^T
&=&
\begin{bmatrix} 0.7483 \\ -0.1439 \end{bmatrix}^T
\begin{bmatrix} 1.3345 & 0 \\ -0.0526 & 1.0921 \end{bmatrix}
\begin{bmatrix} 0.8442 & 0 \\  0.0231 & 0.9500 \end{bmatrix}
=
\begin{bmatrix} 0.8457 \\ -0.1493 \end{bmatrix}^T.\nonumber\\
\end{eqnarray}

\subsection{
Lattice to continuum matching factor and $O(a)$ improvement coefficient in HQET}
\label{SEC:matching_final}

We present one-loop perturbative results for the lattice to continuum
matching factor and $O(a)$ improvement coefficients in HQET.
The expectation value of plaquette $P$ is the value in the chiral limit,
obtained by linear extrapolation in quark masses using the lightest
two degenerate up and down quark mass parameters.
For Configurations (B) the deviation from that using three quark mass
parameters is less than $0.01\%$.

Given the value of $P$, the matching factor and $O(a)$ improvement
coefficients are obtained by the following steps:
\begin{list}{}{}
\item[(1)]
The MF value of link $u_0=P^{1/4}$ and the MF
improved value of the domain-wall height $M_5^{\rm MF}$,
which is calculated from Eq.~(\ref{EQ:MFimp-M5}),
are obtained as presented in Tab.~\ref{TAB:simulation_parameters}.

\item[(2)] 
The value of $\Sigma_w$ at $M_5=M_5^{\rm MF}$ is
taken from Tab.~\ref{TAB:Sigma_w} using cubic spline interpolation.
Then $\hat{z_w}^{\rm MF}(M_5)$ is obtained using Eq.~(\ref{EQ:MFimp-hat_zw}).
The values   are presented in Tab.~\ref{TAB:numerical_one-loop_coefficients}.

\item[(3)]
The one-loop coefficient of the matching factor
(without the renormalization factor of $M_5$) and $O(a)$ improvement
coefficients are read from
Tabs.~\ref{TAB:hat_z_G_1_hat_z_G_pa_IWASAKI}--\ref{TAB:hat_z_L_pa_hat_z_L_ma_IWASAKI}
using cubic spline interpolation.
The values are presented in Tab.~\ref{TAB:numerical_one-loop_coefficients}.

\item[(4)]
The coupling $g_{\overline{\rm MS}}(\mu{=}a^{-1})$
is obtained from Eq.~(\ref{EQ:g_MSbar}) as presented in
Tab.~\ref{TAB:numerical_one-loop_coefficients}.
In the calculation the value of $d_f$ at $M_5=M_5^{\rm MF}$ is used,
which is read from Tab.~\ref{TAB:df} using cubic spline interpolation.

\item[(5)]
The matching factors and $O(a)$ improvement coefficients in
Eqs.~(\ref{EQ:matching_bilinear})--(\ref{EQ:matching_four-quark_S})
are obtained by combining the steps above and are presented in
Tab.~\ref{TAB:numerical_one-loop_factors}.
\end{list}

\begin{table}[t]
\begin{center}
\begin{tabular}{c|cccc}
\hline\hline
\multirow{2}{5mm}{} &
\multicolumn{4}{c}{Configurations (A)} \\
 & unsmear & APE & HYP1 & HYP2 \\
\hline
$g_{\rm MS}^2/4\pi$ &
\multicolumn{4}{c}{$0.1769$} \\
$\hat{z}_w^{\rm MF}$ &
\multicolumn{4}{c}{$5.257$} \\
$\hat{z}_{\Gamma (R)}^{(1)\rm MF}$ &
$-4.87061(56)$ & $-1.58382(56)$ & $-1.50401(56)$ & $0.07768(56)$ \\
$\hat{z}_{\Gamma}^{(pa)\rm MF}$ &
$0.41612(41)$ & $-3.48036(41)$ & $-3.68664(42)$ & $-6.41186(44)$ \\
$\hat{z}_{\Gamma}^{(ma)\rm MF}$ &
$1.70169(57)$ & $-4.12249(57)$ & $-4.43848(57)$ & $-8.52774(57)$ \\
$\hat{z}_{L(R)}^{(1)\rm MF}$ &
$-15.4170(15)$ & $-4.4611(15)$ & $-4.1950(15)$ & $1.0773(15)$ \\
$\hat{z}_L^{(pa)\rm MF}$ &
$-0.83223(82)$ & $6.96072(83)$ & $7.37328(84)$ & $12.82372(87)$ \\
$\hat{z}_L^{(ma)\rm MF}$ &
$-3.4034(11)$ & $8.2450(11)$ & $8.8770(11)$ & $17.0555(11)$ \\
\hline\hline
\multirow{2}{5mm}{} &
\multicolumn{4}{c}{Configurations (B)} \\
 & unsmear & APE & HYP1 & HYP2 \\
\hline
$g_{\rm MS}^2/4\pi$ &
\multicolumn{4}{c}{$0.1683$} \\
$\hat{z}_w^{\rm MF}$ &
\multicolumn{4}{c}{$6.133$} \\
$\hat{z}_{\Gamma (R)}^{(1)\rm MF}$ &
$-4.88040(68)$ & $-1.58901(68)$ & $-1.50890(68)$ & $0.07608(68)$ \\
$\hat{z}_{\Gamma}^{(pa)\rm MF}$ &
$0.41254(50)$ & $-3.48553(51)$ & $-3.69280(51)$ & $-6.42056(53)$ \\
$\hat{z}_{\Gamma}^{(ma)\rm MF}$ &
$2.13243(68)$ & $-3.84087(68)$ & $-4.15412(68)$ & $-8.33056(68)$ \\
$\hat{z}_{L(R)}^{(1)\rm MF}$ &
$-15.3875(18)$ & $-4.4162(18)$ & $-4.1492(18)$ & $1.1340(18)$ \\
$\hat{z}_L^{(pa)\rm MF}$ &
$-0.8251(10)$ & $6.9711(10)$ & $7.3856(10)$ & $12.8411(11)$ \\
$\hat{z}_L^{(ma)\rm MF}$ &
$-4.2649(13)$ & $7.6817(14)$ & $8.3082(14)$ & $16.6611(14)$ \\
\hline\hline
\end{tabular}
\caption{
Numerical values of the one-loop coefficients of matching factors
$g_{\rm MS}^2/4\pi$,
$\hat{z}_w^{\rm MF}$,
$\hat{z}_{\Gamma (R)}^{(1)\rm MF}$,
$\hat{z}_{\Gamma}^{(pa)\rm MF}$, $\hat{z}_{\Gamma}^{(ma)\rm MF}$,
$\hat{z}_{L (R)}^{(1)\rm MF}$,
$\hat{z}_L^{(pa)\rm MF}$ and $\hat{z}_L^{(ma)\rm MF}$.
}
\label{TAB:numerical_one-loop_coefficients}
\end{center}
\end{table}
\begin{table}[t]
\begin{center}
\begin{tabular}{c|cccc|cccc}
\hline\hline
\multirow{2}{5mm}{} &
\multicolumn{4}{c|}{Configurations (A)} &
\multicolumn{4}{c}{Configurations (B)} \\
 & unsmear & APE & HYP1 & HYP2 & unsmear & APE & HYP1 & HYP2 \\
\hline
${\cal Z}_{\Gamma=\gamma_0\gamma_5}$ &
$0.8513$ & $0.9091$ & $0.9105$ & $0.9383$ &
$0.8684$ & $0.9243$ & $0.9256$ & $0.9526$ \\
$c_{\Gamma=\gamma_0\gamma_5}^{(pa)}$ &
$-0.0089$ & $0.0746$ & $0.0790$ & $0.1374$ &
$-0.0083$ & $0.0703$ & $0.0744$ & $0.1294$ \\
$c_{\Gamma=\gamma_0\gamma_5}^{(ma)}$ &
$-0.0331$ & $0.0802$ & $0.0864$ & $0.1660$ &
$-0.0379$ & $0.0683$ & $0.0739$ & $0.1482$ \\
${\cal Z}_L$ &
$0.6873$ & $0.8227$ & $0.8260$ & $0.8911$ &
$0.7184$ & $0.8514$ & $0.8546$ & $0.9187$ \\
$c_L^{(pa)}$ &
$-0.0134$ & $0.1119$ & $0.1185$ & $0.2061$ &
$-0.0125$ & $0.1054$ & $0.1117$ & $0.1942$ \\
$c_L^{(ma)}$ &
$-0.0497$ & $0.1203$ & $0.1296$ & $0.2489$ &
$-0.0569$ & $0.1025$ & $0.1108$ & $0.2222$ \\
${\cal Z}_S$ &
\multicolumn{4}{c|}{$0.9645$} &
\multicolumn{4}{c}{$1.0040$} \\
\hline\hline
\end{tabular}
\caption{
Numerical values of the one-loop lattice to continuum matching factors
and $O(a)$ improvement coefficients in the HQET
(MF-improved value).}
\label{TAB:numerical_one-loop_factors}
\end{center}
\end{table}
As expected, for the $O(a^0)$ matching factors $Z_{\Gamma(R)}^{(1)\rm MF}$
and $Z_{L(R)}^{(1)\rm MF}$ link smearing tends to produce a small correction;
HYP2 smearing gives an especially tiny change.
For the values of the one-loop $O(pa)$ and $O(ma)$ improvement coefficients
$Z_{\Gamma}^{(pa)\rm MF}$, $Z_L^{(pa)\rm MF}$, $Z_{\Gamma}^{(ma)\rm MF}$
and $Z_L^{(ma)\rm MF}$, link smearing, especially HYP2 smearing,
tends to produce large coefficients.
This property depends on the choice of the gluon action, however,
since when we choose the standard plaquette action,
the link smearings tend to give tiny coefficients.

\subsection{$O(a)$ corrections for physical quantities}

In this subsection we estimate the $O(a)$ corrections for $B$ meson
quantities using the results obtained in Sec.~\ref{SEC:matching_final}.
Using~(\ref{EQ:J_O(pa)_2}) the $|B\rangle$ to $|0\rangle$ transition
amplitude leads
\begin{eqnarray}
\frac{\langle 0|A_0^{O(a)\rm impr}|B\rangle_{\rm CHQET}}
{\langle 0|A_0|B\rangle_{\rm CHQET}}
=
1+c_{\Gamma=\gamma_0\gamma_5}^{(pa)}
\left(u_0E_B^{\rm bind}+u_0-1\right)
+\left(c_{\Gamma=\gamma_0\gamma_5}^{(ma)}
-c_{\Gamma=\gamma_0\gamma_5}^{(pa)}\right)m_f,\nonumber\\ \label{EQ:Bto0}
\end{eqnarray}
where $E_B^{\rm bind}$ is the binding energy of the $B$ meson obtained
from the fit of the correlator $\langle A_0(t)A_0^{\dagger}(0)\rangle$.
Note that one does not need to compute
$\langle A_0^{(pa)}(t)A_0^{\dagger}(0)\rangle$.
For the $\Delta B=2$ four-quark operator, we need to calculate the matrix
elements of the operators $O_L^{(pa)}$ and $O_L^{(ma)}$.
For simplicity we use the vacuum saturation approximation (VSA),
where the $|B\rangle$ to $|\overline{B}\rangle$ transition amplitude can be
written in the form:
\begin{eqnarray}
&&\frac{\langle\overline{B}|O_L^{O(a)\rm impr}|B\rangle_{\rm CHQET}^{\rm VSA}}
{\langle\overline{B}|O_L|B\rangle_{\rm CHQET}^{\rm VSA}}
=
1+2c_{\Gamma=\gamma_0\gamma_5}^{(pa)}\left(u_0E_B^{\rm bind}+u_0-1\right)
+2\left(c_{\Gamma=\gamma_0\gamma_5}^{(ma)}
-c_{\Gamma=\gamma_0\gamma_5}^{(pa)}\right)m_f.\nonumber\\
\label{EQ:BtoBb_L} 
\end{eqnarray}
The fact that the overall matching factor in Eqs.~(\ref{EQ:BtoBb_L})
is the square of that in Eq.~(\ref{EQ:Bto0}) is consistent with
the spirit of the VSA.
Using Eqs.~(\ref{EQ:Bto0}) and (\ref{EQ:BtoBb_L}),
The $O(a)$ corrections to the $B$ meson decay constant and $\Delta B=2$
matrix element are evaluated.
The $O(a)$ correction to the $B$ meson decay constant $f_B$ is
given by Eq.~(\ref{EQ:Bto0}).
For the $\Delta B=2$ matrix element ${\cal M}_B$,
we can estimate the $O(a)$ correction using Eq.~(\ref{EQ:BtoBb_L}),
where, of course, the bag parameter $B_B\propto{\cal M}_Bf_B^{-2}$
in the VSA has no $O(a)$ correction.
The $O(a)$ correction to the $SU(3)$ breaking ratio
$\xi=f_{B_s}\sqrt{B_{B_s}}/(f_{B_d}\sqrt{B_{B_d}})$ in the VSA is
\begin{eqnarray}
\left(\frac{\xi+\Delta\xi}{\xi}\right)_{\rm VSA}=
1+c_{\Gamma=\gamma_0\gamma_5}^{(pa)}
u_0(E_{B_s}^{\rm bind}-E_{B_d}^{\rm bind})
+\left(c_{\Gamma=\gamma_0\gamma_5}^{(ma)}
-c_{\Gamma=\gamma_0\gamma_5}^{(pa)}\right)(m_f(s)-m_f(d)).\nonumber\\
\end{eqnarray}

We apply the above equations to the actual numerical
simulations performed in Ref.~\cite{Albertus:2010nm}.
The $\beta$ in the simulation is the same as for configurations (A) in
Tab.~\ref{TAB:simulation_parameters} but the lattice size is
$L^3\times T\times L_s=16^3\times 32\times 16$.
The $B$ meson binding energy is roughly $E_B^{\rm bind}\sim 0.6$~(APE),
$0.5$~(HYP2).
To estimate the $O(a)$ contribution, we omit the $O(ma)$ effect because
the domain-wall mass parameter $m_f$ in the simulation is $O(0.01)$,
which is much smaller than that of $E_B^{\rm bind}$.
The conclusion is that the $O(g^2a)$ effects on $f_B$ are around
$3\%({\rm APE}), ~4\%({\rm HYP2})$, and on
${\cal M}_B$ are around $6\%({\rm APE}), ~8\%({\rm HYP2})$ in the VSA.
Using the assumption $(E_{B_s}^{\rm bind}-E_{B_d}^{\rm bind})\sim
(m_{B_s}-m_{B_d})$, the effect on $\xi$ is less than $2\%$ in the VSA.


\section{Conclusion}
\label{SEC:conclusion}

In this paper, we have calculated renormalization factors for
the heavy-light quark bilinear operator and $\Delta B=2$ four quark
operator in the static heavy and domain-wall light quark system,
including $O(a)$ corrections.
Even for domain-wall fermions, which have good chiral symmetry,
there is an $O(a)$ correction.
The allowed set of $O(pa)$ and $O(ma)$ operators was constrained by symmetries.
We showed the one-loop perturbative calculation of the
lattice to continuum matching factors and the $O(a)$ improvement
coefficients taking into account link smearing (APE, HYP1, HYP2) in
the static heavy action and four types of gluon action
(Plaquette, Symanzik, Iwasaki, DBW2).
The results showed that the $O(pa)$ correction is not negligible and should
be included in simulations used for precise determinations of CKM matrix
elements.


\acknowledgments
The authors would like to thank Sinya Aoki, Peter A. Boyle,
Thomas T. Dumitrescu, Norman H. Christ, Christopher T. Sachrajda,
Amarjit Soni, Ruth Van de Water, Jan Wennekers and Oliver Witzel
for fruitful discussion.
We also thank Ruth Van de Water for carefully reading the manuscript.
Y.A. and T.~Izubuchi are partially supported by JSPS Kakenhi grant
Nos. 22540301, 21540289, 20105002, 20025010.
JMF acknowledges support from STFC Grant ST/G000557/1.
T.~Izubuchi is partially supported by the US DOE under contract
No. DE-AC02-98CH10886.

\appendix

\section{Lattice Feynman rules}
\label{SEC:feynman_rules}

In this appendix we list the Feynman rules used in the lattice
perturbation theory.

\subsection*{Static heavy quark sector}

\begin{itemize}
\item static/anti-static heavy quark ($+/-$) propagator
\begin{eqnarray}
S_{h\pm}(k)
&=&\langle h_{\pm}(-k)\overline{h}_{\pm}(k)\rangle
=\frac{1}{1-e^{\mp ik_0}+\epsilon}P_{\pm},
\label{EQ:static_propagator}
\end{eqnarray}
where $P_{\pm}=(1\pm\gamma_0)/2$.
\item static/anti-static heavy quark ($+/-$) -- one gluon vertex
\begin{eqnarray}
W_{\mu}^{\pm A}(k,k')
&=&\mp igT^A\delta_{0\mu}e^{\mp i(k_0+k'_0)/2}P_{\pm}.
\label{EQ:hhg-vertex}
\end{eqnarray}
\item static /anti-static heavy quark ($+/-$) -- two gluon vertex
\begin{eqnarray}
W_{\mu\nu}^{\pm AB}(k,k')
&=&-\frac{1}{2}g^2
\left\{T^A,T^B\right\}
\delta_{\mu 0}\delta_{\nu 0}e^{\mp i(k_0+k_0')/2}P_{\pm}.
\label{EQ:hhgg-vertex}
\end{eqnarray}
\item link smearing

For the static heavy action, we impose link smearing.
This smearing changes the Feynman rules~\cite{Lee:2002fj,DeGrand:2002va}.
In this case the rules of the static heavy quark -- gluon vertex 
(\ref{EQ:hhg-vertex}) and (\ref{EQ:hhgg-vertex}) are changed as
\begin{eqnarray}
W_{\mu}^{\pm A}(k,k')&\longrightarrow&
\widetilde{W}_{\mu}^{\pm A}(k,k')
=\sum_{\rho}\tilde{h}_{\mu\rho}(k)W_{\rho}^{\pm A}(k,k')
=\tilde{h}_{\mu0}(k)W_0^{\pm A}(k,k'),\;\;\;\\
W_{\mu\nu}^{\pm AB}(k,k')&\longrightarrow&
\widetilde{W}_{\mu\nu}^{\pm AB}(k,k')=\sum_{\rho\sigma}\tilde{h}_{\mu\rho}(k)
W_{\rho\sigma}^{\pm AB}(k,k')\tilde{h}_{\sigma\nu}(k')\\
&&\hspace*{20mm}
=\tilde{h}_{\mu0}(k)W_{00}^{\pm AB}(k,k')\tilde{h}_{0\nu}(k'),\nonumber
\end{eqnarray}
where
\begin{eqnarray}
\tilde{h}_{\mu\nu}(k)
&=&\delta_{\mu\nu}
\left[1-\frac{\alpha_1}{6}\sum_{\rho}\hat{k}_{\rho}^2\Omega_{\mu\rho}(k)\right]
+\frac{\alpha_1}{6}\hat{k}_{\mu}\hat{k}_{\nu}\Omega_{\mu\nu}(k),\\
\Omega_{\mu\nu}(k)&=&1+\alpha_2(1+\alpha_3)
-\frac{\alpha_2}{4}(1+2\alpha_3)(\hat{k}^2-\hat{k}_{\mu}^2-\hat{k}_{\nu}^2)
+\frac{\alpha_2\alpha_3}{4}\prod_{\eta\not=\mu,\nu}\hat{k}_{\eta}^2,\\
\hat{k}_{\mu}
&=&2\sin\left(\frac{k_{\mu}}{2}\right),\;\;\;\;
\hat{k}^2=\sum_{\mu}\hat{k}_{\mu}^2,
\end{eqnarray}
with smearing parameters $(\alpha_1, \alpha_2, \alpha_3)$.
\end{itemize}

\subsection*{Domain-wall light quark sector}

\begin{itemize}
\item light quark propagator
\begin{eqnarray}
S_q(p;m_f)
&=&
\langle q(-p)\overline{q}(p)\rangle\\
&=&
\frac{-i\gamma_{\mu}\sin p_{\mu}+(1-W(p)e^{-\alpha(p)})m_f}
{-(1-W(p)e^{\alpha(p)})+m_f^2(1-W(p)e^{-\alpha(p)})}\nonumber\\
&=&\frac{1-w_0^2}{i\not\!p+(1-w_0^2)m_f}
\left(1+O(p^2,pm_f,m_f^2)\right),\nonumber
\end{eqnarray}
where
\begin{eqnarray}
w_0&=&1-M_5,\\
W(p)&=&w_0+\sum_{\mu}(1-\cos p_{\mu}),\\
W(p)\cosh\alpha(p)
&=&\frac{1}{2}\left(1+W^2(p)+\sum_{\mu}\sin^2p_{\mu}\right) .
\end{eqnarray}

\item light quark to domain-wall fermion propagator

\begin{eqnarray}
\langle q(-p)\overline{\psi}_s(p)\rangle
&=&
S_q(p;m_f)(e^{-\alpha(p)(L_s-s)}P_R+e^{-\alpha(p)(s-1)}P_L)\\
&&
+\left(S_q(p;m_f)m_f-1\right)e^{-\alpha(p)}
(e^{-\alpha(p)(L_s-s)}P_L+e^{-\alpha(p)(s-1)}P_R),
\nonumber\\
\langle\psi_s(-p)\overline{q}(p)\rangle
&=&
(e^{-\alpha(L_s-s)}P_L+e^{-\alpha(s-1)}P_R)S_q(p;m_f)\\
&&
+(e^{-\alpha(L_s-s)}P_R+e^{-\alpha(s-1)}P_L)
e^{-\alpha(p)}\left(S_q(p;m_f)m_f-1\right).\nonumber
\end{eqnarray}

\item external light quark to domain-wall fermion propagator

\begin{eqnarray}
\langle q_{\rm ext}(-p)\overline{\psi}_s(p)\rangle
&=&\frac{1-w_0^2}{i\!\!\not{\!p}+(1-w_0^2)m_f}\overline{H}_s(p;m_f),\\
\langle\psi_s(-p)\overline{q}_{\rm ext}(p)\rangle
&=&H_s(p;m_f)\frac{1-w_0^2}{i\!\!\not{\!p}+(1-w_0^2)m_f},
\end{eqnarray}
where
\begin{eqnarray}
&&\overline{H}_s(p;m_f)\\
&=&
\biggl[(w_0^{L_s-s}P_R+w_0^{s-1}P_L)+m_fw_0(w_0^{L_s-s}P_L+w_0^{s-1}P_R)
+O(p^2,pm_f,m_f^2)\biggr],\nonumber\\
&&H_s(p;m_f)\\
&=&
\biggl[(w_0^{L_s-s}P_L+w_0^{s-1}P_R)+(w_0^{L_s-s}P_R+w_0^{s-1}P_L)w_0m_f
+O(p^2,pm_f,m_f^2)\biggr],\nonumber
\end{eqnarray}
in which equation of motion for light quark is imposed.
\item domain-wall fermions -- one gluon vertex
\begin{eqnarray}
V_{\mu}^A(p,p')_{s,s'}
&=&-igT^A\left(\gamma_{\mu}\cos\frac{(p+p')_{\mu}}{2}
+i\sin\frac{(p+p')_{\mu}}{2}\right)\delta_{s,s'}.
\end{eqnarray}
\item domain-wall fermions -- two gluons vertex
\begin{eqnarray}
V_{\mu\nu}^{AB}(p,p')_{s,s'}
&=&\frac{1}{2}g^2\left\{T^A,T^B\right\}\delta_{\mu\nu}
\left(i\gamma_{\mu}\sin\frac{(p+p')_{\mu}}{2}
+\cos\frac{(p+p')_{\mu}}{2}\right)\delta_{s,s'}.\;\;\;\;\;\;\;
\end{eqnarray}
\item some useful definitions
\begin{eqnarray}
S_{\chi}(p;m_f)
&=&\sum_{s=1}^{L_s}\langle q(-p)\overline{\psi}_s(p)\rangle
   (w_0^{L_s-s}P_R+w_0^{s-1}P_L)\\
&=&-S_q(p;m_f)S_w(p;0)e^{\alpha(p)},\nonumber\\
S_w(p;m_f)
&=&\sum_{s=1}^{L_s}\langle q(-p)\overline{\psi}_s(p)\rangle
   (w_0^{L_s-s}P_L+w_0^{s-1}P_R)\\
&=&(1-S_q(p;m_f)m_f)S_w(p;0),\nonumber
\end{eqnarray}
where
\begin{eqnarray}
S_w(p;0)=\frac{1}{w_0-e^{\alpha(p)}}.
\end{eqnarray}
\end{itemize}

\subsection*{Improved gluon sector}

In our calculation, the RG-improved gluon action is used.
For the perturbation theory, we fix the gauge to Feynman gauge ($\xi=1$).

\begin{itemize}
\item RG-improved gluon propagator \cite{Weisz:1982zw, Weisz:1983bn}
\begin{eqnarray}
D_{\mu\nu}^{c_1}(k)&=&\frac{1}{(\hat{k}^2)^2}\left[
(1-A_{\mu\nu}^{c_1}(k))\hat{k}_{\mu}\hat{k}_{\nu}
+\delta_{\mu\nu}\sum_{\sigma}\hat{k}_{\sigma}^2A_{\nu\sigma}^{c_1}(k)\right],
\end{eqnarray}
where
\begin{eqnarray}
A_{\mu\nu}^{c_1}(k)&=&\frac{1-\delta_{\mu\nu}}{\Delta(k)}\left[
(\hat{k}^2)^2-c_1\hat{k}^2\left(2\sum_{\rho}\hat{k}_{\rho}^4
+\hat{k}^2\sum_{\rho\not=\mu,\nu}\hat{k}_{\rho}^2\right)\right.\nonumber\\
&&+c_1^2\left.\left(\left(\sum_{\rho}\hat{k}_{\rho}^4\right)^2
+\hat{k}^2\sum_{\rho}\hat{k}_{\rho}^4\sum_{\tau\not=\mu,\nu}\hat{k}_{\tau}^2
+(\hat{k}^2)^2\prod_{\rho\not=\mu,\nu}\hat{k}_{\rho}^2\right)\right],\\  
\Delta(k)&=&\left(\hat{k}^2-c_1\sum_{\rho}\hat{k}_{\rho}^4\right)\nonumber\\
&&\times
\left[\hat{k}^2-c_1\left((\hat{k}^2)^2+\sum_{\tau}\hat{k}_{\tau}^4\right)
+\frac{c_1^2}{2}\left((\hat{k}^2)^3+2\sum_{\tau}\hat{k}_{\tau}^6
-\hat{k}^2\sum_{\tau}\hat{k}_{\tau}^4\right)\right]\nonumber\\
&&-4c_1^3\sum_{\rho}\hat{k}_{\rho}^4\prod_{\tau\not=\rho}\hat{k}_{\tau}^2 .
\end{eqnarray}
\item some useful definitions for the smeared link
\begin{eqnarray}
D_{\mu\nu}^{\widetilde{c_1}}(k)&\equiv&
\sum_{\rho}\tilde{h}_{\mu\rho}(k)D_{\rho\nu}^{c_1}(k)
=\sum_{\rho}D_{\mu\rho}^{c_1}(k)\tilde{h}_{\rho\nu}(k),\\
D_{\mu\nu}^{\widetilde{\widetilde{c_1}}}(k)&\equiv&
\sum_{\rho,\sigma}\tilde{h}_{\mu\rho}(k)D_{\rho\sigma}^{c_1}(k)
\tilde{h}_{\sigma\nu}(k).
\end{eqnarray}
\end{itemize}


\section{Calculation of lattice perturbation}
\label{SEC:lattice_perturbation}

\subsection*{Static heavy quark propagator correction
(Eq.~(\ref{EQ:static_RS_TP}))}

The RS and TP contributions to the static heavy quark self-energy can be
written in the one-loop integrals:
\begin{eqnarray}
\Sigma_{h_{\pm}}^{\rm RS}(p)&=&\int_k D_{\mu\nu}^{c_1}(k)
\widetilde{W}_{\nu}^{\pm A}(p,p+k)
S_{\pm h}(p+k)
\widetilde{W}_{\mu}^{\pm A}(p+k,p),\\
\Sigma_{h_{\pm}}^{\rm TP}(p)&=&\frac{1}{2}\int_k D_{\mu\nu}^{c_1}(k)
\widetilde{W}_{\nu\mu}^{\pm AA}(p,p),
\end{eqnarray}
where $\int_k\equiv\int_{-\pi}^{\pi}\frac{d^4 k}{(2\pi)^4}$.
In order to obtain the radiative correction of the static mass and wave function
renormalization, we expand them like
\begin{eqnarray}
\Sigma_{h_{\pm}}^{\rm RS}(p)&=&\Sigma_{h_{\pm}}^{\rm RS}(p_0=0)
+\left.\frac{\partial\Sigma_{h_{\pm}}^{\rm RS}(p_0)}{\partial e^{\mp ip_0}}
\right|_{p_0=0}(e^{\mp ip_0}-1)+O((e^{\mp ip_0}-1)^2)\\
&=&-P_{\pm}\left(\frac{g}{4\pi}\right)^2C_F\frac{(4\pi)^2}{2}
\left(T_3^{\widetilde{c_1}}-T_4^{\widetilde{c_1}}\right)
\nonumber\\
&&-P_{\pm}\left(\frac{g}{4\pi}\right)^2C_F
\left\{R^{\widetilde{c_1}}+\frac{(4\pi)^2}{2}
\left(T_3^{\widetilde{c_1}}
-T_4^{\widetilde{c_1}}\right)\right\}
(e^{\mp ip_0}-1)+O((e^{\mp ip_0}-1)^2),\nonumber\\
\Sigma_{h_{\pm}}^{\rm TP}(p)&=&\Sigma_{h_{\pm}}^{\rm TP}(p_0=0)
+\left.\frac{\partial\Sigma_{h_{\pm}}^{\rm TP}(p_0)}{\partial e^{\mp ip_0}}
\right|_{p_0=0}(e^{\mp ip_0}-1)+O((e^{\mp ip_0}-1)^2)\\
&=&-P_{\pm}\left(\frac{g}{4\pi}\right)^2C_F\frac{(4\pi)^2}{2}
T_4^{\widetilde{c_1}}
-P_{\pm}\left(\frac{g}{4\pi}\right)C_F\frac{(4\pi)^2}{2}T_4^{\widetilde{c_1}}
(e^{\mp ip_0}-1)+O((e^{\mp ip_0}-1)^2),\nonumber
\end{eqnarray}
where we define integrations:
\begin{eqnarray}
R^{\widetilde{c_1}}&=&{\cal R}^{\widetilde{c_1}-{\rm PL}}
+{\cal R}^{{\rm PL}-{\rm div}}+R^{\rm div},\\
{\cal R}^{\widetilde{c_1}-{\rm PL}}&=&(4\pi)^2\int_k
\left[D_{00}^{\widetilde{\widetilde{c_1}}}(k)-D_{00}^{c_1=0}(k)\right]
\frac{e^{-ik_0}}{(1-e^{-ik_0}+a\epsilon)^2},\\
{\cal R}^{{\rm PL}-{\rm div}}&=&(4\pi)^2\int_{\bm{k}}
\left[\frac{1}{(\hat{\bm{k}}^2+\lambda^2)^{3/2}
(4+\hat{\bm{k}}^2+\lambda^2)^{1/2}}
-\frac{\theta(1-\bm{k}^2)}{2(\bm{k}^2+\lambda^2)^{3/2}}\right],\\
R^{\rm div}&=&\frac{(4\pi)^2}{2}
\int_{\bm{k}}\frac{\theta(1-\bm{k}^2)}{(\bm{k}^2+\lambda^2)^{3/2}}
=4\left(\ln 2-\ln\lambda-1\right),
\end{eqnarray}
\begin{eqnarray}
T_4^{\widetilde{c_1}}&=&\int_kD_{00}^{\widetilde{\widetilde{c_1}}}(k),
\label{EQ:T_4}\\
T_3^{\widetilde{c_1}}&=&T_3^{\widetilde{c_1}-{\rm PL}}+T_3^{\rm PL},\\
T_3^{\widetilde{c_1}-{\rm PL}}&=&\int_{\bm k}
\left[D_{00}^{\widetilde{\widetilde{c_1}}}(0,\bm{k})
-D_{00}^{c_1=0}(0,\bm{k})\right],\\
T_3^{\rm PL}&=&\int_{\bm{k}}\frac{1}{\hat{\bm{k}}^2},
\end{eqnarray}
where $\int_{\bm k}\equiv\int_{-\pi}^{\pi}\frac{d^3 k}{(2\pi)^3}$.
The numerical values of ${\cal R}^{{\rm PL}-{\rm div}}$ and $T_3^{\rm PL}$ are
\begin{eqnarray}
{\cal R}^{{\rm PL}-{\rm div}}=5.7531708(67),\;\;\;
T_3^{\rm PL}=0.2527296(13), \label{EQ:R_PL-div_T3_PL-div}
\end{eqnarray}
in which their numerical numbers was also given in Ref.~\cite{Loktik:2006kz}.
The new ingredients regard with regard to these quantities are the calculations
taking into account the link smearing.
The numerical values of ${\cal R}^{\widetilde{c_1}-{\rm PL}}$ and
$T_3^{\widetilde{c_1}-{\rm PL}}$ are listed
in Tabs.~\ref{TAB:R} and \ref{TAB:T_3}.
Using them, the radiative correction to the static quark mass
(\ref{EQ:delta_M}) and the wave function renormalization (\ref{EQ:Z_h}) are
\begin{eqnarray}
\delta\hat{M}=\frac{(4\pi)^2}{2}T_3^{\widetilde{c_1}},\;\;\;
Z_h=1+\left(\frac{g}{4\pi}\right)^2C_F
\left(R^{\widetilde{c_1}}+\delta\hat{M}\right),
\end{eqnarray}
and then $\cal R$ in Eq.~(\ref{EQ:definition_e}),
which is a finite part of $R^{\widetilde{c_1}}$, is given by
\begin{eqnarray}
{\cal R}=R^{\widetilde{c_1}}+4\ln\lambda.\label{EQ:definition_cal-R}
\end{eqnarray}

\subsection*{Heavy-light vertex correction
(Eq.~(\ref{EQ:heavy-light_vertex_correction}))}

The one-loop vertex correction including $O(pa)$ and $O(ma)$ parts is
calculated as
\begin{eqnarray}
&&\delta\Gamma_{h_{\pm}q}(p;m_q)\label{EQ:h-l_vertex_correction}\\
&=&\int_k D_{\mu\nu}^{c_1}(k)
\widetilde{W}_{\nu}^{\pm A}(0,k)
S_{\pm h}(k)\Gamma
\sum_{s=1}^{L_s}\langle q(-p-k)\overline{\psi}_s(p+k)\rangle
V_{\mu}^A(p+k,p)_{s,s}H_s(p;m_f)\nonumber\\
&=&\int_k D_{\mu\nu}^{c_1}(k)
\widetilde{W}_{\nu}^{\pm A}(0,k)
S_{\pm h}(k)\Gamma\nonumber\\
&&\times(-i)gT^A\biggl\{
\left(S_{\chi}(p+k;m_f)+S_w(p+k;0)w_0m_f\right)
\gamma_{\mu}\cos\frac{(k+2p)_{\mu}}{2}\nonumber\\
&&\hspace{+17mm}+\left(S_w(p+k;m_f)+S_{\chi}(p+k;0)w_0m_f\right)
i\sin\frac{(k+2p)_{\mu}}{2}\biggr\}+O(p^2,pm_f,m_f^2)\nonumber\\
&=&\left(\frac{g}{4\pi}\right)^2P_{\pm}C_F\Gamma\left\{
\delta\hat{\Gamma}_{hq}^{(1)}
+G\delta\hat{\Gamma}_{hq}^{(pa)}(i\bm{\gamma}\cdot\bm{p})
+G\delta\hat{\Gamma}_{hq}^{(ma)}m_q\right\}+O(p^2,pm_f,m_f^2),\nonumber
\end{eqnarray}
where we decomposed the vertex correction into $O(1)$, $O(pa)$
and $O(ma)$ parts in the last line.
These $O(pa)$ and $O(ma)$ parts are the new calculations in this work,
as well as the inclusion of link smearings.
The $O(1)$ part is
\begin{eqnarray}
\delta\hat{\Gamma}_{hq}^{(1)}&=&-{\cal I}_{\chi}^{\widetilde{c_1}-{\rm PL}}
-{\cal I}_{\chi}^{{\rm PL}-{\rm div}}-I_{\chi}^{\rm div}-{\cal I}_w^{\rm PL},
\end{eqnarray}
with integrations
\begin{eqnarray}
{\cal I}_{\chi}^{\widetilde{c_1}-{\rm PL}}&=&
(4\pi)^2\int_k\frac{1}{w_0-e^{\alpha(k)}}\frac{1}{W(k)-e^{-\alpha(k)}}
\frac{1}{\hat{k}_0}\\
&&\times\sum_{\rho}
\left[D_{0\rho}^{\widetilde{c_1}}(k)-D_{0\rho}^{\rm PL}(k)\right]
\hat{k}_{\rho}\cos^2\left(\frac{k_{\rho}}{2}\right),\nonumber\\
{\cal I}_{\chi}^{{\rm PL}-{\rm div}}&=&
(4\pi)^2\int_k\biggl[\frac{1}{w_0-e^{\alpha(k)}}\frac{1}{W(k)-e^{-\alpha(k)}}
\frac{1}{\hat{k}^2}\cos^2\left(\frac{k_0}{2}\right)
+\frac{\theta(1-k^2)}{(k^2)^2}\biggr],\\
I_{\chi}^{\rm div}&=&
-(4\pi)^2\int_k\frac{\theta(1-k^2)}{k^2(k^2+(a\lambda)^2)}
=\ln(a\lambda)^2,\\
{\cal I}_w^{\rm PL}&=&(4\pi)^2\frac{1}{2}\int_kS_w(k)D_{00}^{c_1=0}(k).
\end{eqnarray}
The numerical values of $I_{\chi}^{{\rm PL}-{\rm div}}$ and $I_w^{\rm PL}$ are
listed in Tab.~\ref{TAB:VXX_X_pl}, and also that of
${\cal I}_{\chi}^{\widetilde{c_1}-{\rm PL}}$ is listed in
Tabs.~\ref{TAB:V_1_smr-pl_PLAQUETTE}, \ref{TAB:V_1_smr-pl_SYMANZIK}, 
\ref{TAB:V_1_smr-pl_IWASAKI} and \ref{TAB:V_1_smr-pl_DBW2}.
The $O(pa)$ part is
\begin{eqnarray}
\delta\hat{\Gamma}_{hq}^{(pa)}&=&-{\cal J}_{\chi}^{\widetilde{c_1}-{\rm PL}}
-{\cal J}_{\chi}^{{\rm PL}-{\rm div}}-J_{\chi}^{\rm div}
-{\cal J}_w^{\widetilde{c_1}-{\rm PL}}-{\cal J}_w^{\rm PL},
\end{eqnarray}
with integrals
\begin{eqnarray}
{\cal J}_{\chi}^{\widetilde{c_1}-{\rm PL}}
&=&
\frac{(4\pi)^2}{6}\int_{\bm{k}}
\biggl[D_{00}^{\widetilde{c_1}}(\bm{k})-D_{00}^{\rm PL}(\bm{k})\biggr]
\frac{S_w({\bm k};0)e^{\alpha({\bm k})}}{1-W({\bm k})e^{\alpha({\bm k})}}\\
&&\times\Biggl\{2\sum_{\rho\not=0}\cos^2\left(\frac{k_{\rho}}{2}\right)
+\frac{e^{\alpha({\bm k})}}{1-W({\bm k})e^{\alpha({\bm k})}}
\sum_{\rho\not=0}\sin^2k_{\rho}\nonumber\\
&&
+\left(\frac{1}{1-W({\bm k})e^{\alpha({\bm k})}}
+S_w({\bm k};0)e^{\alpha({\bm k})}\right)
\sum_{\rho\not=0}\sin^2k_{\rho}
\frac{W(\bm{k})+\cos k_{\rho}-\cosh\alpha(\bm{k})}
{W(\bm{k})\sinh\alpha(\bm{k})}\Biggr\},\nonumber
\end{eqnarray}
\begin{eqnarray}
{\cal J}_{\chi}^{{\rm PL}-{\rm div}}
&=&
\frac{(4\pi)^2}{6}\int_{\bm{k}}
\Biggl[\frac{1}{\hat{\bm{k}}^2} 
\frac{S_w({\bm k};0)e^{\alpha({\bm k})}}{1-W({\bm k})e^{\alpha({\bm k})}}
\Biggl\{2\sum_{\rho\not=0}\cos^2\left(\frac{k_{\rho}}{2}\right)
+\frac{e^{\alpha({\bm k})}}{1-W({\bm k})e^{\alpha({\bm k})}}
\sum_{\rho\not=0}\sin^2k_{\rho}\nonumber\\
&&
+\left(\frac{1}{1-W({\bm k})e^{\alpha({\bm k})}}
+S_w({\bm k};0)e^{\alpha({\bm k})}\right)
\sum_{\rho\not=0}\sin^2k_{\rho}
\frac{W(\bm{k})+\cos k_{\rho}-\cosh\alpha(\bm{k})}
{W(\bm{k})\sinh\alpha(\bm{k})}\Biggr\}\nonumber\\
&&\hspace*{+5mm}-\frac{4\theta(1-\bm{k}^2)}{(\bm{k}^2)^2}\Biggr],\\
J_{\chi}^{\rm div}&=&\frac{2}{3}(4\pi)^2\int_{\bm{k}}
\frac{\theta(1-\bm{k}^2)}{\bm{k}^2(\bm{k}^2+(a\lambda)^2)}
=\frac{8\pi}{3a\lambda}-\frac{16}{3},\\
{\cal J}_w^{\widetilde{c_1}-{\rm PL}}&=&-\frac{(4\pi)^2}{2}\int_{\bm{k}}
\biggl[D_{00}^{\widetilde{c_1}}(\bm{k})-D_{00}^{\rm PL}(\bm{k})\biggr]
S_w(\bm{k};0),\\
{\cal J}_w^{\rm PL}&=&-\frac{(4\pi)^2}{2}\int_{\bm{k}}\frac{1}{\hat{\bm{k}}^2}
S_w(\bm{k};0).
\end{eqnarray}
The numerical values of ${\cal J}_{\chi}^{{\rm PL}-{\rm div}}$ and
${\cal J}_w^{\rm PL}$ are listed in Tab.~\ref{TAB:VXX_X_pl} and the those of
${\cal J}_{\chi}^{\widetilde{c_1}-{\rm PL}}$ and
${\cal J}_w^{\widetilde{c_1}-{\rm PL}}$
are listed in Tabs.~
\ref{TAB:V_1_smr-pl_PLAQUETTE}, \ref{TAB:V_2_smr-pl_PLAQUETTE}
\ref{TAB:V_1_smr-pl_SYMANZIK}, \ref{TAB:V_2_smr-pl_SYMANZIK}
\ref{TAB:V_1_smr-pl_IWASAKI}, \ref{TAB:V_2_smr-pl_IWASAKI},
\ref{TAB:V_1_smr-pl_DBW2} and \ref{TAB:V_2_smr-pl_DBW2}.
The $O(ma)$ part is
\begin{eqnarray}
\delta\hat{\Gamma}_{hq}^{(ma)}
&=&
-{\cal J}_{\chi'}^{\widetilde{c_1}-{\rm PL}}
-{\cal J}_{\chi'}^{{\rm PL}-{\rm div}}-J_{\chi'}^{\rm div}
-{\cal J}_w^{\widetilde{c_1}-{\rm PL}}-{\cal J}_w^{\rm PL}
\label{EQ:delta_Gamma_ma}\\
&&-{\cal K}_{\chi}^{\widetilde{c_1}-{\rm PL}}
-{\cal K}_{\chi}^{{\rm PL}-{\rm div}}-K_{\chi}^{\rm div},\nonumber
\end{eqnarray}
with integrals
\begin{eqnarray}
{\cal J}_{\chi'}^{\widetilde{c_1}-{\rm PL}}
&=&
\frac{(4\pi)^2}{2}\int_{\bm{k}}
\biggl[D_{00}^{\widetilde{c_1}}(\bm{k})-D_{00}^{\rm PL}(\bm{k})\biggr]
\frac{S_w({\bm k};0)e^{\alpha({\bm k})}}{1-W({\bm k})e^{\alpha({\bm k})}},\\
{\cal J}_{\chi'}^{{\rm PL}-{\rm div}}
&=&
\frac{(4\pi)^2}{2}\int_{\bm{k}}\Biggl[\frac{1}{\hat{\bm k}^2}
\frac{S_w({\bm k};0)e^{\alpha({\bm k})}}{1-W({\bm k})e^{\alpha({\bm k})}}
-\frac{\theta(1-{\bm k}^2)}{({\bm k}^2)^2}\Biggr],\\
J_{\chi'}^{\rm div}
&=&\frac{(4\pi)^2}{2}\int_{\bm{k}}
\frac{\theta(1-\bm{k}^2)}{\bm{k}^2(\bm{k}^2+(a\lambda)^2)}
=\frac{2\pi}{a\lambda}-4,\\
{\cal K}_{\chi}^{\widetilde{c_1}-{\rm PL}}
&=&
\frac{(4\pi)^2}{2}\int_{\bm k}
\biggl[D_{00}^{\widetilde{c_1}}(\bm{k})-D_{00}^{\rm PL}(\bm{k})\biggr]
\left\{1+2S_w({\bm k};0)e^{\alpha({\bm k})}
\frac{1-W({\bm k})\cosh\alpha({\bm k})}{1-W({\bm k})e^{\alpha({\bm k})}}
\right\}\frac{1}{1-w_0^2},\nonumber\\ \\
{\cal K}_{\chi}^{{\rm PL}-{\rm div}}
&=&
\frac{(4\pi)^2}{2}\int_{\bm k}\Biggl[\frac{1}{\hat{\bm k}^2}
\left\{1+2S_w({\bm k};0)e^{\alpha({\bm k})}
\frac{1-W({\bm k})\cosh\alpha({\bm k})}{1-W({\bm k})e^{\alpha({\bm k})}}
\right\}\frac{1}{1-w_0^2}
-\frac{\theta(1-{\bm k}^2)}{({\bm k}^2)^2}\Biggr],\nonumber\\ \\
K_{\chi}^{\rm div}
&=&\frac{(4\pi)^2}{2}
\int_{\bm k}\frac{\theta(1-{\bm k}^2)}{{\bm k}^2({\bm k}^2+(a\lambda)^2)}
=\frac{2\pi}{a\lambda}-4.
\end{eqnarray}
The numerical values of ${\cal J}_{\chi'}^{{\rm PL}-{\rm div}}$ and
${\cal K}_{\chi}^{{\rm PL}-{\rm div}}$
are listed in Tab.~\ref{TAB:VXX_X_pl} and the those of
${\cal J}_{\chi'}^{\widetilde{c_1}-{\rm PL}}$ and
${\cal K}_{\chi}^{\widetilde{c_1}-{\rm PL}}$
are listed in Tabs.~
\ref{TAB:V_2_smr-pl_PLAQUETTE}, \ref{TAB:V_3_smr-pl_PLAQUETTE}
\ref{TAB:V_2_smr-pl_SYMANZIK}, \ref{TAB:V_3_smr-pl_SYMANZIK}
\ref{TAB:V_2_smr-pl_IWASAKI}, \ref{TAB:V_3_smr-pl_IWASAKI},
\ref{TAB:V_2_smr-pl_DBW2} and \ref{TAB:V_3_smr-pl_DBW2}.
Note that the first five terms in the right hand side of
Eq.~(\ref{EQ:delta_Gamma_ma}) come from the $O(pa)$ part by applying
the light quark equation of motion $-i\gamma_0p_0=i{\bm\gamma}\cdot{\bm p}+m_q$ 
in the calculation of the transition amplitudes.


\section{Relations of the Feynman rules and the one-loop amplitudes
         between different domain-wall fermion actions}
\label{SEC:different_DWF_definition}

In this paper, we calculate the one-loop matching factors using
the domain-wall fermions action defined in Sec.~\ref{SEC:dwf_action}.
In the most of numerical simulations, however, the following different
action is employed :
\begin{eqnarray}
S_{\rm DW}&=&\sum_{s, s'=1}^{L_s}\sum_{x, y}\overline{\psi}_{s}(x)
D_{ss'}^{\rm DW}(x, y)\psi_{s'}(y)-\sum_{x}m_f\overline{q}(x)q(x),\\
D_{ss'}^{\rm DW}(x, y)&=&D^4(x, y)\delta_{ss'}+D^5(s, s')\delta_{xy}
+(M_5-5)\delta_{ss'}\delta_{xy},\\
D^4(x, y)&=&\sum_{\mu}\frac{1}{2}
\left[(1-\gamma_{\mu})U_{\mu}(x)\delta_{x+\hat{\mu}, y}
+(1+\gamma_{\mu})U_{\mu}^{\dagger}(y)\delta_{x-\hat{\mu}, y}\right],\\
D^5(s,s')&=&\left\{
\begin{array}{ll}
 P_L\delta_{2, s'} & (s=1) \\
 P_L\delta_{s+1, s'} + P_R\delta_{s-1, s'} & (1<s<L_s) \\
 P_R\delta_{L_s-1, s'} & (s=L_s) \\
\end{array}
\right.,\\
q(x)&=&P_L\psi_1(x)+P_R\psi_{L_s}(x),\\
\overline{q}(x)&=&\overline{\psi}_1(x)P_R+\overline{\psi}_{L_s}(x)P_L.
\end{eqnarray}
When $m_f=+1$, the anti-periodic boundary condition in the fifth
direction, $s$-direction, is imposed for this action while the action
in Sec.~\ref{SEC:dwf_action} imposes the periodic boundary condition.
The Feynman rules for this action are different from those in
Appendix~\ref{SEC:feynman_rules}
(The new rules from the action in this appendix are specified by prime
symbols.):
\begin{itemize}
\item light quark propagator
\begin{eqnarray}
S_q'(p;m_f)=\langle q(-p)\overline{q}(p)\rangle'=-S_q(p;m_f).
\end{eqnarray}
\item light quark to domain-wall fermion propagator
\begin{eqnarray}
\langle q(-p)\overline{\psi}_s(p)\rangle'
&=&
-S_q(p;m_f)(e^{-\alpha(p)(L_s-s)}P_L+e^{-\alpha(p)(s-1)}P_R)\\
&&
+\left(S_q(p;m_f)m_f-1\right)e^{-\alpha(p)}
(e^{-\alpha(p)(L_s-s)}P_R+e^{-\alpha(p)(s-1)}P_L),
\nonumber\\
\langle\psi_s(-p)\overline{q}(p)\rangle'
&=&
-(e^{-\alpha(L_s-s)}P_R+e^{-\alpha(s-1)}P_L)S_q(p;m_f)\\
&&
+(e^{-\alpha(L_s-s)}P_L+e^{-\alpha(s-1)}P_R)
e^{-\alpha(p)}\left(S_q(p;m_f)m_f-1\right).\nonumber
\end{eqnarray}
\item external light quark to domain-wall fermion propagator
\begin{eqnarray}
\langle q_{\rm ext}(-p)\overline{\psi}_s(p)\rangle'
&=&-\frac{1-w_0^2}{i\!\!\not{\!p}+(1-w_0^2)m_f}\overline{H}_s'(p;m_f),\\
\langle\psi_s(-p)\overline{q}_{\rm ext}(p)\rangle'
&=&-H_s'(p;m_f)\frac{1-w_0^2}{i\!\!\not{\!p}+(1-w_0^2)m_f},
\end{eqnarray}
where
\begin{eqnarray}
&&\overline{H}_s'(p;m_f)\\
&=&
\biggl[(w_0^{L_s-s}P_L+w_0^{s-1}P_R)-m_fw_0(w_0^{L_s-s}P_R+w_0^{s-1}P_L)
+O(p^2,pm_f,m_f^2)\biggr],\nonumber\\
&&H_s'(p;m_f)\\
&=&
\biggl[(w_0^{L_s-s}P_R+w_0^{s-1}P_L)-(w_0^{L_s-s}P_L+w_0^{s-1}P_R)w_0m_f
+O(p^2,pm_f,m_f^2)\biggr].\nonumber
\end{eqnarray}
\item domain-wall fermions -- one gluon vertex
\begin{eqnarray}
V_{\mu}'^A(p,p')_{s,s'}
&=&-igT^A\left(-\gamma_{\mu}\cos\frac{(p+p')_{\mu}}{2}
+i\sin\frac{(p+p')_{\mu}}{2}\right)\delta_{s,s'}.
\end{eqnarray}
\item domain-wall fermions -- two gluons vertex
\begin{eqnarray}
V_{\mu\nu}'^{AB}(p,p')_{s,s'}
&=&\frac{1}{2}g^2\left\{T^A,T^B\right\}\delta_{\mu\nu}
\left(-i\gamma_{\mu}\sin\frac{(p+p')_{\mu}}{2}
+\cos\frac{(p+p')_{\mu}}{2}\right)\delta_{s,s'}.\;\;\;\;\;\;\;\;\;
\end{eqnarray}
\item some useful definitions
\begin{eqnarray}
S_{\chi}'(p;m_f)
&=&\sum_{s=1}^{L_s}\langle q(-p)\overline{\psi}_s(p)\rangle'
   (w_0^{L_s-s}P_L+w_0^{s-1}P_R)
=-S_{\chi}(p;m_f),\\
S_w'(p;m_f)
&=&\sum_{s=1}^{L_s}\langle q(-p)\overline{\psi}_s(p)\rangle'
   (w_0^{L_s-s}P_R+w_0^{s-1}P_L)
=S_w(p;m_f).
\end{eqnarray}
\end{itemize}
We can, however, see that the one-loop heavy-light vertex correction
(\ref{EQ:heavy-light_vertex_correction}) with the above Feynman rules
is exactly same as Eq.~(\ref{EQ:h-l_vertex_correction}):
\begin{eqnarray}
&&\delta\Gamma_{h_{\pm}q}'(p;m_q)\\
&=&\int_k D_{\mu\nu}^{c_1}(k)
\widetilde{W}_{\nu}^{\pm A}(0,k)
S_{\pm h}(k)\Gamma
\sum_{s=1}^{L_s}\langle q(-p-k)\overline{\psi}_s(p+k)\rangle'
V_{\mu}'^A(p+k,p)_{s,s}H_s'(p;m_f)\nonumber\\
&=&\int_k D_{\mu\nu}^{c_1}(k)
\widetilde{W}_{\nu}^{\pm A}(0,k)
S_{\pm h}(k)\Gamma\nonumber\\
&&\times(-i)gT^A\biggl\{
-\left(S_{\chi}'(p+k;m_f)-S_w'(p+k;0)w_0m_f\right)
\gamma_{\mu}\cos\frac{(k+2p)_{\mu}}{2}\nonumber\\
&&\hspace{+17mm}+\left(S_w'(p+k;m_f)-S_{\chi}'(p+k;0)w_0m_f\right)
i\sin\frac{(k+2p)_{\mu}}{2}\biggr\}+O(p^2,pm_f,m_f^2)\nonumber\\
&=&\delta\Gamma_{h_{\pm}q}(p;m_q).\nonumber
\end{eqnarray}
The expressions of the light quark wave function renormalization and
the light-light vertex correction are also not altered
by the change of the action convention.


\section{Power divergence structure in equations of motion}
\label{SEC:power-div_eom}

In this appendix, we look into the power divergence structure in the
equations of motion diagrammatically as the way in
Sec.~\ref{SEC:power_divergence}.
In the analysis, we consider the heavy-light transition amplitude of
operators $\overline{h}_{\pm}\Gamma\overrightarrow{\not\!\!D}q$ for the
light quark equation of motion and
$\overline{h}_{\pm}\Gamma(\gamma_0\overleftarrow{D}_0)q$ for the static heavy
quark equation of motion.
Although these operators are identically vanished in the on-shell condition,
each Feynman diagrams can have the divergences which are eventually
canceled out. 

\subsubsection*{Equation of motion for light quark}

The light quark mass in the equation of motion is omitted, since this
part does not cause the power divergence.
Owing to the anisotropy of the static heavy-light system, the power
divergent structure is different between temporal and spatial direction.
The structure in the temporal direction is:
\begin{eqnarray}
\parbox[c][10mm][c]{43mm}{
\includegraphics[scale=0.28, viewport = 0 0 400 180, clip]
                {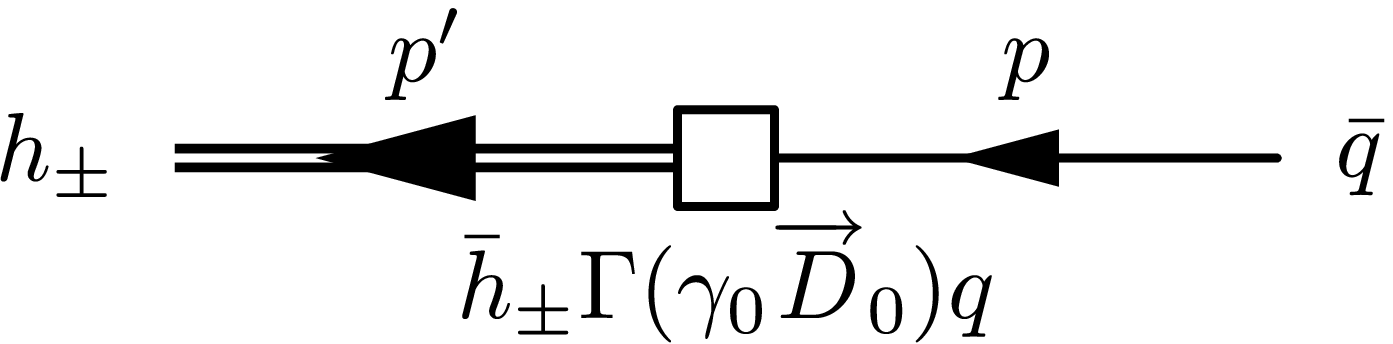}
		\vspace*{+7mm}}
&=&\frac{1}{a}\left[0+O(pa)\right],\\
\parbox[b][16mm][b]{43mm}{
\includegraphics[scale=0.28, viewport = 0 0 400 190, clip]
                {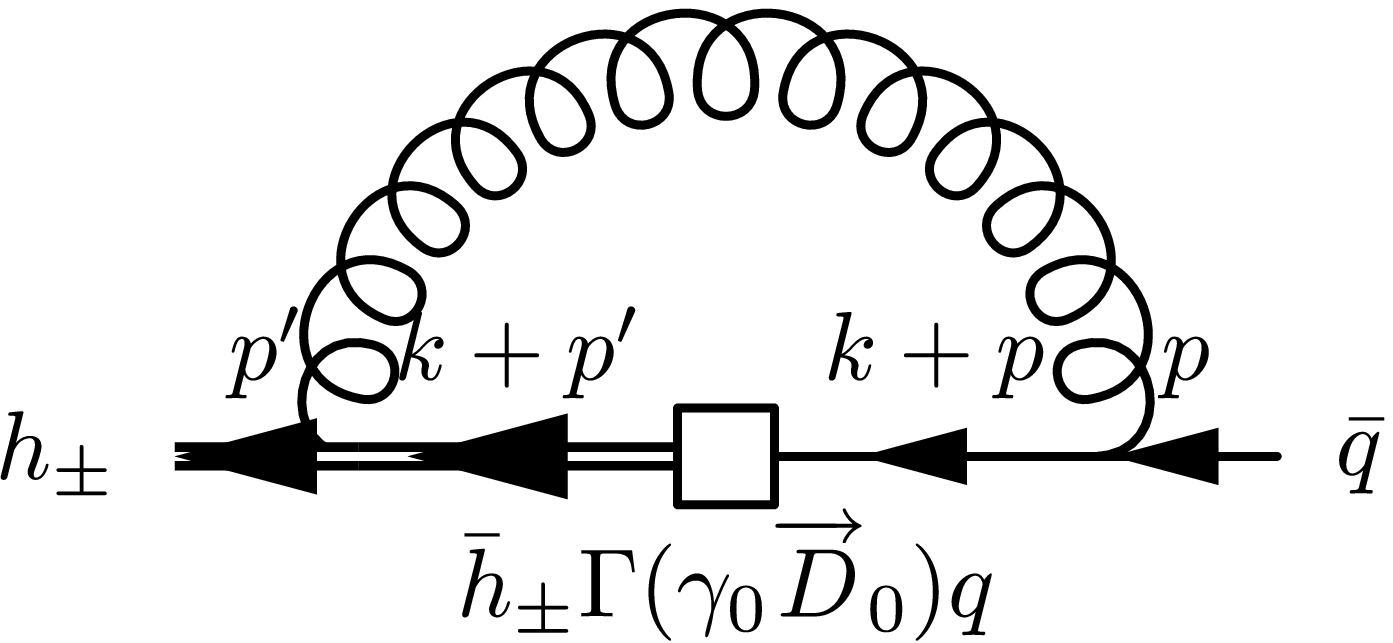}
		\vspace*{-5mm}}
&=&\frac{1}{a}\left[0+O(g^4, pa)\right],\\
\parbox[b][16mm][b]{43mm}{
\includegraphics[scale=0.28, viewport = 0 0 400 190, clip]
                {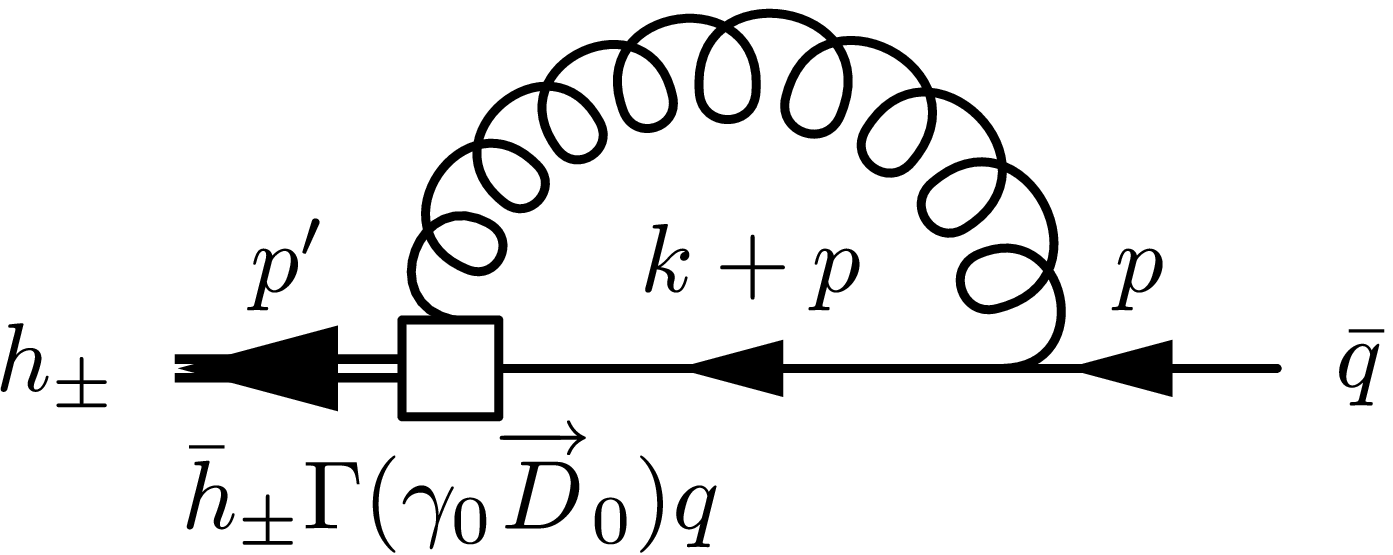}
		\vspace*{-5mm}}
&=&\frac{1}{a}\left[0+O(g^4, pa)\right],\\
\parbox[b][16mm][b]{43mm}{
\includegraphics[scale=0.28, viewport = 0 0 400 190, clip]
                {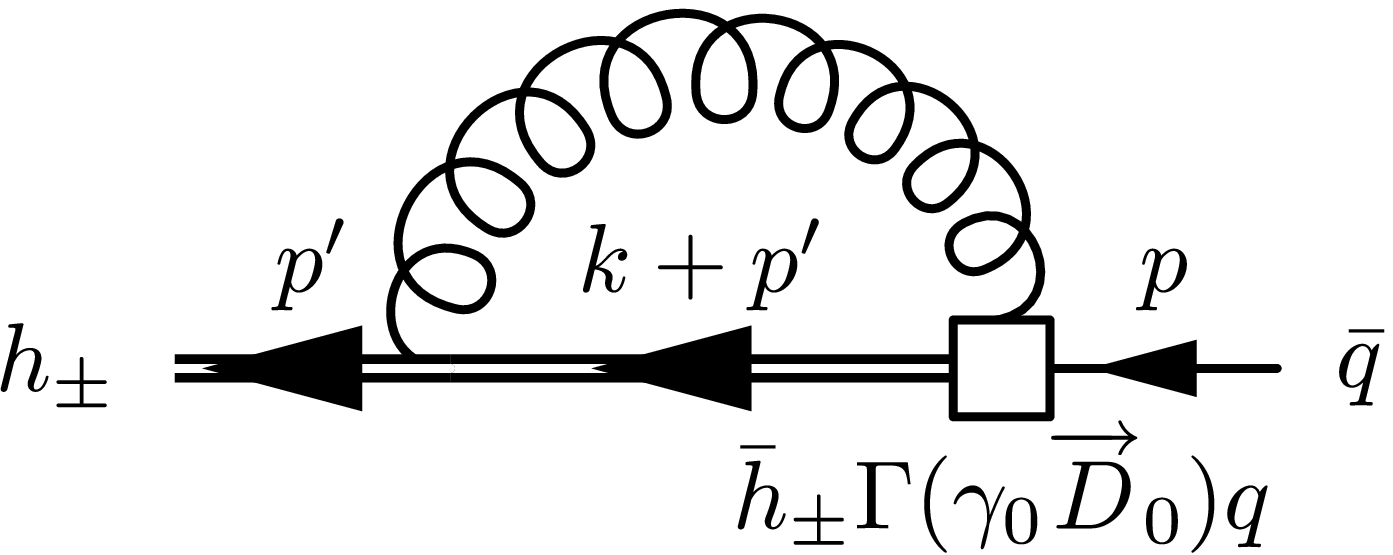}
		\vspace*{-5mm}}
&=&\frac{1}{a}\left[+G\delta M+O(g^4, pa)\right],
\label{EQ:heavy-light_Opa_type1_loop3}\\
\parbox[b][25mm][b]{43mm}{
\includegraphics[scale=0.28, viewport = 0 0 400 300, clip]
                {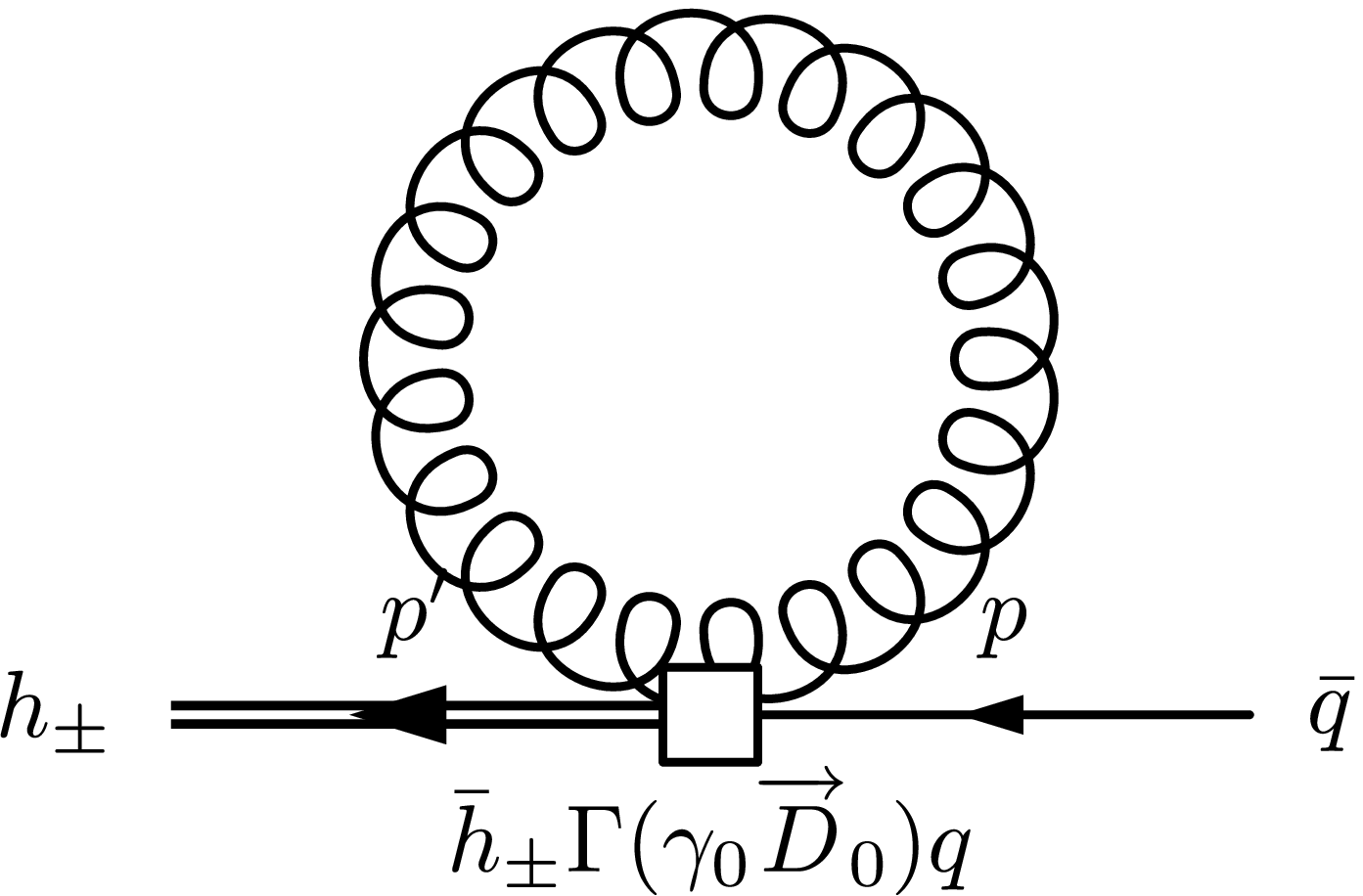}
		\vspace*{-5mm}}
&=&\frac{1}{a}\left[0+O(g^4, pa)\right],\\
\nonumber
\end{eqnarray}
while in the spatial direction $i$ ($=1, 2, 3$) is:
\begin{eqnarray}
\parbox[c][12mm][c]{43mm}{
\includegraphics[scale=0.28, viewport = 0 0 400 180, clip]
                {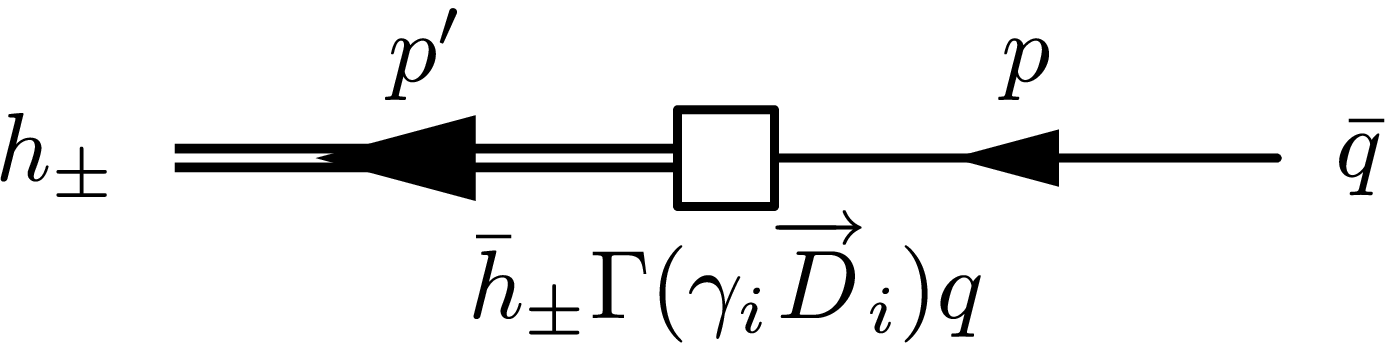}
		\vspace*{+7mm}}
&=&\frac{1}{a}\left[0+O(pa)\right],\\
\parbox[b][15mm][b]{43mm}{
\includegraphics[scale=0.28, viewport = 0 0 400 190, clip]
                {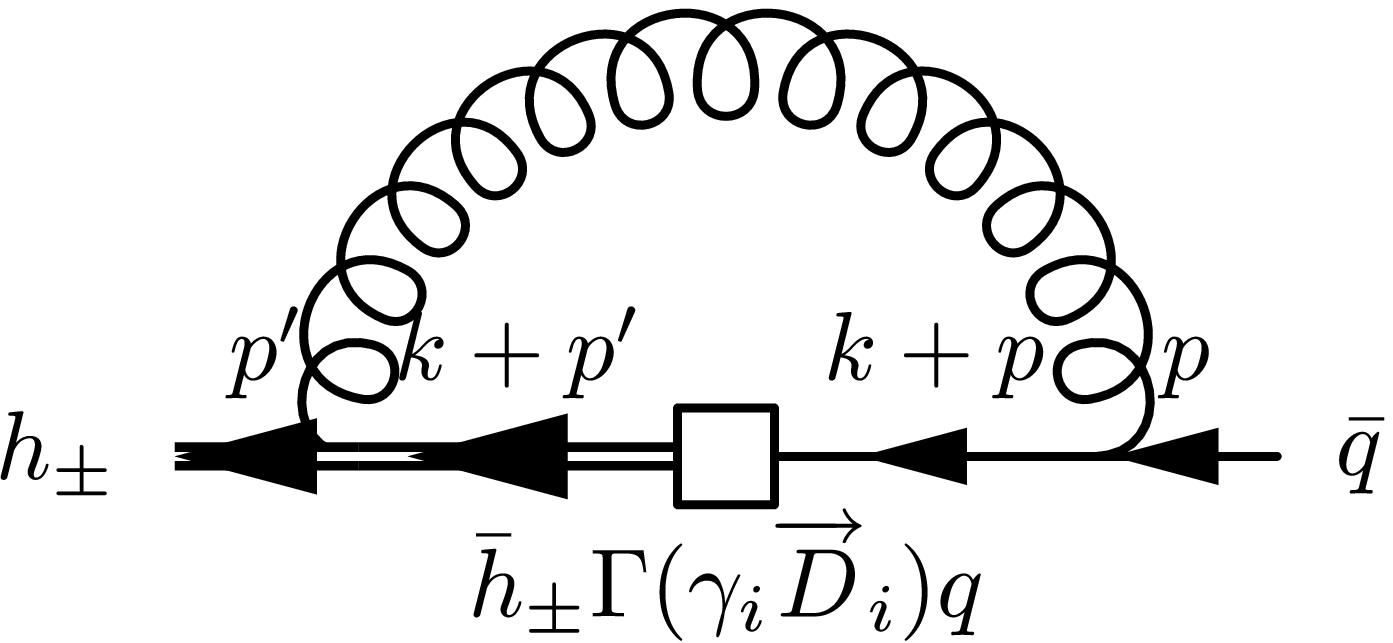}
		\vspace*{-5mm}}
&=&\frac{1}{a}\left[-\frac{1}{3}G\delta M+O(g^4, pa)\right],
\label{EQ:heavy-light_Opa_type5_loop2}
\end{eqnarray}
\begin{eqnarray}
\parbox[b][15mm][b]{43mm}{
\includegraphics[scale=0.28, viewport = 0 0 400 190, clip]
                {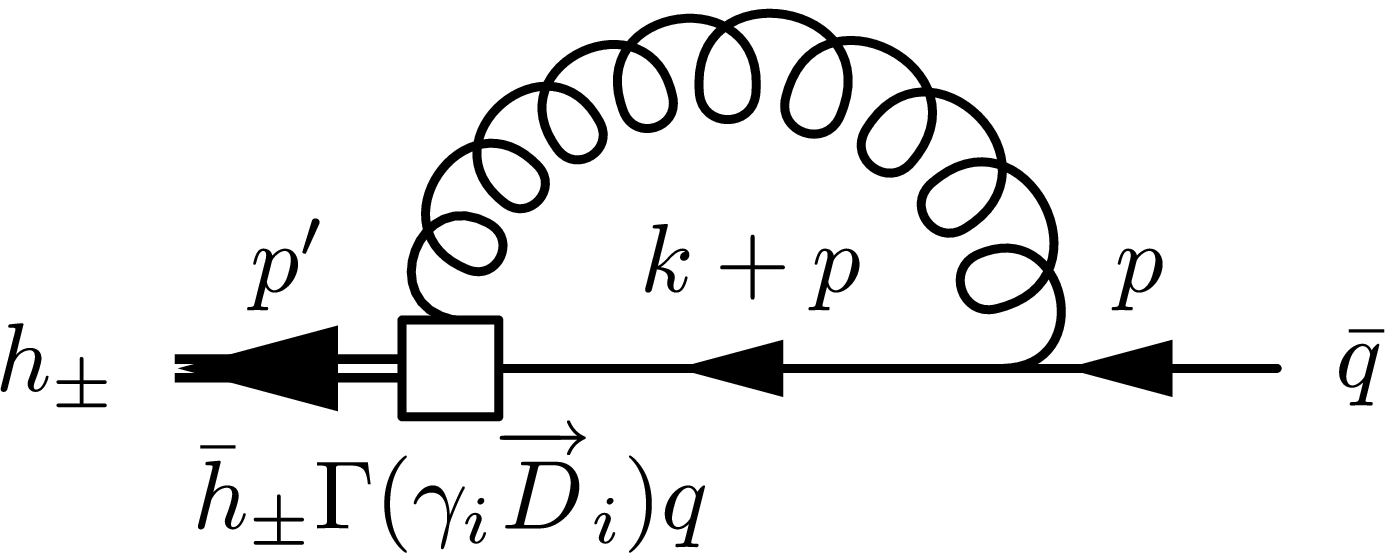}
		\vspace*{-5mm}}
&=&\frac{1}{a}\left[0+O(g^4, pa)\right],\\
\parbox[b][15mm][b]{43mm}{
\includegraphics[scale=0.28, viewport = 0 0 400 190, clip]
                {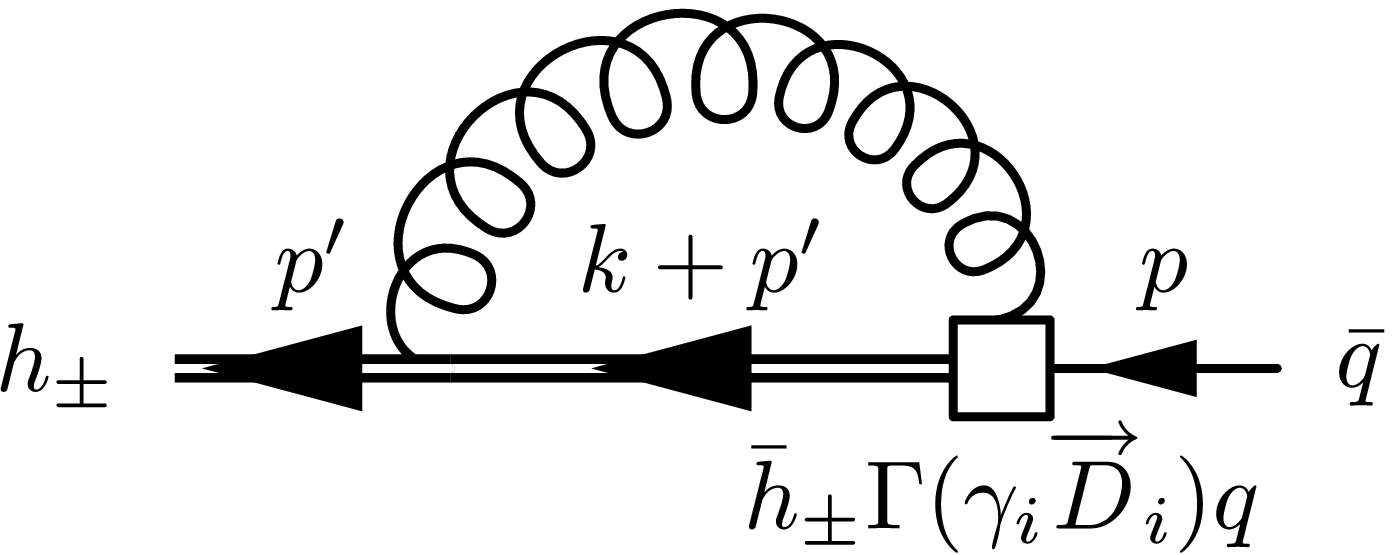}
		\vspace*{-5mm}}
&=&\frac{1}{a}\left[0+O(g^4, pa)\right],\\
\parbox[b][25mm][b]{43mm}{
\includegraphics[scale=0.28, viewport = 0 0 400 300, clip]
                {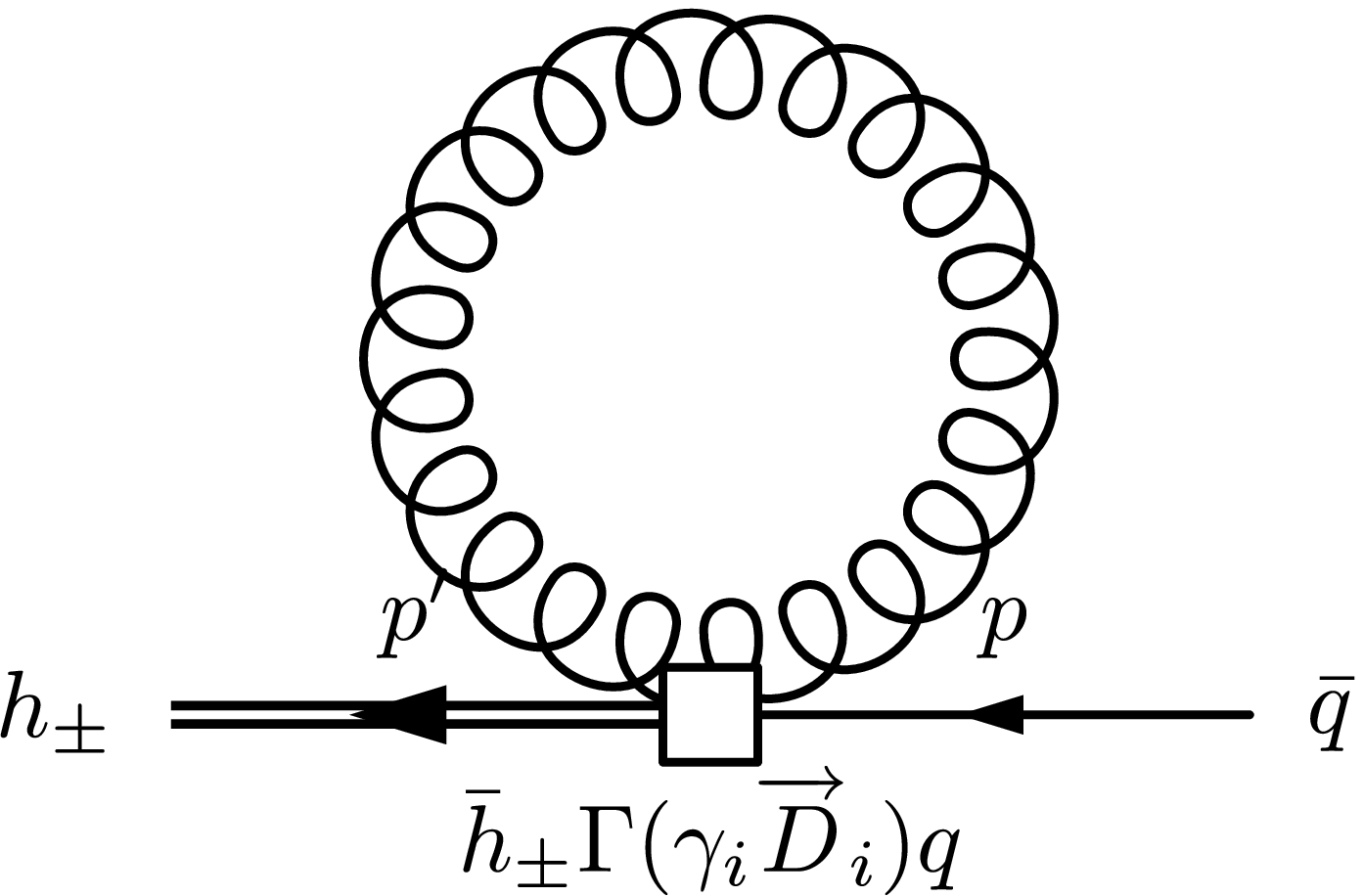}
		\vspace*{-5mm}}
&=&\frac{1}{a}\left[0+O(g^4, pa)\right].\\
\nonumber
\end{eqnarray}
In the temporal direction, the divergence comes from the loop diagram
(\ref{EQ:heavy-light_Opa_type1_loop3}), but from the diagram
(\ref{EQ:heavy-light_Opa_type5_loop2}) in the spatial direction.
These divergences are totally cancelled out in their summation.

\subsubsection*{Equation of motion for static quark}

Because the static quark action employed in this work does not use
the symmetric covariant derivative,
we consider forward and backward derivatives separately to analyze the
divergence structure.
One definition of the covariant derivative, which is read directly from
the static quark action (\ref{EQ:static_action}), is
\begin{eqnarray}
\overline{h}_+\overleftarrow{D}_0&=&
\overline{h}_+(x+\hat{0})U_0^{\dagger}(x)-\overline{h}_+(x),\\
\overline{h}_-\overleftarrow{D}_0&=&
\overline{h}_-(x)-\overline{h}_-(x-\hat{0})U_0(x-\hat{0}).
\end{eqnarray}
This definition has following power divergence structure:
\begin{eqnarray}
\parbox[c][12mm][c]{43mm}{
\includegraphics[scale=0.28, viewport = 0 0 400 180, clip]
                {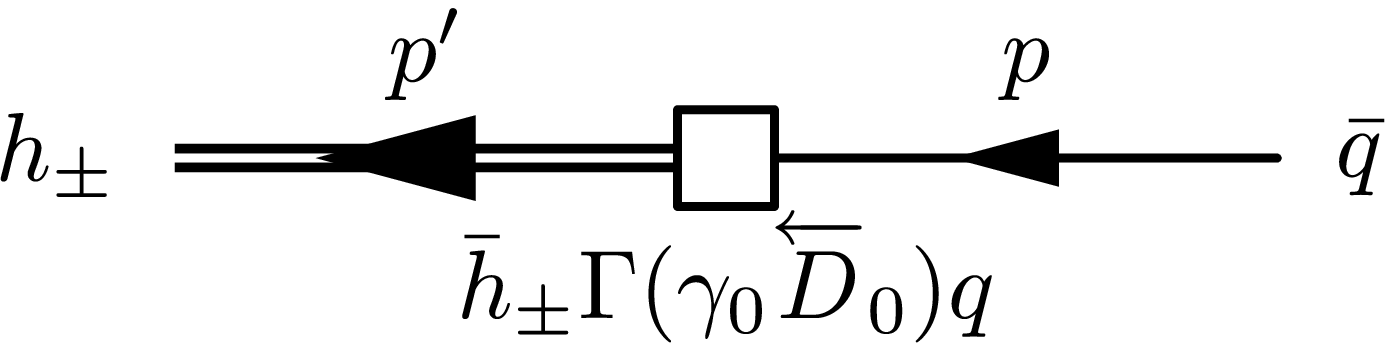}
\vspace*{+7mm}}
&=&\frac{1}{a}\left[G\delta M+O(g^4)\right],
\label{EQ:heavy-light_Opa_type2_tree}\\
&&\hspace*{-49mm}
\parbox[b][15mm][b]{43mm}{
\includegraphics[scale=0.28, viewport = 0 0 400 190, clip]
                {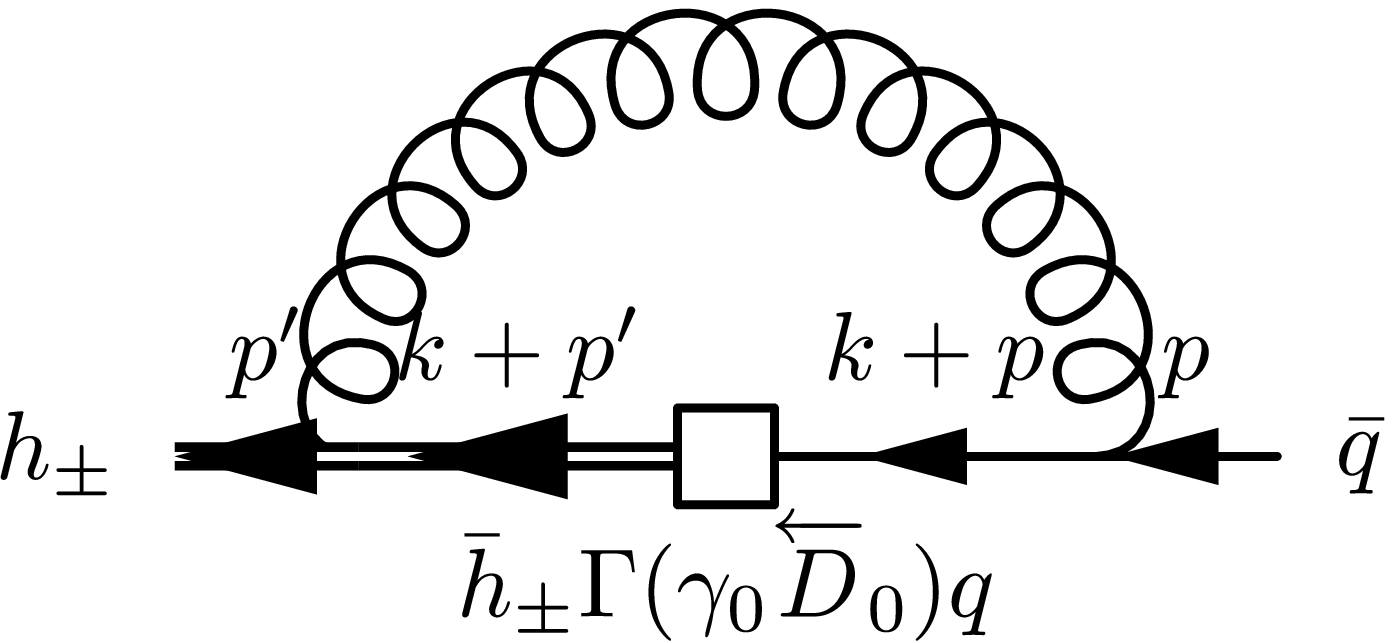}
		\vspace*{-5mm}}
+\;
\parbox[b][15mm][b]{43mm}{
\includegraphics[scale=0.28, viewport = 0 0 400 190, clip]
                {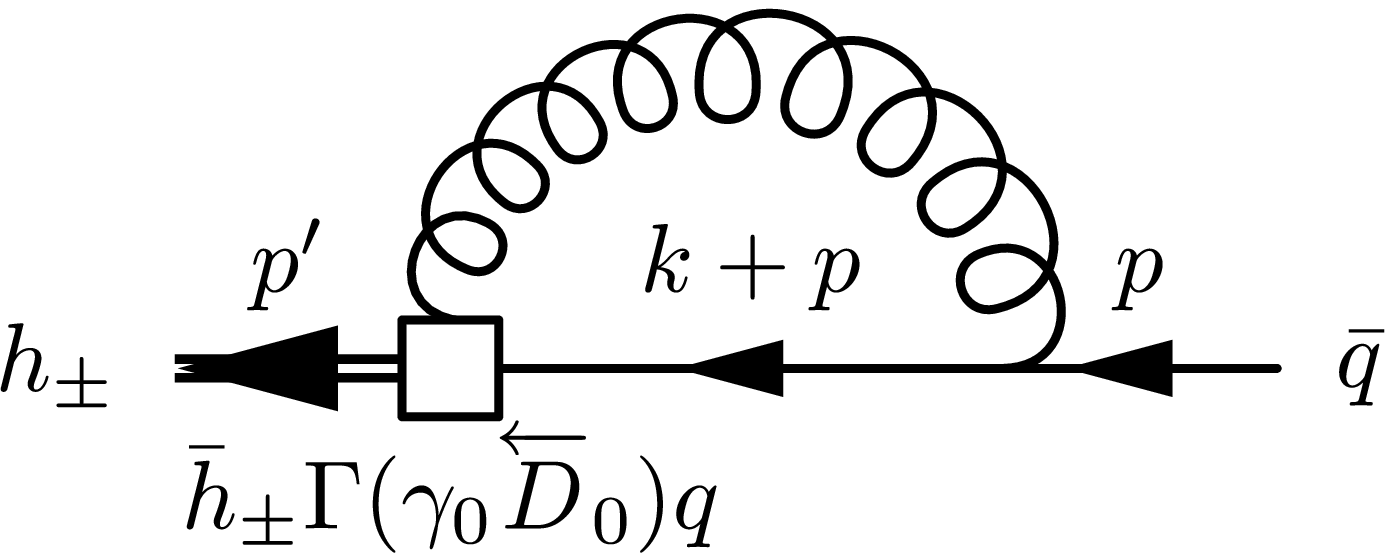}
		\vspace*{-5mm}}
=\frac{1}{a}\left[0+O(g^4)\right],
\label{EQ:heavy-light_Opa_type2_loop12}\\
\parbox[b][15mm][b]{43mm}{
\includegraphics[scale=0.28, viewport = 0 0 400 190, clip]
                {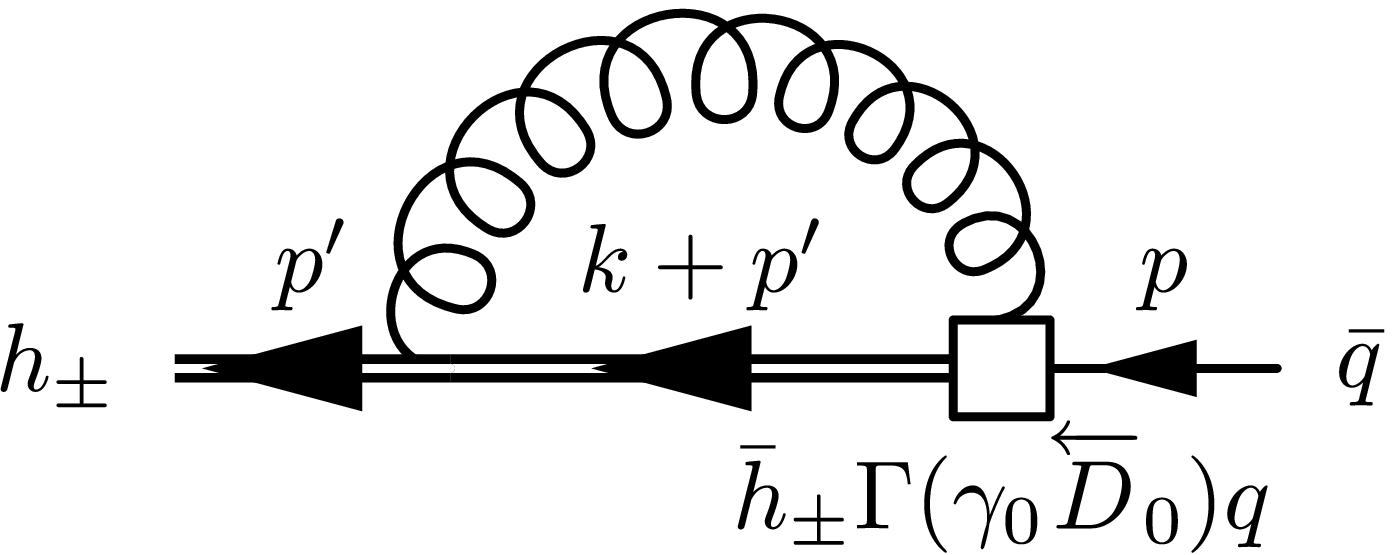}
\vspace*{-5mm}}
&=&\frac{1}{a}\left[-G\delta M+G\frac{g^2}{2}C_FT_4+O(g^4)\right],
\label{EQ:heavy-light_Opa_type2_loop3}\\
\parbox[b][25mm][b]{43mm}{
\includegraphics[scale=0.28, viewport = 0 0 400 300, clip]
                {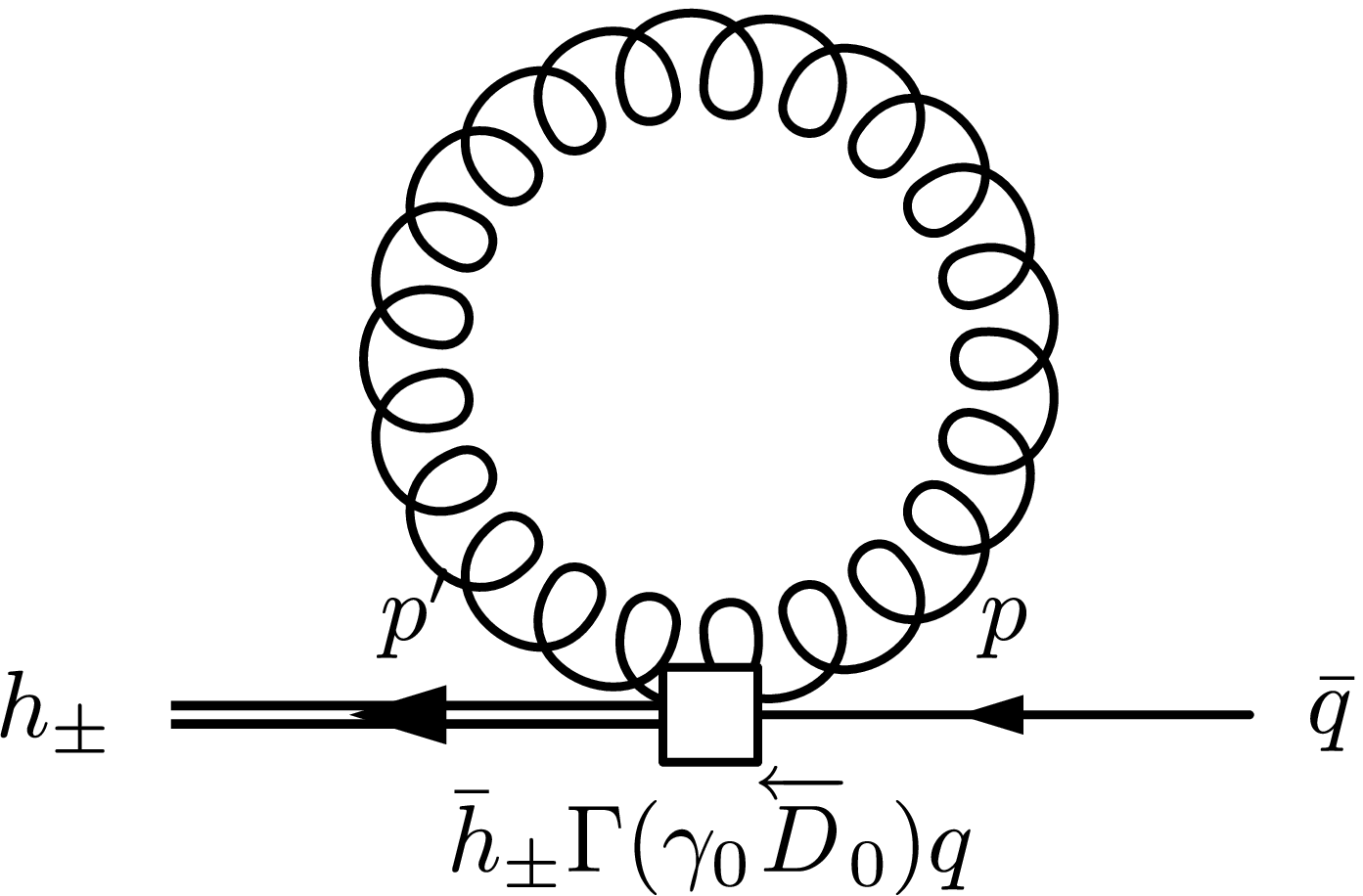}
		\vspace*{-5mm}}
&=&\frac{1}{a}\left[-G\frac{g^2}{2}C_FT_4+O(g^4)\right].
\label{EQ:heavy-light_Opa_type2_loop4}
\end{eqnarray}
The tree diagram (\ref{EQ:heavy-light_Opa_type2_tree}) generates
a $1/a$ power divergence when the one-loop renormalized
on-shell condition for the static quark is imposed.
The loop diagrams also contain power divergences.
An interesting fact is that another $1/a$ power divergence $T_4$,
which is defined by Eq.~(\ref{EQ:T_4}), emerges in the diagrams
(\ref{EQ:heavy-light_Opa_type2_loop3}) and
(\ref{EQ:heavy-light_Opa_type2_loop4}).
In total, all $1/a$ divergences in the operator cancel,
as is consistent with the equation of motion
$\overline{h}\overleftarrow{D}_0=0$.

Another definition of the covariant derivative is
\begin{eqnarray}
\overline{h}_+\overleftarrow{D}'_0
&=&\overline{h}_+(x)-\overline{h}_+(x-\hat{0})U_0(x-\hat{0}),\\
\overline{h}_-\overleftarrow{D}'_0
&=&\overline{h}_-(x+\hat{0})U_0^{\dagger}(x)-\overline{h}_-(x).
\end{eqnarray}
In this definition, we have the divergence structure
\begin{eqnarray}
\parbox[c][12mm][c]{43mm}{
\includegraphics[scale=0.28, viewport = 0 0 400 180, clip]
                {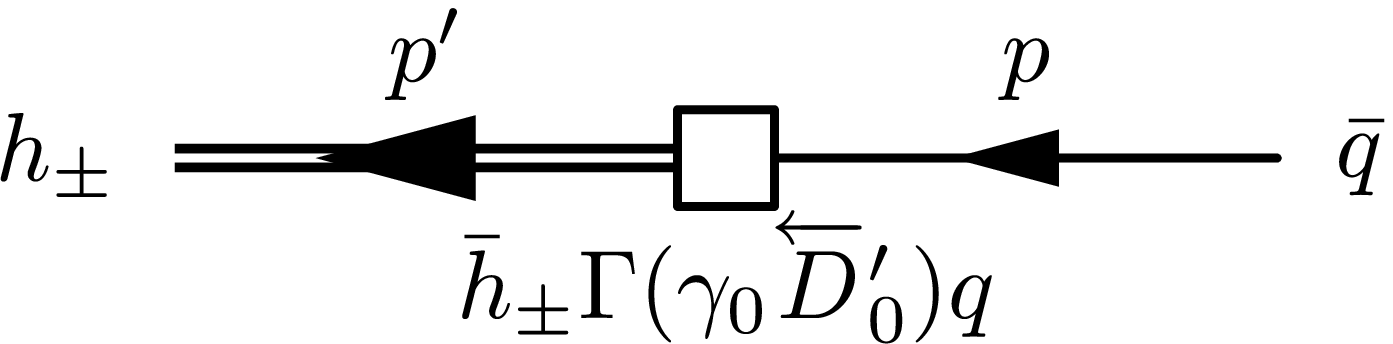}
		\vspace*{+7mm}}
&=&\frac{1}{a}\left[G\delta M+O(g^4)\right],\\
&&\hspace*{-49mm}
\parbox[b][15mm][b]{43mm}{
\includegraphics[scale=0.28, viewport = 0 0 400 190, clip]
                {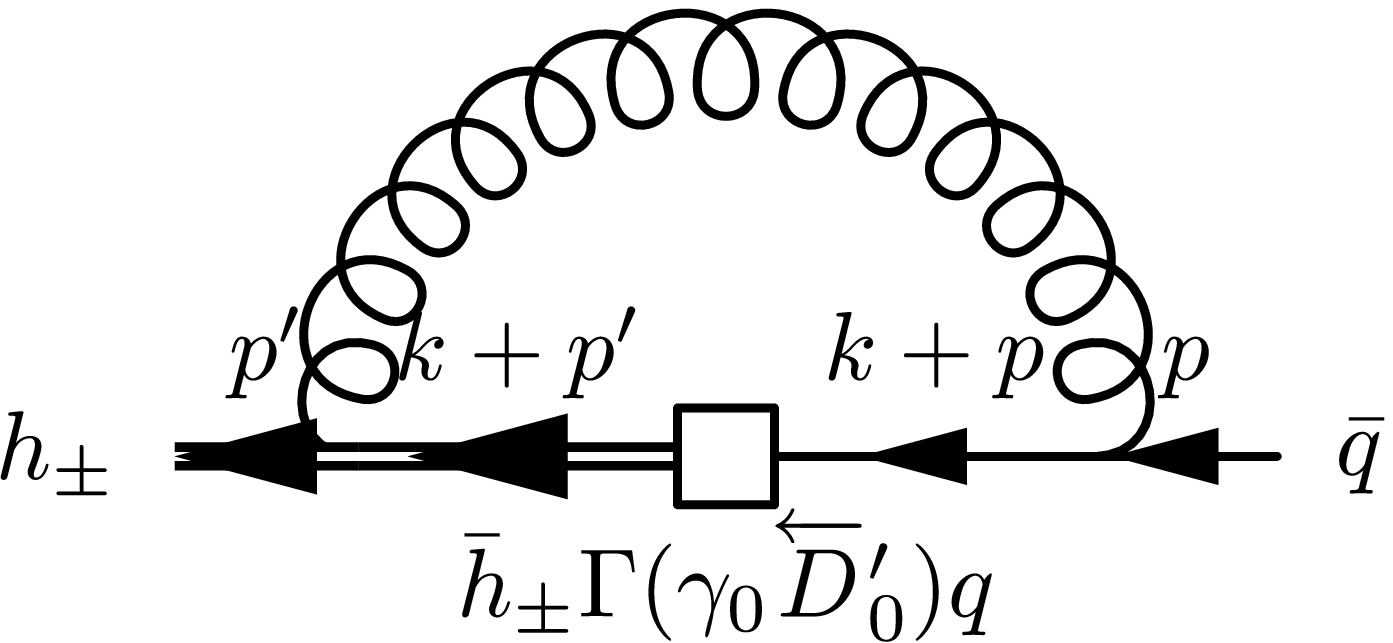}
		\vspace*{-5mm}}
+\;
\parbox[b][15mm][b]{43mm}{
\includegraphics[scale=0.28, viewport = 0 0 400 190, clip]
                {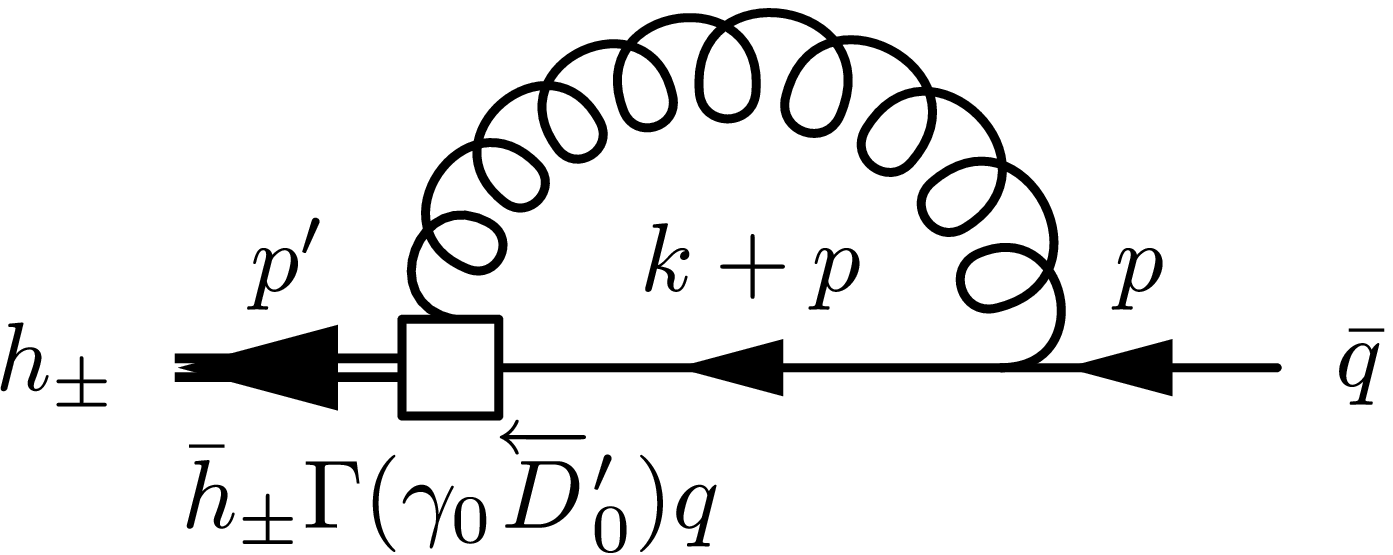}
		\vspace*{-5mm}}
=\frac{1}{a}\left[0+O(g^4)\right],\\
\parbox[b][15mm][b]{43mm}{
\includegraphics[scale=0.28, viewport = 0 0 400 190, clip]
                {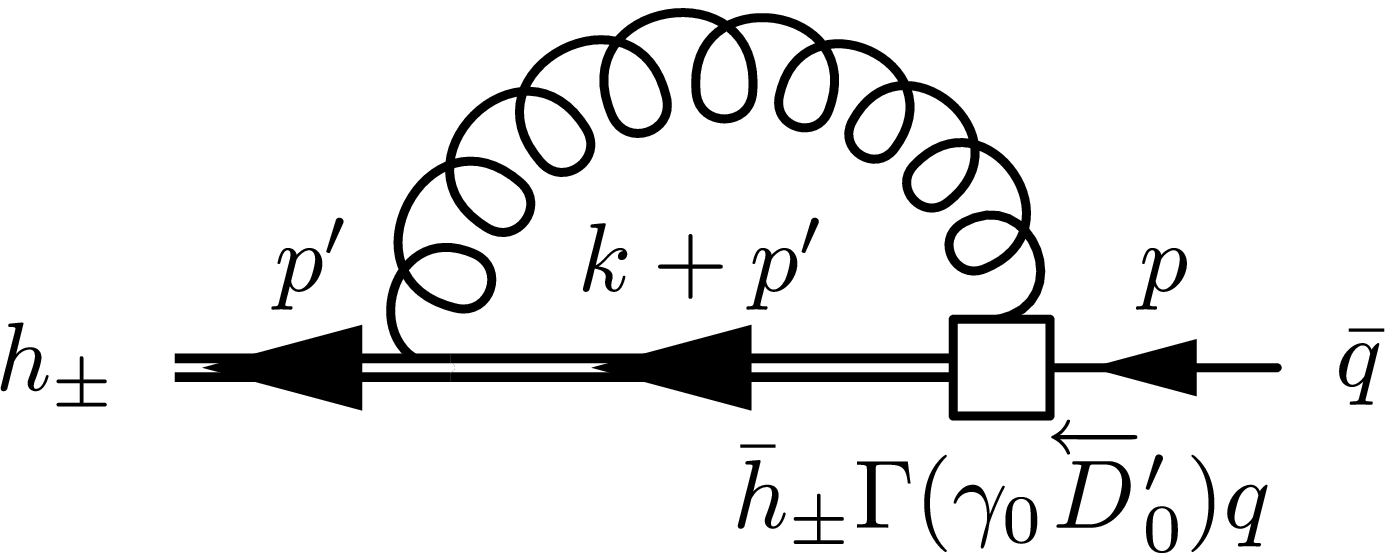}
\vspace*{-5mm}}
&=&\frac{1}{a}\left[-G\delta M-G\frac{g^2}{2}C_FT_4+O(g^4)\right],
\label{EQ:heavy-light_Opa_type3_loop3}\\
\parbox[b][25mm][b]{43mm}{
\includegraphics[scale=0.28, viewport = 0 0 400 300, clip]
                {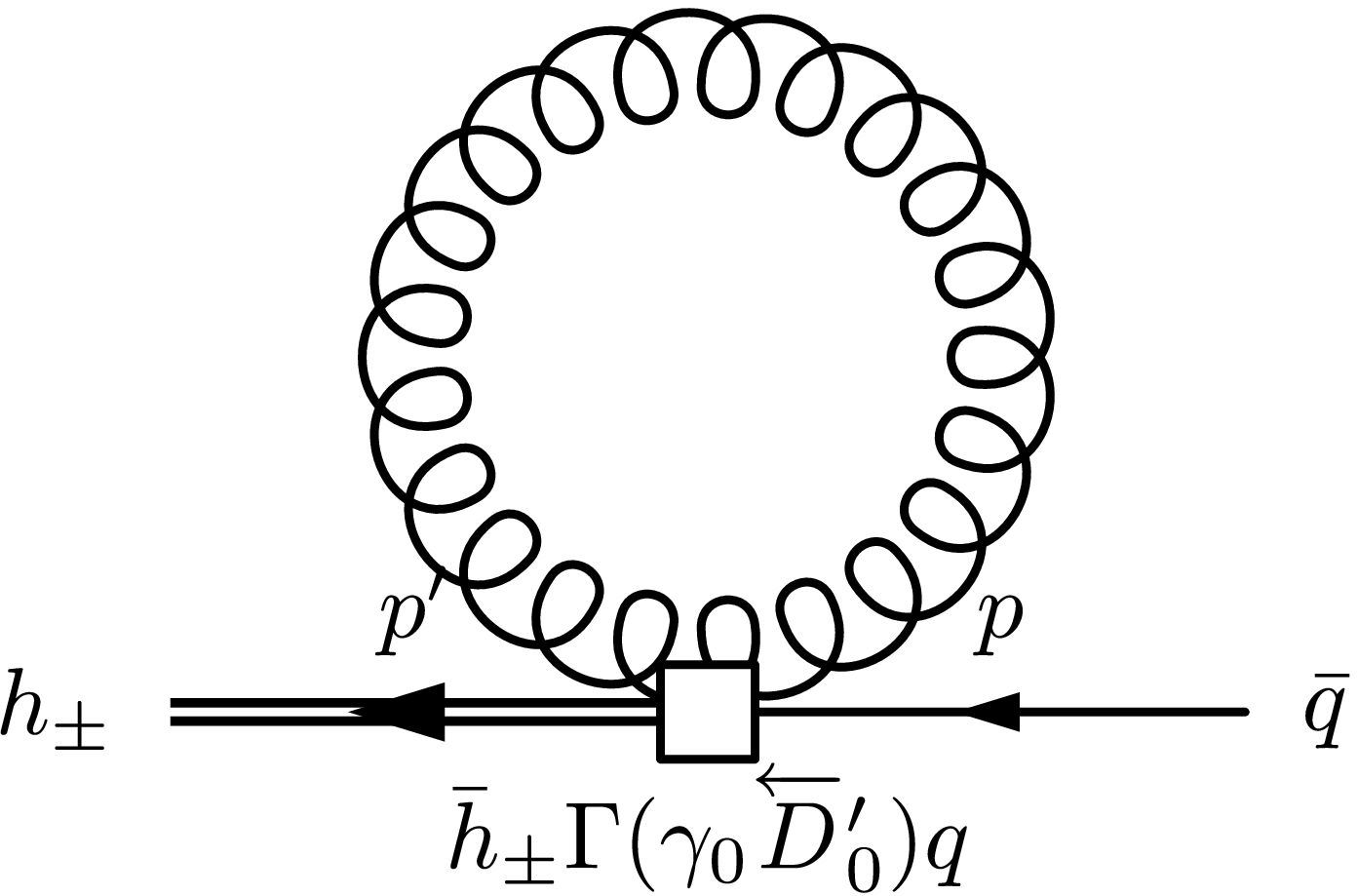}
\vspace*{-5mm}}
&=&\frac{1}{a}\left[G\frac{g^2}{2}C_FT_4+O(g^4)\right].
\label{EQ:heavy-light_Opa_type3_loop4}
\end{eqnarray}
Again, the $T_4$ divergences appear in the diagrams
(\ref{EQ:heavy-light_Opa_type3_loop3}) and
(\ref{EQ:heavy-light_Opa_type3_loop4}).
Note that if we separately
combine diagram (\ref{EQ:heavy-light_Opa_type2_loop3}) with
(\ref{EQ:heavy-light_Opa_type3_loop3}) and
(\ref{EQ:heavy-light_Opa_type2_loop4}) with
(\ref{EQ:heavy-light_Opa_type3_loop4}) respectively,
the divergences $T_4$ vanish.
The symmetric derivative does not induce divergences proportional to $T_4$,
which is consistent with the ${\cal P}\cdot{\cal T}$ symmetry.


\section{Operator mixing in the static heavy and Wilson light quark system}
\label{SEC:Wilson_quark}

In the paper we have used a domain-wall fermion action with chiral symmetry
for the light quarks.
In this appendix we briefly describe the pattern of operator mixing
for Wilson fermions, for which there is explicit chiral symmetry breaking.

Operator mixing for Wilson fermions with Wilson parameter $r$ can
conveniently be investigated by extending $r$ to become a matrix $R$ and
by requiring consistency with the extended chiral symmetry,
$SU(N_F)\times SU(N_F)$. 
The  chiral transformation of $R$ matrix is defined as
\begin{eqnarray}
R\longrightarrow U_L R U_R^{\dagger},\;\;\;
R^{\dagger}\longrightarrow U_R R^{\dagger} U_L^{\dagger},
\label{EQ:ar_spurion}
\end{eqnarray}
similar to Eq.~(\ref{EQ:mass_transformation}).

Analogous to Eq.~(\ref{EQ:matching_bilinear_npr}),
the matching of bilinear operators is
\begin{eqnarray}
\langle{\rm f}|J_{\pm\Gamma}|{\rm i}\rangle_{\rm CHQET}=
\langle{\rm f}|{\cal Z}_{\Gamma}^T{\cal J}_{\pm\Gamma}
|{\rm i}\rangle_{\rm LHQET},
\label{EQ:bilinear_matching_wilson}
\end{eqnarray}
with
\begin{eqnarray}
{\cal Z}_{\Gamma}=
\begin{bmatrix}
Z_{\Gamma}(r^2)+Z_{\Gamma}'(r^2)rG \\
Z_{\Gamma D}(r^2)G+Z_{\Gamma D}'(r^2)r \\
Z_{\Gamma M}(r^2)G+Z_{\Gamma M}'(r^2)r
\end{bmatrix},
\;\;\;
{\cal J}_{\pm\Gamma}=
\begin{bmatrix}
J_{\pm\Gamma} \\
aJ_{\pm\Gamma D} \\
aJ_{\pm\Gamma M}
\end{bmatrix},
\label{EQ:wilson_bilinear_full}
\end{eqnarray}
where each $Z_{\ast}^{\ast}(r^2)$ denotes  functions of $r^2$.
For the matching of the $\Delta B=2$ four quark operators,
we need following operators:
\begin{eqnarray}
O_L&=&[\overline{h}\gamma_{\mu}^Lq][\overline{h}\gamma_{\mu}^Lq],\\
O_N&=&2[\overline{h}\gamma_{\mu}^Rq][\overline{h}\gamma_{\mu}^Lq]
+4[\overline{h}P_Rq][\overline{h}P_Lq],\\
O_{\overline{N}}&=&2[\overline{h}\gamma_{\mu}^Rq][\overline{h}\gamma_{\mu}^Lq]
-4[\overline{h}P_Rq][\overline{h}P_Lq],\\
O_R&=&[\overline{h}\gamma_{\mu}^Rq][\overline{h}\gamma_{\mu}^Rq],\\
O_S&=&[\overline{h}P_Lq][\overline{h}P_Lq],\\
O_T&=&[\overline{h}P_Rq][\overline{h}P_Rq],\\
O_{LD}&=&[\overline{h}\gamma_{\mu}^L(\bm{\gamma}\cdot\overrightarrow{\bm{D}})q]
[\overline{h}\gamma_{\mu}^Lq],\\
O_{SD}&=&[\overline{h}P_L(\bm{\gamma}\cdot\overrightarrow{\bm{D}})q]
[\overline{h}P_Lq],\\
O_{ND}&=&
2[\overline{h}\gamma_{\mu}^R(\bm{\gamma}\cdot\overrightarrow{\bm{D}})q]
[\overline{h}\gamma_{\mu}^Lq]
+4[\overline{h}P_R(\bm{\gamma}\cdot\overrightarrow{\bm{D}})q]
[\overline{h}P_Lq],\\
O_{\overline{ND}}&=&
2[\overline{h}\gamma_{\mu}^R(\bm{\gamma}\cdot\overrightarrow{\bm{D}})q]
[\overline{h}\gamma_{\mu}^Lq]
-4[\overline{h}P_R(\bm{\gamma}\cdot\overrightarrow{\bm{D}})q]
[\overline{h}P_Lq],\\
O_{ND'}&=&
2[\overline{h}\gamma_{\mu}^L(\bm{\gamma}\cdot\overrightarrow{\bm{D}})q]
[\overline{h}\gamma_{\mu}^Rq]
+4[\overline{h}P_L(\bm{\gamma}\cdot\overrightarrow{\bm{D}})q]
[\overline{h}P_Rq],\\
O_{\overline{ND}'}&=&
2[\overline{h}\gamma_{\mu}^L(\bm{\gamma}\cdot\overrightarrow{\bm{D}})q]
[\overline{h}\gamma_{\mu}^Rq]
-4[\overline{h}P_L(\bm{\gamma}\cdot\overrightarrow{\bm{D}})q]
[\overline{h}P_Rq],\\
O_{RD}&=&[\overline{h}\gamma_{\mu}^R(\bm{\gamma}\cdot\overrightarrow{\bm{D}})q]
[\overline{h}\gamma_{\mu}^Rq],\\
O_{TD}&=&[\overline{h}P_R(\bm{\gamma}\cdot\overrightarrow{\bm{D}})q]
[\overline{h}P_Rq],\\
O_{LM}&=&m_q[\overline{h}\gamma_{\mu}^Lq][\overline{h}\gamma_{\mu}^Lq],\\
O_{SM}&=&m_q[\overline{h}P_Lq][\overline{h}P_Lq],\\
O_{NM}&=&m_q\left(2[\overline{h}\gamma_{\mu}^Rq]
[\overline{h}\gamma_{\mu}^Lq]
+4[\overline{h}P_Rq][\overline{h}P_Lq]\right),\\
O_{\overline{NM}}&=&m_q\left(2[\overline{h}\gamma_{\mu}^Rq]
[\overline{h}\gamma_{\mu}^Lq]
-4[\overline{h}P_Rq][\overline{h}P_Lq]\right),\\
O_{RM}&=&m_q[\overline{h}\gamma_{\mu}^Rq][\overline{h}\gamma_{\mu}^Rq],\\
O_{TM}&=&m_q[\overline{h}P_Rq][\overline{h}P_Rq],
\end{eqnarray}
and the matching relations are modified from
 Eqs.~(\ref{EQ:matching_four-quark1_npr}) and
(\ref{EQ:matching_four-quark2_npr}) to
\begin{eqnarray}
\langle{\rm f}|O_L|{\rm i}\rangle_{\rm CHQET}
&=&
\langle{\rm f}|{\cal Z}_1^T{\cal O}_1|{\rm i}\rangle_{\rm LHQET},
\label{EQ:four-quark_matching_wilson}\\
\langle{\rm f}|O_L+4O_S|{\rm i}\rangle_{\rm CHQET}
&=&
\langle{\rm f}|{\cal Z}_2^T{\cal O}_2|{\rm i}\rangle_{\rm LHQET},
\end{eqnarray}
with
\begin{eqnarray}
{\cal Z}_1=
\begin{bmatrix}
Z_L(r^2) \\
Z_N(r^2)r \\
Z_R(r^2)r^2 \\
Z_{LD}(r^2)r \\
Z_{ND}(r^2) \\
Z_{ND'}(r^2)r^2 \\
Z_{RD}(r^2)r \\
Z_{LM}(r^2)r \\
Z_{NM}(r^2) \\
Z_{RM}(r^2)r \\
\end{bmatrix},
\;\;\;&&
{\cal O}_1=
\begin{bmatrix}
O_L \\
O_N \\
O_R \\
aO_{LD} \\
aO_{ND} \\
aO_{ND'} \\
aO_{RD} \\
aO_{LM} \\
aO_{NM} \\
aO_{RM}
\end{bmatrix},
\label{EQ:wilson_four-quark_full}\\
{\cal Z}_2=
\begin{bmatrix}
Z_{L+4S}(r^2) \\
Z_{\overline{N}}(r^2)r \\
Z_{R+4T}(r^2)r^2 \\
Z_{LD+4SD}(r^2)r \\
Z_{\overline{ND}}(r^2) \\
Z_{\overline{ND}'}(r^2)r^2 \\
Z_{RD+4TD}(r^2)r \\
Z_{LM+4SM}(r^2)r \\
Z_{\overline{NM}}(r^2) \\
Z_{RM+4TM}(r^2)r
\end{bmatrix},
\;\;\;&&
{\cal O}_2=
\begin{bmatrix}
O_L+4O_S \\
O_{\overline{N}} \\
O_R+4O_T \\
a(O_{LD}+4O_{SD}) \\
aO_{\overline{ND}} \\
aO_{\overline{ND}'} \\
a(O_{RD}+4O_{TD}) \\
a(O_{LM}+4O_{SM}) \\
aO_{\overline{NM}} \\
a(O_{RM}+4O_{TM})
\end{bmatrix}.
\end{eqnarray}

The one-loop matching factor of the static heavy and clover-Wilson light
quark system was calculated by Ishikawa et al.~\cite{Ishikawa:1998rv},
in which the $O(pa)$ correction
was included but the $O(ma)$ correction was not.
Note that operators $O_{ND'}$ and $O_{RD}$ in
Eq.~(\ref{EQ:wilson_four-quark_full}) are not seen
in Ref.~\cite{Ishikawa:1998rv}, because these operators arise at $O(g^4)$.
We can partly see that these operators actually arise from cross terms
of $O(g^2)$ contributions in Eqs.~(\ref{EQ:bilinear_matching_wilson}) and
(\ref{EQ:wilson_bilinear_full}), which leads to $O(g^4)$.

In the domain-wall quark formalism, similar intrinsic chiral symmetry
breaking occurs for finite extent of the fifth dimension $L_s$.
The size of the breaking effect is characterized by the
residual mass $am_{res}$, which is suppressed as $L_s$ increases.
In principle, similar analysis for the operator mixing for the domain-wall
fermion with finite $L_s$ would be possible by introducing extended
chiral symmetry for $am_{res}\neq 0$ similar to Eq.~(\ref{EQ:ar_spurion}).
In actual numerical simulations using the domain-wall fermion,
the residual mass is well controlled, $am_{res}\sim O(10^{-3})$,
which is negligible in practice.

\newpage
\section{Tables of numerical values of domain-wall light quark part}
\label{SEC:tables_light}

The values in this appendix are all cited from
Refs.~\cite{Aoki:2002iq, Aoki:2003uf}.

\begin{table}[H]
\begin{center}
\begin{tabular}{c|cccc}
\hline\hline
\multirow{2}{5mm}{$M_5$} &
\multicolumn{4}{c}{$\Sigma_w$} \\
 & Plaquette & Symanzik & Iwasaki & DBW2 \\
\hline
$0.1$ & $-51.048195$   & $-40.074665$   & $-25.740748$   & $-11.263725$   \\
$0.2$ & $-50.744997$   & $-39.788806$   & $-25.489672$   & $-11.080847$   \\
$0.3$ & $-50.488509$   & $-39.548978$   & $-25.282889$   & $-10.936985$   \\
$0.4$ & $-50.266419$   & $-39.342775$   & $-25.107854$   & $-10.819677$   \\
$0.5$ & $-50.072616$   & $-39.163976$   & $-24.958176$   & $-10.722539$   \\
$0.6$ & $-49.903778$   & $-39.009135$   & $-24.830209$   & $-10.641842$   \\
$0.7$ & $-49.758186$   & $-38.876383$   & $-24.721839$   & $-10.575304$   \\
$0.8$ & $-49.635208$   & $-38.764914$   & $-24.631970$   & $-10.521558$   \\
$0.9$ & $-49.535083$   & $-38.674758$   & $-24.560281$   & $-10.479897$   \\
$1.0$ & $-49.458848$   & $-38.606702$   & $-24.507131$   & $-10.450158$   \\
$1.1$ & $-49.408369$   & $-38.562312$   & $-24.473568$   & $-10.432689$   \\
$1.2$ & $-49.386472$   & $-38.544048$   & $-24.461408$   & $-10.428393$   \\
$1.3$ & $-49.397179$   & $-38.555483$   & $-24.473426$   & $-10.438846$   \\
$1.4$ & $-49.446110$   & $-38.601678$   & $-24.513680$   & $-10.466525$   \\
$1.5$ & $-49.541138$   & $-38.689808$   & $-24.588063$   & $-10.515219$   \\
$1.6$ & $-49.693506$   & $-38.830222$   & $-24.705264$   & $-10.590776$   \\
$1.7$ & $-49.919841$   & $-39.038393$   & $-24.878568$   & $-10.702549$   \\
$1.8$ & $-50.246184$   & $-39.338831$   & $-25.129527$   & $-10.866515$   \\
$1.9$ & $-50.717592$   & $-39.774527$   & $-25.497027$   & $-11.113350$   \\
\hline\hline
\end{tabular}
\caption{Numerical values of $\Sigma_w$.
The values are cited from Ref.~\cite{Aoki:2002iq}.}
\label{TAB:Sigma_w}
\end{center}
\end{table}

\newpage

\begin{table}[H]
\begin{center}
\begin{tabular}{c|cccc}
\hline\hline
\multirow{2}{5mm}{$M_5$} &
\multicolumn{4}{c}{$f$} \\
 & Plaquette & Symanzik & Iwasaki & DBW2 \\
\hline
$0.1$ & $13.1613(11)$  & $9.5458(11)$   & $4.6519(11)$   & $-0.5152(11)$  \\
$0.2$ & $13.0109(11)$  & $9.4009(11)$   & $4.5193(11)$   & $-0.6190(11)$  \\
$0.3$ & $12.8838(11)$  & $9.2791(11)$   & $4.4093(11)$   & $-0.7020(11)$  \\
$0.4$ & $12.7739(11)$  & $9.1743(11)$   & $4.3158(11)$   & $-0.7704(11)$  \\
$0.5$ & $12.6781(11)$  & $9.0834(11)$   & $4.2354(11)$   & $-0.8276(11)$  \\
$0.6$ & $12.5948(11)$  & $9.0046(11)$   & $4.1665(11)$   & $-0.8754(11)$  \\
$0.7$ & $12.5230(11)$  & $8.9370(11)$   & $4.1079(11)$   & $-0.9151(11)$  \\
$0.8$ & $12.4625(11)$  & $8.8803(11)$   & $4.0593(11)$   & $-0.9474(11)$  \\
$0.9$ & $12.4133(11)$  & $8.8344(11)$   & $4.0204(11)$   & $-0.9725(11)$  \\
$1.0$ & $12.3760(11)$  & $8.7998(11)$   & $3.9915(11)$   & $-0.9905(11)$  \\
$1.1$ & $12.3513(11)$  & $8.7773(11)$   & $3.9731(11)$   & $-1.0012(11)$  \\
$1.2$ & $12.3408(11)$  & $8.7681(11)$   & $3.9664(11)$   & $-1.0041(11)$  \\
$1.3$ & $12.3464(11)$  & $8.7740(11)$   & $3.9727(11)$   & $-0.9981(11)$  \\
$1.4$ & $12.3708(11)$  & $8.7976(11)$   & $3.9943(11)$   & $-0.9818(11)$  \\
$1.5$ & $12.4179(11)$  & $8.8426(11)$   & $4.0343(11)$   & $-0.9531(11)$  \\
$1.6$ & $12.4932(11)$  & $8.9141(11)$   & $4.0974(11)$   & $-0.9085(11)$  \\
$1.7$ & $12.6050(11)$  & $9.0200(11)$   & $4.1905(11)$   & $-0.8427(11)$  \\
$1.8$ & $12.7661(11)$  & $9.1726(11)$   & $4.3249(11)$   & $-0.7469(11)$  \\
$1.9$ & $12.9989(11)$  & $9.3936(11)$   & $4.5209(11)$   & $-0.6041(11)$  \\
\hline\hline
\end{tabular}
\end{center}
\begin{center}
\begin{tabular}{c|cccc}
\hline\hline
\multirow{2}{5mm}{$M_5$} &
\multicolumn{4}{c}{$v$} \\
 & Plaquette & Symanzik & Iwasaki & DBW2 \\
\hline
$0.1$ & $-3.8182(35)$  & $-3.4782(35)$  & $-2.7691(35)$  & $-1.1915(35)$  \\
$0.2$ & $-4.6031(37)$  & $-4.2403(36)$  & $-3.4827(37)$  & $-1.7967(36)$  \\
$0.3$ & $-5.2388(34)$  & $-4.8520(35)$  & $-4.0444(34)$  & $-2.2502(34)$  \\
$0.4$ & $-5.7981(36)$  & $-5.3863(36)$  & $-4.5265(36)$  & $-2.6240(36)$  \\
$0.5$ & $-6.3138(36)$  & $-5.8754(36)$  & $-4.9616(36)$  & $-2.9496(36)$  \\
$0.6$ & $-6.8035(36)$  & $-6.3371(36)$  & $-5.3668(36)$  & $-3.2439(36)$  \\
$0.7$ & $-7.2795(35)$  & $-6.7834(36)$  & $-5.7538(35)$  & $-3.5176(35)$  \\
$0.8$ & $-7.7511(35)$  & $-7.2234(35)$  & $-6.1314(35)$  & $-3.7789(35)$  \\
$0.9$ & $-8.2259(36)$  & $-7.6643(37)$  & $-6.5064(36)$  & $-4.0335(36)$  \\
$1.0$ & $-8.7112(36)$  & $-8.1136(35)$  & $-6.8854(36)$  & $-4.2871(35)$  \\
$1.1$ & $-9.2169(19)$  & $-8.5801(19)$  & $-7.2772(19)$  & $-4.5470(19)$  \\
$1.2$ & $-9.7463(30)$  & $-9.0671(30)$  & $-7.6839(30)$  & $-4.8141(30)$  \\
$1.3$ & $-10.3140(34)$ & $-9.5888(33)$  & $-8.1186(34)$  & $-5.0994(34)$  \\
$1.4$ & $-10.9324(33)$ & $-10.1564(37)$ & $-8.5918(33)$  & $-5.4109(34)$  \\
$1.5$ & $-11.6187(37)$ & $-10.7872(37)$ & $-9.1182(37)$  & $-5.7606(37)$  \\
$1.6$ & $-12.3986(37)$ & $-11.5052(34)$ & $-9.7205(37)$  & $-6.1668(36)$  \\
$1.7$ & $-13.3112(37)$ & $-12.3479(37)$ & $-10.4338(37)$ & $-6.6591(37)$  \\
$1.8$ & $-14.4234(37)$ & $-13.3811(36)$ & $-11.3197(37)$ & $-7.2913(38)$  \\
$1.9$ & $-15.8762(35)$ & $-14.7426(37)$ & $-12.5115(35)$ & $-8.1852(35)$  \\
\hline\hline
\end{tabular}
\caption{Numerical values of $f$ and $v$.
The values are produced using the data in Ref.~\cite{Aoki:2002iq}.}
\label{TAB:f_v}
\end{center}
\end{table}

\newpage

\begin{table}[H]
\begin{center}
\begin{tabular}{c|c||c|c}
\hline\hline
$M_5$ & $d_f$ & $M_5$ & $d_f$ \\
\hline
$0.05$ & $-0.028209(17)$  & $1.05$ & $-0.0037322(49)$ \\
$0.10$ & $-0.023030(14)$  & $1.10$ & $-0.0032266(45)$ \\
$0.15$ & $-0.020075(13)$  & $1.15$ & $-0.0027422(43)$ \\
$0.20$ & $-0.017981(11)$  & $1.20$ & $-0.0022896(41)$ \\
$0.25$ & $-0.016332(11)$  & $1.25$ & $-0.0018718(39)$ \\
$0.30$ & $-0.0149658(97)$ & $1.30$ & $-0.0014846(37)$ \\
$0.35$ & $-0.0137846(96)$ & $1.35$ & $-0.0011441(36)$ \\
$0.40$ & $-0.0127360(87)$ & $1.40$ & $-0.0008650(35)$ \\
$0.45$ & $-0.0117805(83)$ & $1.45$ & $-0.0006560(34)$ \\
$0.50$ & $-0.0109029(79)$ & $1.50$ & $-0.0005360(33)$ \\
$0.55$ & $-0.0100864(76)$ & $1.55$ & $-0.0005236(33)$ \\
$0.60$ & $-0.0093153(73)$ & $1.60$ & $-0.0006566(34)$ \\
$0.65$ & $-0.0085901(67)$ & $1.65$ & $-0.0009842(36)$ \\
$0.70$ & $-0.0078955(64)$ & $1.70$ & $-0.0015704(39)$ \\
$0.75$ & $-0.0072294(63)$ & $1.75$ & $-0.0025200(43)$ \\
$0.80$ & $-0.0065893(62)$ & $1.80$ & $-0.0040140(51)$ \\
$0.85$ & $-0.0059729(55)$ & $1.85$ & $-0.0063612(63)$ \\
$0.90$ & $-0.0053788(55)$ & $1.90$ & $-0.0102019(82)$ \\
$0.95$ & $-0.0048118(51)$ & $1.95$ & $-0.017386(12)$  \\
$1.00$ & $-0.0042608(51)$ &        &                 \\
\hline\hline
\end{tabular}
\caption{Numerical values of $d_f$.
The values are cited from Ref.~\cite{Aoki:2003uf}.}
\label{TAB:df}
\end{center}
\end{table}

\newpage
\section{Tables of numerical values of integrals}
\label{SEC:tables}

To perform the numerical integration,
we use the Monte Carlo integration routine VEGAS \cite{Lepage:1977sw}.

\begin{table}[H]
\begin{center}
\begin{tabular}{c|cccc}
\hline\hline
\multirow{2}{22mm}{gluon action} &
\multicolumn{4}{c}{${\cal R}^{\widetilde{c_1}-{\rm PL}}$} \\
 & unsmear & APE & HYP1 & HYP2 \\
\hline
Plaquette & $0.0$            & $-8.155358(22)$  & $-7.768507(29)$  & $-12.346005(55)$ \\
Symanzik  & $-0.595480(4)$   & $-8.157309(24)$  & $-8.061365(30)$  & $-11.862191(52)$ \\
Iwasaki   & $-2.620439(12)$  & $-8.610915(27)$  & $-8.769236(31)$  & $-11.574859(51)$ \\
DBW2      & $-7.020924(38)$  & $-10.287034(43)$ & $-10.528146(45)$ & $-12.174154(55)$ \\
\hline\hline
\end{tabular}
\caption{Numerical values of integrals ${\cal R}^{\widetilde{c_1}-{\rm PL}}$.}
\label{TAB:R}
\end{center}
\end{table}
\begin{table}[H]
\begin{center}
\begin{tabular}{c|cccc}
\hline\hline
\multirow{2}{22mm}{gluon action} &
\multicolumn{4}{c}{$T_3^{\widetilde{c_1}-{\rm PL}}$} \\
 & unsmear & APE & HYP1 & HYP2 \\
\hline
Plaquette & $0.0$            & $-0.166666(1)$   & $-0.179744(1)$   & $-0.199563(1)$   \\
Symanzik  & $-0.033703(0)$   & $-0.172377(0)$   & $-0.183290(1)$   & $-0.201904(1)$   \\
Iwasaki   & $-0.088361(0)$   & $-0.182888(1)$   & $-0.190549(1)$   & $-0.206241(1)$   \\
DBW2      & $-0.158892(1)$   & $-0.201084(1)$   & $-0.204909(1)$   & $-0.214596(1)$   \\
\hline\hline
\end{tabular}
\caption{Numerical values of integrals $T_3^{\widetilde{c_1}-{\rm PL}}$.}
\label{TAB:T_3}
\end{center}
\end{table}
\begin{table}[H]
\begin{center}
\begin{tabular}{c|cccc}
\hline\hline
\multirow{2}{22mm}{gluon action} &
\multicolumn{4}{c}{$\delta\hat{M}$} \\
 & unsmear & APE & HYP1 & HYP2 \\
\hline
Plaquette & $19.95473(10)$   & $6.79531(11)$    & $5.76272(11)$    & $4.19788(13)$    \\
Symanzik  & $17.29363(10)$   & $6.34435(11)$    & $5.48274(11)$    & $4.01300(13)$    \\
Iwasaki   & $12.97799(11)$   & $5.51444(11)$    & $4.90959(11)$    & $3.67056(12)$    \\
DBW2      & $7.40912(12)$    & $4.07778(12)$    & $3.77575(12)$    & $3.01091(13)$    \\
\hline\hline
\end{tabular}
\caption{Numerical values of $\delta\hat{M}$.}
\label{TAB:delta_M}
\end{center}
\end{table}
\begin{table}[H]
\begin{center}
\begin{tabular}{c|cccc}
\hline\hline
\multirow{2}{22mm}{gluon action} &
\multicolumn{4}{c}{$e$} \\
 & unsmear & APE & HYP1 & HYP2 \\
\hline
Plaquette & $24.48049(10)$   & $3.16571(11)$    & $2.51997(12)$    & $-3.62237(14)$   \\
Symanzik  & $21.22391(10)$   & $2.71280(12)$    & $1.94714(12)$    & $-3.32344(14)$   \\
Iwasaki   & $14.88332(11)$   & $1.42929(12)$    & $0.66611(12)$    & $-3.37854(14)$   \\
DBW2      & $4.91396(13)$    & $-1.68349(13)$   & $-2.22663(13)$   & $-4.63749(14)$   \\
\hline\hline
\end{tabular}
\caption{Numerical values of $e$.}
\label{TAB:e}
\end{center}
\end{table}

\newpage

\begin{table}[H]
\begin{center}
\begin{tabular}{c|cc|cc}
\hline\hline
$M_5$ &
${\cal I}_{\chi}^{{\rm PL}-{\rm div}}$ &
${\cal I}_w^{\rm PL}$ &
${\cal J}_{\chi}^{{\rm PL}-{\rm div}}$ &
${\cal J}_w^{\rm PL}$ \\
\hline
$0.1$ & $-1.317357(54)$  & $-4.034531(33)$  & $-2.55733(63)$   & $13.49853(12)$   \\
$0.2$ & $-1.496882(49)$  & $-3.872305(25)$  & $-0.47980(56)$   & $11.072832(82)$  \\
$0.3$ & $-1.648560(46)$  & $-3.738575(21)$  & $0.73748(38)$    & $9.680365(61)$   \\
$0.4$ & $-1.786090(44)$  & $-3.620014(18)$  & $1.62142(33)$    & $8.689872(48)$   \\
$0.5$ & $-1.915172(44)$  & $-3.510747(16)$  & $2.33301(29)$    & $7.907941(39)$   \\
$0.6$ & $-2.039462(43)$  & $-3.407449(14)$  & $2.94407(27)$    & $7.249261(31)$   \\
$0.7$ & $-2.161204(43)$  & $-3.307920(12)$  & $3.49325(30)$    & $6.668012(21)$   \\
$0.8$ & $-2.282065(43)$  & $-3.210533(10)$  & $4.00481(32)$    & $6.135881(16)$   \\
$0.9$ & $-2.403557(45)$  & $-3.113948(9)$   & $4.49564(27)$    & $5.633217(11)$   \\
$1.0$ & $-2.527097(45)$  & $-3.016961(6)$   & $4.97879(29)$    & $5.144634(4)$    \\
$1.1$ & $-2.654088(46)$  & $-2.918368(8)$   & $5.46667(30)$    & $4.656455(10)$   \\
$1.2$ & $-2.785922(47)$  & $-2.816917(9)$   & $5.97114(39)$    & $4.154848(13)$   \\
$1.3$ & $-2.924429(48)$  & $-2.711123(10)$  & $6.50702(41)$    & $3.623851(16)$   \\
$1.4$ & $-3.071762(49)$  & $-2.599160(11)$  & $7.09195(44)$    & $3.042663(22)$   \\
$1.5$ & $-3.230735(51)$  & $-2.478608(12)$  & $7.75540(40)$    & $2.380852(30)$   \\
$1.6$ & $-3.405303(53)$  & $-2.346024(14)$  & $8.53794(43)$    & $1.588916(37)$   \\
$1.7$ & $-3.601408(58)$  & $-2.196048(16)$  & $9.52090(47)$    & $0.574365(48)$   \\
$1.8$ & $-3.829124(59)$  & $-2.019452(20)$  & $10.88105(65)$   & $-0.871646(65)$  \\
$1.9$ & $-4.109490(66)$  & $-1.796311(27)$  & $13.18814(87)$   & $-3.42206(10)$   \\
\hline\hline
\end{tabular}
\end{center}
\begin{center}
\begin{tabular}{c|cc}
\hline\hline
$M_5$ &
${\cal J}_{\chi'}^{{\rm PL}-{\rm div}}$ &
${\cal K}_{\chi}^{{\rm PL}-{\rm div}}$ \\
\hline
$0.1$ & $-0.03163(56)$   & $-41.6877(16)$   \\
$0.2$ & $1.23722(41)$    & $-19.57253(70)$  \\
$0.3$ & $1.99022(35)$    & $-11.74501(46)$  \\
$0.4$ & $2.54399(30)$    & $-7.55285(36)$   \\
$0.5$ & $2.99506(27)$    & $-4.82466(43)$   \\
$0.6$ & $3.38651(26)$    & $-2.82047(29)$   \\
$0.7$ & $3.74227(26)$    & $-1.21451(20)$   \\
$0.8$ & $4.07642(27)$    & $0.16622(17)$    \\
$0.9$ & $4.39987(29)$    & $1.42852(17)$    \\
$1.0$ & $4.72007(34)$    & $2.64870(20)$    \\
$1.1$ & $5.04472(32)$    & $3.89320(23)$    \\
$1.2$ & $5.38204(35)$    & $5.23129(33)$    \\
$1.3$ & $5.74066(34)$    & $6.75210(44)$    \\
$1.4$ & $6.13289(37)$    & $8.58591(35)$    \\
$1.5$ & $6.57496(39)$    & $10.95370(62)$   \\
$1.6$ & $7.09382(45)$    & $14.28337(50)$   \\
$1.7$ & $7.73863(46)$    & $19.55246(79)$   \\
$1.8$ & $8.61765(53)$    & $29.65130(89)$   \\
$1.9$ & $10.07462(69)$   & $58.8508(17)$    \\
\hline\hline
\end{tabular}
\caption{Numerical values of integrals
${\cal I}_{\chi}^{{\rm PL}-{\rm div}}$, ${\cal I}_w^{\rm PL}$,
${\cal J}_{\chi}^{{\rm PL}-{\rm div}}$, ${\cal J}_w^{\rm PL}$,
${\cal J}_{\chi'}^{{\rm PL}-{\rm div}}$ and
${\cal K}_{\chi}^{{\rm PL}-{\rm div}}$.}
\label{TAB:VXX_X_pl}
\end{center}
\end{table}

\newpage

\begin{table}[H]
\begin{center}
\begin{tabular}{c|cccc}
\hline\hline
\multirow{2}{5mm}{$M_5$} &
\multicolumn{4}{c}{${\cal I}_{\chi}^{\widetilde{c_1}-{\rm PL}}$} \\
 & unsmear & APE & HYP1 & HYP2 \\
\hline
$0.1$ & $0.0$            & $0.407534(12)$   & $0.393862(14)$   & $0.621219(25)$   \\
$0.2$ & $0.0$            & $0.418636(13)$   & $0.405266(15)$   & $0.639854(26)$   \\
$0.3$ & $0.0$            & $0.430385(13)$   & $0.417365(15)$   & $0.659656(27)$   \\
$0.4$ & $0.0$            & $0.442837(13)$   & $0.430221(16)$   & $0.680726(28)$   \\
$0.5$ & $0.0$            & $0.456053(14)$   & $0.443901(16)$   & $0.703179(29)$   \\
$0.6$ & $0.0$            & $0.470102(14)$   & $0.458481(17)$   & $0.727146(30)$   \\
$0.7$ & $0.0$            & $0.485064(15)$   & $0.474051(17)$   & $0.752777(31)$   \\
$0.8$ & $0.0$            & $0.501027(16)$   & $0.490709(18)$   & $0.780244(32)$   \\
$0.9$ & $0.0$            & $0.518093(16)$   & $0.508571(19)$   & $0.809745(34)$   \\
$1.0$ & $0.0$            & $0.536381(17)$   & $0.527771(20)$   & $0.841510(35)$   \\
$1.1$ & $0.0$            & $0.556028(18)$   & $0.548465(21)$   & $0.875811(37)$   \\
$1.2$ & $0.0$            & $0.577192(18)$   & $0.570837(22)$   & $0.912966(39)$   \\
$1.3$ & $0.0$            & $0.600063(19)$   & $0.595104(23)$   & $0.953355(41)$   \\
$1.4$ & $0.0$            & $0.624863(20)$   & $0.621527(24)$   & $0.997435(43)$   \\
$1.5$ & $0.0$            & $0.651860(22)$   & $0.650423(26)$   & $1.045763(46)$   \\
$1.6$ & $0.0$            & $0.681381(23)$   & $0.682178(27)$   & $1.099024(49)$   \\
$1.7$ & $0.0$            & $0.713828(25)$   & $0.717277(30)$   & $1.158079(53)$   \\
$1.8$ & $0.0$            & $0.749705(27)$   & $0.756331(32)$   & $1.224021(58)$   \\
$1.9$ & $0.0$            & $0.789656(30)$   & $0.800130(36)$   & $1.298275(65)$   \\
\hline\hline
\end{tabular}
\end{center}
\begin{center}
\begin{tabular}{c|cccc}
\hline\hline
\multirow{2}{5mm}{$M_5$} &
\multicolumn{4}{c}{${\cal J}_{\chi}^{\widetilde{c_1}-{\rm PL}}$} \\
 & unsmear & APE & HYP1 & HYP2 \\
\hline
$0.1$ & $0.0$            & $-2.037596(54)$  & $-2.365827(68)$  & $-4.13518(14)$   \\
$0.2$ & $0.0$            & $-2.100353(55)$  & $-2.440650(69)$  & $-4.26767(14)$   \\
$0.3$ & $0.0$            & $-2.163706(56)$  & $-2.515804(70)$  & $-4.40038(14)$   \\
$0.4$ & $0.0$            & $-2.228348(57)$  & $-2.592118(80)$  & $-4.53494(14)$   \\
$0.5$ & $0.0$            & $-2.294787(58)$  & $-2.670352(81)$  & $-4.67253(15)$   \\
$0.6$ & $0.0$            & $-2.363474(59)$  & $-2.751008(83)$  & $-4.81415(15)$   \\
$0.7$ & $0.0$            & $-2.434800(67)$  & $-2.834629(84)$  & $-4.96078(15)$   \\
$0.8$ & $0.0$            & $-2.509381(61)$  & $-2.921776(85)$  & $-5.11343(16)$   \\
$0.9$ & $0.0$            & $-2.587563(62)$  & $-3.013065(87)$  & $-5.27319(16)$   \\
$1.0$ & $0.0$            & $-2.669967(63)$  & $-3.109187(89)$  & $-5.44131(16)$   \\
$1.1$ & $0.0$            & $-2.757255(64)$  & $-3.210948(90)$  & $-5.61923(16)$   \\
$1.2$ & $0.0$            & $-2.850204(65)$  & $-3.319302(92)$  & $-5.80866(17)$   \\
$1.3$ & $0.0$            & $-2.949756(66)$  & $-3.435408(93)$  & $-6.01168(17)$   \\
$1.4$ & $0.0$            & $-3.057067(68)$  & $-3.560702(95)$  & $-6.23089(17)$   \\
$1.5$ & $0.0$            & $-3.173600(69)$  & $-3.697012(97)$  & $-6.46961(18)$   \\
$1.6$ & $0.0$            & $-3.301186(79)$  & $-3.84673(10)$   & $-6.73217(18)$   \\
$1.7$ & $0.0$            & $-3.442518(81)$  & $-4.01310(10)$   & $-7.02452(19)$   \\
$1.8$ & $0.0$            & $-3.601161(83)$  & $-4.20078(10)$   & $-7.35520(19)$   \\
$1.9$ & $0.0$            & $-3.782775(87)$  & $-4.41709(11)$   & $-7.73773(20)$   \\
\hline\hline
\end{tabular}
\caption{Numerical values of integrals
${\cal I}_{\chi}^{\widetilde{c_1}-{\rm PL}}$ and
${\cal J}_{\chi}^{\widetilde{c_1}-{\rm PL}}$ (Plaquette).}
\label{TAB:V_1_smr-pl_PLAQUETTE}
\end{center}
\end{table}

\newpage

\begin{table}[H]
\begin{center}
\begin{tabular}{c|cccc}
\hline\hline
\multirow{2}{5mm}{$M_5$} &
\multicolumn{4}{c}{${\cal J}_w^{\widetilde{c_1}-{\rm PL}}$} \\
 & unsmear & APE & HYP1 & HYP2 \\
\hline
$0.1$ & $0.0$            & $-4.283384(10)$  & $-3.984201(16)$  & $-6.147964(32)$  \\
$0.2$ & $0.0$            & $-4.209373(9)$   & $-3.895649(14)$  & $-5.990706(28)$  \\
$0.3$ & $0.0$            & $-4.137273(8)$   & $-3.810173(13)$  & $-5.839628(28)$  \\
$0.4$ & $0.0$            & $-4.065997(7)$   & $-3.726307(12)$  & $-5.692018(25)$  \\
$0.5$ & $0.0$            & $-3.994823(7)$   & $-3.643141(12)$  & $-5.546171(24)$  \\
$0.6$ & $0.0$            & $-3.923205(6)$   & $-3.559974(11)$  & $-5.400810(23)$  \\
$0.7$ & $0.0$            & $-3.850673(6)$   & $-3.476212(11)$  & $-5.254859(23)$  \\
$0.8$ & $0.0$            & $-3.776781(6)$   & $-3.391307(11)$  & $-5.107324(22)$  \\
$0.9$ & $0.0$            & $-3.701087(7)$   & $-3.304709(11)$  & $-4.957219(22)$  \\
$1.0$ & $0.0$            & $-3.623098(7)$   & $-3.215843(11)$  & $-4.803507(22)$  \\
$1.1$ & $0.0$            & $-3.542290(6)$   & $-3.124055(12)$  & $-4.645032(22)$  \\
$1.2$ & $0.0$            & $-3.458053(7)$   & $-3.028608(12)$  & $-4.480485(23)$  \\
$1.3$ & $0.0$            & $-3.369660(7)$   & $-2.928625(12)$  & $-4.308282(23)$  \\
$1.4$ & $0.0$            & $-3.276227(7)$   & $-2.823037(11)$  & $-4.126494(24)$  \\
$1.5$ & $0.0$            & $-3.176639(8)$   & $-2.710462(12)$  & $-3.932684(23)$  \\
$1.6$ & $0.0$            & $-3.069451(8)$   & $-2.589120(12)$  & $-3.723605(24)$  \\
$1.7$ & $0.0$            & $-2.952709(9)$   & $-2.456580(13)$  & $-3.494856(24)$  \\
$1.8$ & $0.0$            & $-2.823657(13)$  & $-2.309344(14)$  & $-3.240068(26)$  \\
$1.9$ & $0.0$            & $-2.677956(12)$  & $-2.141922(19)$  & $-2.949198(38)$  \\
\hline\hline
\end{tabular}
\end{center}
\begin{center}
\begin{tabular}{c|cccc}
\hline\hline
\multirow{2}{5mm}{$M_5$} &
\multicolumn{4}{c}{${\cal J}_{\chi'}^{\widetilde{c_1}-{\rm PL}}$} \\
 & unsmear & APE & HYP1 & HYP2 \\
\hline
$0.1$ & $0.0$            & $-2.037576(16)$  & $-2.224503(20)$  & $-3.773845(37)$  \\
$0.2$ & $0.0$            & $-2.100332(16)$  & $-2.295041(21)$  & $-3.895183(38)$  \\
$0.3$ & $0.0$            & $-2.163685(16)$  & $-2.365678(21)$  & $-4.016136(38)$  \\
$0.4$ & $0.0$            & $-2.228325(17)$  & $-2.437291(21)$  & $-4.138302(39)$  \\
$0.5$ & $0.0$            & $-2.294764(17)$  & $-2.510505(22)$  & $-4.262808(40)$  \\
$0.6$ & $0.0$            & $-2.363451(17)$  & $-2.585858(22)$  & $-4.390612(40)$  \\
$0.7$ & $0.0$            & $-2.434827(18)$  & $-2.663872(22)$  & $-4.522630(41)$  \\
$0.8$ & $0.0$            & $-2.509355(18)$  & $-2.745082(23)$  & $-4.659803(42)$  \\
$0.9$ & $0.0$            & $-2.587537(18)$  & $-2.830075(23)$  & $-4.803152(42)$  \\
$1.0$ & $0.0$            & $-2.669943(18)$  & $-2.919507(24)$  & $-4.953826(43)$  \\
$1.1$ & $0.0$            & $-2.757232(19)$  & $-3.014145(24)$  & $-5.113162(44)$  \\
$1.2$ & $0.0$            & $-2.850182(19)$  & $-3.114897(24)$  & $-5.282749(44)$  \\
$1.3$ & $0.0$            & $-2.949748(26)$  & $-3.222887(33)$  & $-5.464530(45)$  \\
$1.4$ & $0.0$            & $-3.057059(26)$  & $-3.339448(34)$  & $-5.660927(46)$  \\
$1.5$ & $0.0$            & $-3.173594(27)$  & $-3.466349(34)$  & $-5.875094(63)$  \\
$1.6$ & $0.0$            & $-3.301250(28)$  & $-3.605891(35)$  & $-6.111047(64)$  \\
$1.7$ & $0.0$            & $-3.442586(28)$  & $-3.761213(36)$  & $-6.374430(49)$  \\
$1.8$ & $0.0$            & $-3.601226(22)$  & $-3.936824(28)$  & $-6.673536(51)$  \\
$1.9$ & $0.0$            & $-3.782849(23)$  & $-4.139930(29)$  & $-7.021409(54)$  \\
\hline\hline
\end{tabular}
\caption{Numerical values of integrals
${\cal J}_w^{\widetilde{c_1}-{\rm PL}}$ and
${\cal J}_{\chi'}^{\widetilde{c_1}-{\rm PL}}$ (Plaquette).}
\label{TAB:V_2_smr-pl_PLAQUETTE}
\end{center}
\end{table}

\newpage

\begin{table}[H]
\begin{center}
\begin{tabular}{c|cccc}
\hline\hline
\multirow{2}{5mm}{$M_5$} &
\multicolumn{4}{c}{${\cal K}_{\chi}^{\widetilde{c_1}-{\rm PL}}$} \\
 & unsmear & APE & HYP1 & HYP2 \\
\hline
$0.1$ & $0.0$            & $17.853525(31)$  & $15.988954(48)$  & $24.027654(94)$  \\
$0.2$ & $0.0$            & $7.811064(21)$   & $6.811607(32)$   & $10.035191(61)$  \\
$0.3$ & $0.0$            & $4.392120(18)$   & $3.679576(26)$   & $5.251770(48)$   \\
$0.4$ & $0.0$            & $2.621649(16)$   & $2.052067(22)$   & $2.760193(42)$   \\
$0.5$ & $0.0$            & $1.502881(15)$   & $1.019094(20)$   & $1.174087(38)$   \\
$0.6$ & $0.0$            & $0.701465(14)$   & $0.275394(19)$   & $0.028250(36)$   \\
$0.7$ & $0.0$            & $0.071833(15)$   & $-0.312087(19)$  & $-0.880199(34)$  \\
$0.8$ & $0.0$            & $-0.461561(15)$  & $-0.812625(19)$  & $-1.657020(47)$  \\
$0.9$ & $0.0$            & $-0.944408(16)$  & $-1.268104(20)$  & $-2.366378(37)$  \\
$1.0$ & $0.0$            & $-1.408989(17)$  & $-1.708448(21)$  & $-3.054217(38)$  \\
$1.1$ & $0.0$            & $-1.882758(18)$  & $-2.159268(22)$  & $-3.760153(41)$  \\
$1.2$ & $0.0$            & $-2.394296(18)$  & $-2.647479(24)$  & $-4.526036(43)$  \\
$1.3$ & $0.0$            & $-2.979644(20)$  & $-3.207259(25)$  & $-5.405258(46)$  \\
$1.4$ & $0.0$            & $-3.692200(28)$  & $-3.889483(37)$  & $-6.477516(67)$  \\
$1.5$ & $0.0$            & $-4.622672(30)$  & $-4.780726(40)$  & $-7.878588(74)$  \\
$1.6$ & $0.0$            & $-5.947873(34)$  & $-6.049948(45)$  & $-9.873656(83)$  \\
$1.7$ & $0.0$            & $-8.075885(38)$  & $-8.087229(52)$  & $-13.075106(70)$ \\
$1.8$ & $0.0$            & $-12.226002(47)$ & $-12.058228(48)$ & $-19.312790(89)$ \\
$1.9$ & $0.0$            & $-24.487147(52)$ & $-23.783853(74)$ & $-37.72516(14)$  \\
\hline\hline
\end{tabular}
\caption{Numerical values of integrals
${\cal K}_{\chi}^{\widetilde{c_1}-{\rm PL}}$ (Plaquette).}
\label{TAB:V_3_smr-pl_PLAQUETTE}
\end{center}
\end{table}

\newpage

\begin{table}[H]
\begin{center}
\begin{tabular}{c|cccc}
\hline\hline
\multirow{2}{5mm}{$M_5$} &
\multicolumn{4}{c}{${\cal I}_{\chi}^{\widetilde{c_1}-{\rm PL}}$} \\
 & unsmear & APE & HYP1 & HYP2 \\
\hline
$0.1$ & $0.102249(1)$    & $0.420466(9)$    & $0.410991(11)$   & $0.590402(19)$   \\
$0.2$ & $0.105228(1)$    & $0.432189(10)$   & $0.422982(11)$   & $0.608137(20)$   \\
$0.3$ & $0.108389(1)$    & $0.444603(10)$   & $0.435703(12)$   & $0.626973(20)$   \\
$0.4$ & $0.111749(2)$    & $0.457769(10)$   & $0.449216(12)$   & $0.647004(21)$   \\
$0.5$ & $0.115326(2)$    & $0.471749(11)$   & $0.463592(12)$   & $0.668334(22)$   \\
$0.6$ & $0.119140(2)$    & $0.486618(11)$   & $0.478909(13)$   & $0.691085(23)$   \\
$0.7$ & $0.123214(2)$    & $0.502459(11)$   & $0.495258(13)$   & $0.715394(24)$   \\
$0.8$ & $0.127578(2)$    & $0.519371(12)$   & $0.512744(14)$   & $0.741421(25)$   \\
$0.9$ & $0.132258(2)$    & $0.537456(12)$   & $0.531482(14)$   & $0.769343(26)$   \\
$1.0$ & $0.137290(2)$    & $0.556844(13)$   & $0.551610(15)$   & $0.799375(27)$   \\
$1.1$ & $0.142717(2)$    & $0.577681(13)$   & $0.573292(16)$   & $0.831763(28)$   \\
$1.2$ & $0.148589(2)$    & $0.600138(14)$   & $0.596714(17)$   & $0.866801(29)$   \\
$1.3$ & $0.154962(2)$    & $0.624417(15)$   & $0.622103(17)$   & $0.904837(31)$   \\
$1.4$ & $0.161907(2)$    & $0.650759(16)$   & $0.649728(18)$   & $0.946289(33)$   \\
$1.5$ & $0.169510(2)$    & $0.679455(17)$   & $0.679915(19)$   & $0.991668(35)$   \\
$1.6$ & $0.177874(3)$    & $0.710859(18)$   & $0.713064(21)$   & $1.041601(37)$   \\
$1.7$ & $0.187131(3)$    & $0.745409(19)$   & $0.749676(22)$   & $1.096875(40)$   \\
$1.8$ & $0.197446(3)$    & $0.783657(20)$   & $0.790383(24)$   & $1.158490(44)$   \\
$1.9$ & $0.209036(3)$    & $0.826311(23)$   & $0.836005(27)$   & $1.227747(48)$   \\
\hline\hline
\end{tabular}
\end{center}
\begin{center}
\begin{tabular}{c|cccc}
\hline\hline
\multirow{2}{5mm}{$M_5$} &
\multicolumn{4}{c}{${\cal J}_{\chi}^{\widetilde{c_1}-{\rm PL}}$} \\
 & unsmear & APE & HYP1 & HYP2 \\
\hline
$0.1$ & $-0.553407(3)$   & $-2.390646(52)$  & $-2.708685(65)$  & $-4.34123(13)$   \\
$0.2$ & $-0.571158(3)$   & $-2.465179(53)$  & $-2.794633(67)$  & $-4.48021(14)$   \\
$0.3$ & $-0.589055(3)$   & $-2.540081(54)$  & $-2.880630(68)$  & $-4.61890(12)$   \\
$0.4$ & $-0.607304(3)$   & $-2.616213(55)$  & $-2.967729(69)$  & $-4.75889(13)$   \\
$0.5$ & $-0.626058(3)$   & $-2.694200(56)$  & $-3.056681(70)$  & $-4.90152(13)$   \\
$0.6$ & $-0.645451(3)$   & $-2.774581(57)$  & $-3.148126(72)$  & $-5.04784(13)$   \\
$0.7$ & $-0.665616(3)$   & $-2.857880(58)$  & $-3.242680(73)$  & $-5.19885(13)$   \\
$0.8$ & $-0.686693(3)$   & $-2.944636(59)$  & $-3.340975(74)$  & $-5.35558(13)$   \\
$0.9$ & $-0.708836(3)$   & $-3.035439(60)$  & $-3.443695(75)$  & $-5.51912(14)$   \\
$1.0$ & $-0.732219(3)$   & $-3.130949(61)$  & $-3.551606(76)$  & $-5.69062(16)$   \\
$1.1$ & $-0.757046(3)$   & $-3.231926(62)$  & $-3.665600(77)$  & $-5.87172(16)$   \\
$1.2$ & $-0.783559(3)$   & $-3.339278(63)$  & $-3.786732(79)$  & $-6.06401(16)$   \\
$1.3$ & $-0.812055(4)$   & $-3.454095(64)$  & $-3.916224(90)$  & $-6.26956(16)$   \\
$1.4$ & $-0.842901(4)$   & $-3.577726(65)$  & $-4.055774(92)$  & $-6.49095(17)$   \\
$1.5$ & $-0.876565(4)$   & $-3.711877(66)$  & $-4.207342(94)$  & $-6.73145(17)$   \\
$1.6$ & $-0.913662(4)$   & $-3.858768(68)$  & $-4.373567(96)$  & $-6.99535(17)$   \\
$1.7$ & $-0.955035(4)$   & $-4.021407(69)$  & $-4.558045(98)$  & $-7.28853(18)$   \\
$1.8$ & $-1.001875(4)$   & $-4.203997(80)$  & $-4.76591(10)$   & $-7.61944(18)$   \\
$1.9$ & $-1.056084(4)$   & $-4.413346(83)$  & $-5.00527(10)$   & $-8.00142(19)$   \\
\hline\hline
\end{tabular}
\caption{Numerical values of integrals
${\cal I}_{\chi}^{\widetilde{c_1}-{\rm PL}}$ and
${\cal J}_{\chi}^{\widetilde{c_1}-{\rm PL}}$ (Symanzik).}
\label{TAB:V_1_smr-pl_SYMANZIK}
\end{center}
\end{table}

\newpage

\begin{table}[H]
\begin{center}
\begin{tabular}{c|cccc}
\hline\hline
\multirow{2}{5mm}{$M_5$} &
\multicolumn{4}{c}{${\cal J}_w^{\widetilde{c_1}-{\rm PL}}$} \\
 & unsmear & APE & HYP1 & HYP2 \\
\hline
$0.1$ & $-0.950580(2)$   & $-4.515816(10)$  & $-4.306330(15)$  & $-6.168544(28)$  \\
$0.2$ & $-0.930449(1)$   & $-4.427096(9)$   & $-4.203564(13)$  & $-6.001124(25)$  \\
$0.3$ & $-0.910991(1)$   & $-4.341257(8)$   & $-4.104837(12)$  & $-5.841055(23)$  \\
$0.4$ & $-0.891870(1)$   & $-4.256863(7)$   & $-4.008368(11)$  & $-5.685305(21)$  \\
$0.5$ & $-0.872877(1)$   & $-4.173009(6)$   & $-3.913053(10)$  & $-5.531974(20)$  \\
$0.6$ & $-0.853858(1)$   & $-4.089013(6)$   & $-3.818055(10)$  & $-5.379683(19)$  \\
$0.7$ & $-0.834676(1)$   & $-4.004291(5)$   & $-3.722670(9)$   & $-5.227248(19)$  \\
$0.8$ & $-0.815203(1)$   & $-3.918296(5)$   & $-3.626248(9)$   & $-5.073594(18)$  \\
$0.9$ & $-0.795315(1)$   & $-3.830492(6)$   & $-3.528146(9)$   & $-4.917661(18)$  \\
$1.0$ & $-0.774875(1)$   & $-3.740285(6)$   & $-3.427696(9)$   & $-4.758349(17)$  \\
$1.1$ & $-0.753736(1)$   & $-3.647050(6)$   & $-3.324136(9)$   & $-4.594419(17)$  \\
$1.2$ & $-0.731726(1)$   & $-3.550047(6)$   & $-3.216611(10)$  & $-4.424484(19)$  \\
$1.3$ & $-0.708640(1)$   & $-3.448411(6)$   & $-3.104112(10)$  & $-4.246869(20)$  \\
$1.4$ & $-0.684228(1)$   & $-3.341069(6)$   & $-2.985382(10)$  & $-4.059524(20)$  \\
$1.5$ & $-0.658175(1)$   & $-3.226676(7)$   & $-2.858835(10)$  & $-3.859845(21)$  \\
$1.6$ & $-0.630065(1)$   & $-3.103476(8)$   & $-2.722385(10)$  & $-3.644406(20)$  \\
$1.7$ & $-0.599338(1)$   & $-2.969070(9)$   & $-2.573161(12)$  & $-3.408438(21)$  \\
$1.8$ & $-0.565185(1)$   & $-2.820029(10)$  & $-2.407055(14)$  & $-3.145095(23)$  \\
$1.9$ & $-0.526356(1)$   & $-2.651005(10)$  & $-2.217538(15)$  & $-2.843478(25)$  \\
\hline\hline
\end{tabular}
\end{center}
\begin{center}
\begin{tabular}{c|cccc}
\hline\hline
\multirow{2}{5mm}{$M_5$} &
\multicolumn{4}{c}{${\cal J}_{\chi'}^{\widetilde{c_1}-{\rm PL}}$} \\
 & unsmear & APE & HYP1 & HYP2 \\
\hline
$0.1$ & $-0.518474(4)$   & $-2.321568(16)$  & $-2.516149(20)$  & $-3.934517(36)$  \\
$0.2$ & $-0.535073(4)$   & $-2.393807(16)$  & $-2.595982(20)$  & $-4.060531(36)$  \\
$0.3$ & $-0.551753(4)$   & $-2.466300(17)$  & $-2.675549(21)$  & $-4.185501(37)$  \\
$0.4$ & $-0.568716(4)$   & $-2.539902(17)$  & $-2.755889(21)$  & $-4.311167(38)$  \\
$0.5$ & $-0.586110(4)$   & $-2.615232(17)$  & $-2.837732(21)$  & $-4.438731(38)$  \\
$0.6$ & $-0.604063(4)$   & $-2.692823(17)$  & $-2.921698(22)$  & $-4.569195(39)$  \\
$0.7$ & $-0.622701(4)$   & $-2.773186(18)$  & $-3.008372(22)$  & $-4.703504(40)$  \\
$0.8$ & $-0.642157(5)$   & $-2.856852(18)$  & $-3.098354(22)$  & $-4.842612(40)$  \\
$0.9$ & $-0.662575(5)$   & $-2.944397(18)$  & $-3.192292(23)$  & $-4.987546(41)$  \\
$1.0$ & $-0.684118(5)$   & $-3.036465(19)$  & $-3.290912(23)$  & $-5.139455(42)$  \\
$1.1$ & $-0.706978(5)$   & $-3.133803(19)$  & $-3.395053(24)$  & $-5.299665(42)$  \\
$1.2$ & $-0.731382(5)$   & $-3.237294(19)$  & $-3.505709(24)$  & $-5.469750(43)$  \\
$1.3$ & $-0.757609(5)$   & $-3.348003(19)$  & $-3.624090(24)$  & $-5.651628(44)$  \\
$1.4$ & $-0.786005(5)$   & $-3.467249(20)$  & $-3.751695(25)$  & $-5.847688(44)$  \\
$1.5$ & $-0.817022(6)$   & $-3.596708(20)$  & $-3.890449(25)$  & $-6.060999(45)$  \\
$1.6$ & $-0.851228(6)$   & $-3.738542(21)$  & $-4.042855(26)$  & $-6.295587(46)$  \\
$1.7$ & $-0.889431(6)$   & $-3.895707(21)$  & $-4.212360(26)$  & $-6.557014(47)$  \\
$1.8$ & $-0.932788(6)$   & $-4.072415(22)$  & $-4.403936(27)$  & $-6.853349(48)$  \\
$1.9$ & $-0.983140(7)$   & $-4.275259(23)$  & $-4.625453(28)$  & $-7.197456(51)$  \\
\hline\hline
\end{tabular}
\caption{Numerical values of integrals
${\cal J}_w^{\widetilde{c_1}-{\rm PL}}$ and
${\cal J}_{\chi'}^{\widetilde{c_1}-{\rm PL}}$ (Symanzik).}
\label{TAB:V_2_smr-pl_SYMANZIK}
\end{center}
\end{table}

\newpage

\begin{table}[H]
\begin{center}
\begin{tabular}{c|cccc}
\hline\hline
\multirow{2}{5mm}{$M_5$} &
\multicolumn{4}{c}{${\cal K}_{\chi}^{\widetilde{c_1}-{\rm PL}}$} \\
 & unsmear & APE & HYP1 & HYP2 \\
\hline
$0.1$ & $3.849344(10)$   & $18.488439(31)$  & $17.061540(45)$  & $23.805797(84)$  \\
$0.2$ & $1.649442(7)$    & $7.986934(21)$   & $7.196058(30)$   & $9.838490(56)$   \\
$0.3$ & $0.898962(5)$    & $4.407619(18)$   & $3.826504(24)$   & $5.060109(45)$   \\
$0.4$ & $0.509197(5)$    & $2.551179(21)$   & $2.073722(29)$   & $2.568738(40)$   \\
$0.5$ & $0.261965(4)$    & $1.375696(15)$   & $0.959767(20)$   & $0.980994(36)$   \\
$0.6$ & $0.084081(4)$    & $0.531777(14)$   & $0.156666(19)$   & $-0.167351(34)$  \\
$0.7$ & $-0.056353(4)$   & $-0.132810(14)$  & $-0.478629(18)$  & $-1.078794(33)$  \\
$0.8$ & $-0.175939(4)$   & $-0.697215(15)$  & $-1.020603(25)$  & $-1.858881(33)$  \\
$0.9$ & $-0.284736(4)$   & $-1.209231(16)$  & $-1.514365(20)$  & $-2.571793(35)$  \\
$1.0$ & $-0.389905(4)$   & $-1.702809(17)$  & $-1.992117(21)$  & $-3.263345(37)$  \\
$1.1$ & $-0.497591(4)$   & $-2.206883(18)$  & $-2.481506(22)$  & $-3.973186(40)$  \\
$1.2$ & $-0.614251(5)$   & $-2.751683(19)$  & $-3.011611(23)$  & $-4.743187(42)$  \\
$1.3$ & $-0.748088(5)$   & $-3.375423(20)$  & $-3.619401(25)$  & $-5.626782(45)$  \\
$1.4$ & $-0.911311(6)$   & $-4.134800(21)$  & $-4.359904(27)$  & $-6.703685(49)$  \\
$1.5$ & $-1.124723(6)$   & $-5.126245(30)$  & $-5.326878(39)$  & $-8.109874(71)$  \\
$1.6$ & $-1.428914(7)$   & $-6.537665(34)$  & $-6.703115(44)$  & $-10.110511(79)$ \\
$1.7$ & $-1.917625(8)$   & $-8.802843(39)$  & $-8.910722(50)$  & $-13.318052(92)$ \\
$1.8$ & $-2.871027(11)$  & $-13.217740(47)$ & $-13.210794(62)$ & $-19.562256(84)$ \\
$1.9$ & $-5.688473(19)$  & $-26.253637(71)$ & $-25.900451(71)$ & $-37.97904(13)$  \\
\hline\hline
\end{tabular}
\caption{Numerical values of integrals
${\cal K}_{\chi}^{\widetilde{c_1}-{\rm PL}}$ (Symanzik).}
\label{TAB:V_3_smr-pl_SYMANZIK}
\end{center}
\end{table}

\newpage

\begin{table}[H]
\begin{center}
\begin{tabular}{c|cccc}
\hline\hline
\multirow{2}{5mm}{$M_5$} &
\multicolumn{4}{c}{${\cal I}_{\chi}^{\widetilde{c_1}-{\rm PL}}$} \\
 & unsmear & APE & HYP1 & HYP2 \\
\hline
$0.1$ & $0.249872(3)$    & $0.446767(7)$    & $0.442480(7)$    & $0.555974(12)$   \\
$0.2$ & $0.257249(3)$    & $0.459632(7)$    & $0.455554(8)$    & $0.572719(12)$   \\
$0.3$ & $0.265073(3)$    & $0.473262(7)$    & $0.469417(8)$    & $0.590484(13)$   \\
$0.4$ & $0.273383(4)$    & $0.487718(7)$    & $0.484134(8)$    & $0.609353(13)$   \\
$0.5$ & $0.282219(4)$    & $0.503071(7)$    & $0.499776(8)$    & $0.629417(14)$   \\
$0.6$ & $0.291629(4)$    & $0.519399(8)$    & $0.516425(9)$    & $0.650783(14)$   \\
$0.7$ & $0.301668(4)$    & $0.536793(8)$    & $0.534176(9)$    & $0.673573(15)$   \\
$0.8$ & $0.312397(4)$    & $0.555354(8)$    & $0.553135(9)$    & $0.697925(16)$   \\
$0.9$ & $0.323888(4)$    & $0.575201(9)$    & $0.573426(10)$   & $0.724001(16)$   \\
$1.0$ & $0.336225(4)$    & $0.596472(9)$    & $0.595194(10)$   & $0.751989(17)$   \\
$1.1$ & $0.349505(5)$    & $0.619326(9)$    & $0.618606(11)$   & $0.782107(18)$   \\
$1.2$ & $0.363843(5)$    & $0.643951(10)$   & $0.643862(11)$   & $0.814615(18)$   \\
$1.3$ & $0.379375(5)$    & $0.670570(10)$   & $0.671196(12)$   & $0.849823(19)$   \\
$1.4$ & $0.396267(5)$    & $0.699449(11)$   & $0.700891(12)$   & $0.888101(20)$   \\
$1.5$ & $0.414717(6)$    & $0.730910(11)$   & $0.733292(13)$   & $0.929902(21)$   \\
$1.6$ & $0.434971(6)$    & $0.765348(12)$   & $0.768820(14)$   & $0.975784(23)$   \\
$1.7$ & $0.457337(6)$    & $0.803254(13)$   & $0.808001(15)$   & $1.026442(24)$   \\
$1.8$ & $0.482204(7)$    & $0.845247(14)$   & $0.851503(16)$   & $1.082764(26)$   \\
$1.9$ & $0.510080(8)$    & $0.892127(15)$   & $0.900194(17)$   & $1.145901(29)$   \\
\hline\hline
\end{tabular}
\end{center}
\begin{center}
\begin{tabular}{c|cccc}
\hline\hline
\multirow{2}{5mm}{$M_5$} &
\multicolumn{4}{c}{${\cal J}_{\chi}^{\widetilde{c_1}-{\rm PL}}$} \\
 & unsmear & APE & HYP1 & HYP2 \\
\hline
$0.1$ & $-1.775035(10)$  & $-3.268508(50)$  & $-3.562292(62)$  & $-4.94892(11)$   \\
$0.2$ & $-1.831856(10)$  & $-3.371661(50)$  & $-3.675559(63)$  & $-5.10681(11)$   \\
$0.3$ & $-1.888377(10)$  & $-3.474103(51)$  & $-3.787699(64)$  & $-5.26262(12)$   \\
$0.4$ & $-1.945338(11)$  & $-3.577181(52)$  & $-3.900243(65)$  & $-5.41855(12)$   \\
$0.5$ & $-2.003257(11)$  & $-3.681825(53)$  & $-4.014239(66)$  & $-5.57608(12)$   \\
$0.6$ & $-2.062569(11)$  & $-3.788808(54)$  & $-4.130550(67)$  & $-5.73644(12)$   \\
$0.7$ & $-2.123681(11)$  & $-3.898850(55)$  & $-4.249976(68)$  & $-5.90075(12)$   \\
$0.8$ & $-2.187011(12)$  & $-4.012679(56)$  & $-4.373320(69)$  & $-6.07012(12)$   \\
$0.9$ & $-2.253006(12)$  & $-4.131072(56)$  & $-4.501438(70)$  & $-6.24573(13)$   \\
$1.0$ & $-2.322167(12)$  & $-4.254895(57)$  & $-4.635279(71)$  & $-6.42890(13)$   \\
$1.1$ & $-2.395072(12)$  & $-4.385139(58)$  & $-4.775936(73)$  & $-6.62112(13)$   \\
$1.2$ & $-2.472405(12)$  & $-4.522975(59)$  & $-4.924695(74)$  & $-6.82418(13)$   \\
$1.3$ & $-2.554998(13)$  & $-4.669821(60)$  & $-5.083112(75)$  & $-7.04020(14)$   \\
$1.4$ & $-2.643881(13)$  & $-4.827424(61)$  & $-5.253118(76)$  & $-7.27187(14)$   \\
$1.5$ & $-2.740369(13)$  & $-4.998013(62)$  & $-5.437176(78)$  & $-7.52257(14)$   \\
$1.6$ & $-2.846189(13)$  & $-5.184499(63)$  & $-5.638514(79)$  & $-7.79677(14)$   \\
$1.7$ & $-2.963693(14)$  & $-5.390845(65)$  & $-5.861540(81)$  & $-8.10050(16)$   \\
$1.8$ & $-3.096272(14)$  & $-5.622737(66)$  & $-6.112531(92)$  & $-8.44275(17)$   \\
$1.9$ & $-3.249289(14)$  & $-5.889063(76)$  & $-6.401622(95)$  & $-8.83741(17)$   \\
\hline\hline
\end{tabular}
\caption{Numerical values of integrals
${\cal I}_{\chi}^{\widetilde{c_1}-{\rm PL}}$ and
${\cal J}_{\chi}^{\widetilde{c_1}-{\rm PL}}$ (Iwasaki).}
\label{TAB:V_1_smr-pl_IWASAKI}
\end{center}
\end{table}

\newpage

\begin{table}[H]
\begin{center}
\begin{tabular}{c|cccc}
\hline\hline
\multirow{2}{5mm}{$M_5$} &
\multicolumn{4}{c}{${\cal J}_w^{\widetilde{c_1}-{\rm PL}}$} \\
 & unsmear & APE & HYP1 & HYP2 \\
\hline
$0.1$ & $-2.592328(5)$   & $-5.035844(11)$  & $-4.947384(15)$  & $-6.309869(24)$  \\
$0.2$ & $-2.524721(4)$   & $-4.910064(9)$   & $-4.808736(12)$  & $-6.114771(21)$  \\
$0.3$ & $-2.460350(4)$   & $-4.790210(8)$   & $-4.677204(10)$  & $-5.930445(19)$  \\
$0.4$ & $-2.397926(4)$   & $-4.673907(7)$   & $-4.550068(9)$   & $-5.752916(17)$  \\
$0.5$ & $-2.336642(3)$   & $-4.559672(6)$   & $-4.425635(9)$   & $-5.579716(16)$  \\
$0.6$ & $-2.275903(3)$   & $-4.446413(6)$   & $-4.302662(8)$   & $-5.409067(14)$  \\
$0.7$ & $-2.215218(3)$   & $-4.333223(5)$   & $-4.180126(7)$   & $-5.239489(13)$  \\
$0.8$ & $-2.154132(3)$   & $-4.219272(5)$   & $-4.057093(7)$   & $-5.069649(13)$  \\
$0.9$ & $-2.092214(3)$   & $-4.103759(5)$   & $-3.932667(7)$   & $-4.898273(12)$  \\
$1.0$ & $-2.029001(3)$   & $-3.985839(5)$   & $-3.805908(6)$   & $-4.724030(12)$  \\
$1.1$ & $-1.963993(3)$   & $-3.864597(5)$   & $-3.675807(8)$   & $-4.545491(12)$  \\
$1.2$ & $-1.896617(3)$   & $-3.738979(5)$   & $-3.541177(8)$   & $-4.360992(13)$  \\
$1.3$ & $-1.826186(3)$   & $-3.607730(5)$   & $-3.400642(7)$   & $-4.168585(15)$  \\
$1.4$ & $-1.751883(3)$   & $-3.469338(5)$   & $-3.252525(9)$   & $-3.965903(15)$  \\
$1.5$ & $-1.672630(3)$   & $-3.321845(5)$   & $-3.094636(7)$   & $-3.749859(15)$  \\
$1.6$ & $-1.587022(3)$   & $-3.162681(6)$   & $-2.924114(8)$   & $-3.516375(14)$  \\
$1.7$ & $-1.493120(3)$   & $-2.988297(6)$   & $-2.736979(8)$   & $-3.259815(15)$  \\
$1.8$ & $-1.388092(3)$   & $-2.793513(7)$   & $-2.527403(10)$  & $-2.971830(19)$  \\
$1.9$ & $-1.267425(4)$   & $-2.570075(8)$   & $-2.286043(13)$  & $-2.639040(25)$  \\
\hline\hline
\end{tabular}
\end{center}
\begin{center}
\begin{tabular}{c|cccc}
\hline\hline
\multirow{2}{5mm}{$M_5$} &
\multicolumn{4}{c}{${\cal J}_{\chi'}^{\widetilde{c_1}-{\rm PL}}$} \\
 & unsmear & APE & HYP1 & HYP2 \\
\hline
$0.1$ & $-1.615358(14)$  & $-3.028342(20)$  & $-3.226072(23)$  & $-4.411409(36)$  \\
$0.2$ & $-1.666709(15)$  & $-3.123267(21)$  & $-3.328015(24)$  & $-4.551286(37)$  \\
$0.3$ & $-1.717513(15)$  & $-3.217135(21)$  & $-3.428345(24)$  & $-4.688303(37)$  \\
$0.4$ & $-1.768504(15)$  & $-3.311288(22)$  & $-3.528584(24)$  & $-4.824649(38)$  \\
$0.5$ & $-1.820189(16)$  & $-3.406644(22)$  & $-3.629759(25)$  & $-4.961784(39)$  \\
$0.6$ & $-1.872985(16)$  & $-3.503956(22)$  & $-3.732706(25)$  & $-5.100881(39)$  \\
$0.7$ & $-1.927281(16)$  & $-3.603921(23)$  & $-3.838190(26)$  & $-5.243002(40)$  \\
$0.8$ & $-1.983471(16)$  & $-3.707237(23)$  & $-3.946969(26)$  & $-5.389194(40)$  \\
$0.9$ & $-2.041971(17)$  & $-3.814645(23)$  & $-4.059846(26)$  & $-5.540555(41)$  \\
$1.0$ & $-2.103248(17)$  & $-3.926964(24)$  & $-4.177706(27)$  & $-5.698292(42)$  \\
$1.1$ & $-2.167837(17)$  & $-4.045132(24)$  & $-4.301562(27)$  & $-5.863785(42)$  \\
$1.2$ & $-2.236373(17)$  & $-4.170254(24)$  & $-4.432607(28)$  & $-6.038655(43)$  \\
$1.3$ & $-2.309624(18)$  & $-4.303666(25)$  & $-4.572285(28)$  & $-6.224868(43)$  \\
$1.4$ & $-2.388547(18)$  & $-4.447019(25)$  & $-4.722389(29)$  & $-6.424869(44)$  \\
$1.5$ & $-2.474366(18)$  & $-4.602419(26)$  & $-4.885212(29)$  & $-6.641789(45)$  \\
$1.6$ & $-2.568694(19)$  & $-4.772629(26)$  & $-5.063771(30)$  & $-6.879766(45)$  \\
$1.7$ & $-2.673747(19)$  & $-4.961423(27)$  & $-5.262215(30)$  & $-7.144503(46)$  \\
$1.8$ & $-2.792740(20)$  & $-5.174249(27)$  & $-5.486561(31)$  & $-7.444311(47)$  \\
$1.9$ & $-2.930820(23)$  & $-5.419755(30)$  & $-5.746415(33)$  & $-7.792505(50)$  \\
\hline\hline
\end{tabular}
\caption{Numerical values of integrals
${\cal J}_w^{\widetilde{c_1}-{\rm PL}}$ and
${\cal J}_{\chi'}^{\widetilde{c_1}-{\rm PL}}$ (Iwasaki).}
\label{TAB:V_2_smr-pl_IWASAKI}
\end{center}
\end{table}

\newpage

\begin{table}[H]
\begin{center}
\begin{tabular}{c|cccc}
\hline\hline
\multirow{2}{5mm}{$M_5$} &
\multicolumn{4}{c}{${\cal K}_{\chi}^{\widetilde{c_1}-{\rm PL}}$} \\
 & unsmear & APE & HYP1 & HYP2 \\
\hline
$0.1$ & $10.096015(32)$  & $19.775670(48)$  & $18.955169(49)$  & $23.600213(75)$  \\
$0.2$ & $4.194416(21)$   & $8.271905(28)$   & $7.773551(32)$   & $9.484512(51)$   \\
$0.3$ & $2.176515(18)$   & $4.341016(27)$   & $3.946890(27)$   & $4.646750(42)$   \\
$0.4$ & $1.125308(16)$   & $2.295156(25)$   & $1.951130(30)$   & $2.118736(49)$   \\
$0.5$ & $0.456155(14)$   & $0.994411(20)$   & $0.678979(23)$   & $0.503499(35)$   \\
$0.6$ & $-0.027125(14)$  & $0.056320(19)$   & $-0.241092(22)$  & $-0.667694(33)$  \\
$0.7$ & $-0.410081(14)$  & $-0.685851(18)$  & $-0.971164(21)$  & $-1.599469(32)$  \\
$0.8$ & $-0.737285(14)$  & $-1.318921(19)$  & $-1.595696(21)$  & $-2.398537(33)$  \\
$0.9$ & $-1.035791(14)$  & $-1.895458(20)$  & $-2.166002(23)$  & $-3.129847(35)$  \\
$1.0$ & $-1.324912(15)$  & $-2.452964(21)$  & $-2.718689(24)$  & $-3.839838(37)$  \\
$1.1$ & $-1.621273(16)$  & $-3.023576(23)$  & $-3.285352(26)$  & $-4.568746(40)$  \\
$1.2$ & $-1.942401(17)$  & $-3.641067(24)$  & $-3.899304(27)$  & $-5.359136(42)$  \\
$1.3$ & $-2.310611(18)$  & $-4.348303(26)$  & $-4.602970(29)$  & $-6.265350(45)$  \\
$1.4$ & $-2.759141(20)$  & $-5.209025(28)$  & $-5.459556(31)$  & $-7.368470(49)$  \\
$1.5$ & $-3.344628(22)$  & $-6.331731(30)$  & $-6.576745(34)$  & $-8.806717(53)$  \\
$1.6$ & $-4.177540(24)$  & $-7.927926(33)$  & $-8.164523(38)$  & $-10.849676(59)$ \\
$1.7$ & $-5.512888(28)$  & $-10.485721(45)$ & $-10.707531(53)$ & $-14.119546(86)$ \\
$1.8$ & $-8.112468(35)$  & $-15.462945(55)$ & $-15.653132(65)$ & $-20.47423(11)$  \\
$1.9$ & $-15.779753(55)$ & $-30.137524(83)$ & $-30.226655(98)$ & $-39.18792(12)$  \\
\hline\hline
\end{tabular}
\caption{Numerical values of integrals
${\cal K}_{\chi}^{\widetilde{c_1}-{\rm PL}}$ (Iwasaki).}
\label{TAB:V_3_smr-pl_IWASAKI}
\end{center}
\end{table}

\newpage

\begin{table}[H]
\begin{center}
\begin{tabular}{c|cccc}
\hline\hline
\multirow{2}{5mm}{$M_5$} &
\multicolumn{4}{c}{${\cal I}_{\chi}^{\widetilde{c_1}-{\rm PL}}$} \\
 & unsmear & APE & HYP1 & HYP2 \\
\hline
$0.1$ & $0.414157(5)$    & $0.491310(6)$    & $0.490776(6)$    & $0.537070(7)$    \\
$0.2$ & $0.426612(6)$    & $0.505961(6)$    & $0.505541(6)$    & $0.553354(7)$    \\
$0.3$ & $0.439797(6)$    & $0.521465(6)$    & $0.521169(6)$    & $0.570591(8)$    \\
$0.4$ & $0.453766(6)$    & $0.537884(7)$    & $0.537723(7)$    & $0.588849(8)$    \\
$0.5$ & $0.468581(6)$    & $0.555291(7)$    & $0.555275(7)$    & $0.608210(8)$    \\
$0.6$ & $0.484312(6)$    & $0.573767(7)$    & $0.573909(7)$    & $0.628763(9)$    \\
$0.7$ & $0.501044(7)$    & $0.593408(7)$    & $0.593721(7)$    & $0.650617(9)$    \\
$0.8$ & $0.518870(7)$    & $0.614325(8)$    & $0.614823(8)$    & $0.673894(9)$    \\
$0.9$ & $0.537902(7)$    & $0.636644(8)$    & $0.637346(8)$    & $0.698737(10)$   \\
$1.0$ & $0.558268(7)$    & $0.660516(8)$    & $0.661439(8)$    & $0.725315(10)$   \\
$1.1$ & $0.580121(8)$    & $0.686114(8)$    & $0.687283(9)$    & $0.753824(10)$   \\
$1.2$ & $0.603641(8)$    & $0.713647(9)$    & $0.715086(9)$    & $0.784496(11)$   \\
$1.3$ & $0.629041(8)$    & $0.743359(9)$    & $0.745101(9)$    & $0.817612(11)$   \\
$1.4$ & $0.656579(9)$    & $0.775548(10)$   & $0.777629(10)$   & $0.853506(12)$   \\
$1.5$ & $0.686569(9)$    & $0.810572(10)$   & $0.813038(10)$   & $0.892585(12)$   \\
$1.6$ & $0.719398(10)$   & $0.848874(11)$   & $0.851780(11)$   & $0.935354(13)$   \\
$1.7$ & $0.755548(10)$   & $0.891007(11)$   & $0.894421(11)$   & $0.982442(14)$   \\
$1.8$ & $0.795635(11)$   & $0.937670(12)$   & $0.941681(12)$   & $1.034650(15)$   \\
$1.9$ & $0.840457(12)$   & $0.989774(13)$   & $0.994493(13)$   & $1.093020(16)$   \\
\hline\hline
\end{tabular}
\end{center}
\begin{center}
\begin{tabular}{c|cccc}
\hline\hline
\multirow{2}{5mm}{$M_5$} &
\multicolumn{4}{c}{${\cal J}_{\chi}^{\widetilde{c_1}-{\rm PL}}$} \\
 & unsmear & APE & HYP1 & HYP2 \\
\hline
$0.1$ & $-4.956465(34)$  & $-5.948112(53)$  & $-6.184557(61)$  & $-7.175375(98)$  \\
$0.2$ & $-5.111478(35)$  & $-6.133504(54)$  & $-6.377501(62)$  & $-7.39916(10)$   \\
$0.3$ & $-5.260412(36)$  & $-6.311567(55)$  & $-6.562570(63)$  & $-7.61340(10)$   \\
$0.4$ & $-5.406144(36)$  & $-6.485741(56)$  & $-6.743383(64)$  & $-7.82233(10)$   \\
$0.5$ & $-5.550525(37)$  & $-6.658231(57)$  & $-6.922257(65)$  & $-8.02870(10)$   \\
$0.6$ & $-5.694973(38)$  & $-6.830728(57)$  & $-7.100969(66)$  & $-8.23457(11)$   \\
$0.7$ & $-5.840719(38)$  & $-7.004694(58)$  & $-7.281045(67)$  & $-8.44173(11)$   \\
$0.8$ & $-5.988924(39)$  & $-7.181506(59)$  & $-7.463925(68)$  & $-8.65183(11)$   \\
$0.9$ & $-6.140762(40)$  & $-7.362553(60)$  & $-7.651053(69)$  & $-8.86656(11)$   \\
$1.0$ & $-6.297486(40)$  & $-7.549317(61)$  & $-7.843973(70)$  & $-9.08770(11)$   \\
$1.1$ & $-6.460494(41)$  & $-7.743445(62)$  & $-8.044400(71)$  & $-9.31722(11)$   \\
$1.2$ & $-6.631408(41)$  & $-7.946852(63)$  & $-8.254323(72)$  & $-9.55742(12)$   \\
$1.3$ & $-6.812174(42)$  & $-8.161828(64)$  & $-8.476125(73)$  & $-9.81103(12)$   \\
$1.4$ & $-7.005202(43)$  & $-8.391209(64)$  & $-8.712756(75)$  & $-10.08146(12)$  \\
$1.5$ & $-7.213574(43)$  & $-8.638618(66)$  & $-8.967989(76)$  & $-10.37305(12)$  \\
$1.6$ & $-7.441377(44)$  & $-8.908857(67)$  & $-9.246828(77)$  & $-10.69157(12)$  \\
$1.7$ & $-7.694270(45)$  & $-9.208571(68)$  & $-9.556204(78)$  & $-11.04503(12)$  \\
$1.8$ & $-7.980554(46)$  & $-9.547508(69)$  & $-9.906295(80)$  & $-11.44519(13)$  \\
$1.9$ & $-8.313591(47)$  & $-9.941352(71)$  & $-10.313452(89)$ & $-11.91096(14)$  \\
\hline\hline
\end{tabular}
\caption{Numerical values of integrals
${\cal I}_{\chi}^{\widetilde{c_1}-{\rm PL}}$ and
${\cal J}_{\chi}^{\widetilde{c_1}-{\rm PL}}$ (DBW2).}
\label{TAB:V_1_smr-pl_DBW2}
\end{center}
\end{table}

\newpage

\begin{table}[H]
\begin{center}
\begin{tabular}{c|cccc}
\hline\hline
\multirow{2}{5mm}{$M_5$} &
\multicolumn{4}{c}{${\cal J}_w^{\widetilde{c_1}-{\rm PL}}$} \\
 & unsmear & APE & HYP1 & HYP2 \\
\hline
$0.1$ & $-5.079951(12)$  & $-6.202083(14)$  & $-6.210530(15)$  & $-6.914069(20)$  \\
$0.2$ & $-4.883963(11)$  & $-5.966157(12)$  & $-5.964620(13)$  & $-6.627483(17)$  \\
$0.3$ & $-4.704460(10)$  & $-5.749961(11)$  & $-5.739652(12)$  & $-6.365839(15)$  \\
$0.4$ & $-4.535904(9)$   & $-5.546866(10)$  & $-5.528637(10)$  & $-6.120876(13)$  \\
$0.5$ & $-4.374955(8)$   & $-5.352871(9)$   & $-5.327356(9)$   & $-5.887615(12)$  \\
$0.6$ & $-4.219247(7)$   & $-5.165142(8)$   & $-5.132825(8)$   & $-5.662537(11)$  \\
$0.7$ & $-4.066914(6)$   & $-4.981442(7)$   & $-4.942690(7)$   & $-5.442871(10)$  \\
$0.8$ & $-3.916351(6)$   & $-4.799849(6)$   & $-4.754935(7)$   & $-5.226247(9)$   \\
$0.9$ & $-3.766078(5)$   & $-4.618585(6)$   & $-4.567695(6)$   & $-5.010482(8)$   \\
$1.0$ & $-3.614629(5)$   & $-4.435897(5)$   & $-4.379133(6)$   & $-4.793423(7)$   \\
$1.1$ & $-3.460464(5)$   & $-4.249928(5)$   & $-4.187317(5)$   & $-4.572813(7)$   \\
$1.2$ & $-3.301862(5)$   & $-4.058623(5)$   & $-3.990094(6)$   & $-4.346134(9)$   \\
$1.3$ & $-3.136830(5)$   & $-3.859588(5)$   & $-3.784955(6)$   & $-4.110471(8)$   \\
$1.4$ & $-2.962890(5)$   & $-3.649849(6)$   & $-3.568808(6)$   & $-3.862214(8)$   \\
$1.5$ & $-2.776871(6)$   & $-3.425602(6)$   & $-3.337669(7)$   & $-3.596696(9)$   \\
$1.6$ & $-2.574495(6)$   & $-3.181723(7)$   & $-3.086179(7)$   & $-3.307664(9)$   \\
$1.7$ & $-2.349691(7)$   & $-2.910932(8)$   & $-2.806716(8)$   & $-2.986193(11)$  \\
$1.8$ & $-2.093217(9)$   & $-2.602158(10)$  & $-2.487666(10)$  & $-2.618670(13)$  \\
$1.9$ & $-1.789501(9)$   & $-2.236747(11)$  & $-2.109458(11)$  & $-2.182115(15)$  \\
\hline\hline
\end{tabular}
\end{center}
\begin{center}
\begin{tabular}{c|cccc}
\hline\hline
\multirow{2}{5mm}{$M_5$} &
\multicolumn{4}{c}{${\cal J}_{\chi'}^{\widetilde{c_1}-{\rm PL}}$} \\
 & unsmear & APE & HYP1 & HYP2 \\
\hline
$0.1$ & $-4.281982(49)$  & $-5.161542(50)$  & $-5.335295(51)$  & $-6.156654(56)$  \\
$0.2$ & $-4.413021(50)$  & $-5.319005(51)$  & $-5.498172(52)$  & $-6.344465(57)$  \\
$0.3$ & $-4.537621(51)$  & $-5.468743(52)$  & $-5.652733(53)$  & $-6.522188(58)$  \\
$0.4$ & $-4.658679(52)$  & $-5.614217(53)$  & $-5.802622(54)$  & $-6.694118(59)$  \\
$0.5$ & $-4.778024(52)$  & $-5.757616(54)$  & $-5.950136(55)$  & $-6.862948(60)$  \\
$0.6$ & $-4.897031(53)$  & $-5.900582(55)$  & $-6.096996(56)$  & $-7.030691(61)$  \\
$0.7$ & $-5.016865(54)$  & $-6.044503(56)$  & $-6.244650(56)$  & $-7.199031(62)$  \\
$0.8$ & $-5.138600(55)$  & $-6.190657(56)$  & $-6.394429(57)$  & $-7.369510(62)$  \\
$0.9$ & $-5.263302(56)$  & $-6.340312(57)$  & $-6.547649(58)$  & $-7.543645(63)$  \\
$1.0$ & $-5.392091(56)$  & $-6.494800(58)$  & $-6.705689(59)$  & $-7.723024(64)$  \\
$1.1$ & $-5.526209(57)$  & $-6.655589(59)$  & $-6.870071(60)$  & $-7.909392(65)$  \\
$1.2$ & $-5.667084(58)$  & $-6.824374(59)$  & $-7.042546(60)$  & $-8.104757(66)$  \\
$1.3$ & $-5.816428(58)$  & $-7.003175(60)$  & $-7.225207(61)$  & $-8.311515(67)$  \\
$1.4$ & $-5.976357(59)$  & $-7.194495(61)$  & $-7.420644(62)$  & $-8.532634(68)$  \\
$1.5$ & $-6.149592(60)$  & $-7.401542(62)$  & $-7.632181(63)$  & $-8.771926(68)$  \\
$1.6$ & $-6.339756(61)$  & $-7.628587(63)$  & $-7.864251(64)$  & $-9.034481(69)$  \\
$1.7$ & $-6.551909(62)$  & $-7.881590(64)$  & $-8.123045(65)$  & $-9.327421(71)$  \\
$1.8$ & $-6.793565(63)$  & $-8.169390(65)$  & $-8.417768(66)$  & $-9.661352(72)$  \\
$1.9$ & $-7.077040(65)$  & $-8.506461(67)$  & $-8.763516(68)$  & $-10.053716(74)$ \\
\hline\hline
\end{tabular}
\caption{Numerical values of integrals
${\cal J}_w^{\widetilde{c_1}-{\rm PL}}$ and
${\cal J}_{\chi'}^{\widetilde{c_1}-{\rm PL}}$ (DBW2).}
\label{TAB:V_2_smr-pl_DBW2}
\end{center}
\end{table}

\newpage

\begin{table}[H]
\begin{center}
\input{./Tables/table_K_chi_smr-pl_DBW2.tab}
\caption{Numerical values of integrals
${\cal K}_{\chi}^{\widetilde{c_1}-{\rm PL}}$ (DBW2).}
\label{TAB:V_3_smr-pl_DBW2}
\end{center}
\end{table}

\newpage

\begin{table}[H]
\begin{center}
\input{./Tables/table_zR_Gamma_1_MF_WILSON.tab}
\end{center}
\begin{center}
\input{./Tables/table_z_Gamma_pa_WILSON.tab}
\caption{Numerical values of
$\hat{z}_{\Gamma(R)}^{(1)\rm MF}$ and $\hat{z}_{\Gamma}^{(pa)\rm MF}$ (Plaquette).}
\label{TAB:hat_z_GR_1_hat_z_G_pa_plaquette}
\end{center}
\end{table}

\newpage

\begin{table}[H]
\begin{center}
\input{./Tables/table_z_Gamma_ma_WILSON.tab}
\end{center}
\begin{center}
\input{./Tables/table_zR_L_1_MF_WILSON.tab}
\caption{Numerical values of
$\hat{z}_{\Gamma}^{(ma)\rm MF}$ and $\hat{z}_{L(R)}^{(1)\rm MF}$ (Plaquette).}
\end{center}
\end{table}

\newpage

\begin{table}[H]
\begin{center}
\input{./Tables/table_z_L_pa_WILSON.tab}
\end{center}
\begin{center}
\input{./Tables/table_z_L_ma_WILSON.tab}
\caption{Numerical values of
$\hat{z}_L^{(pa)\rm MF}$ and $\hat{z}_L^{(ma)\rm MF}$ (Plaquette).}
\end{center}
\end{table}

\newpage

\begin{table}[H]
\begin{center}
\input{./Tables/table_zR_Gamma_1_MF_SYMANZIK.tab}
\end{center}
\begin{center}
\input{./Tables/table_z_Gamma_pa_SYMANZIK.tab}
\caption{Numerical values of
$\hat{z}_{\Gamma(R)}^{(1)\rm MF}$ and $\hat{z}_{\Gamma}^{(pa)\rm MF}$ (Symanzik).}
\end{center}
\end{table}

\newpage

\begin{table}[H]
\begin{center}
\input{./Tables/table_z_Gamma_ma_SYMANZIK.tab}
\end{center}
\begin{center}
\input{./Tables/table_zR_L_1_MF_SYMANZIK.tab}
\caption{Numerical values of
$\hat{z}_{\Gamma}^{(ma)\rm MF}$ and $\hat{z}_{L(R)}^{(1)\rm MF}$ (Symanzik).}
\end{center}
\end{table}

\newpage

\begin{table}[H]
\begin{center}
\input{./Tables/table_z_L_pa_SYMANZIK.tab}
\end{center}
\begin{center}
\input{./Tables/table_z_L_ma_SYMANZIK.tab}
\caption{Numerical values of
$\hat{z}_L^{(pa)\rm MF}$ and $\hat{z}_L^{(ma)\rm MF}$ (Symanzik).}
\end{center}
\end{table}

\newpage

\begin{table}[H]
\begin{center}
\input{./Tables/table_zR_Gamma_1_MF_IWASAKI.tab}
\end{center}
\begin{center}
\input{./Tables/table_z_Gamma_pa_IWASAKI.tab}
\caption{Numerical values of
$\hat{z}_{\Gamma(R)}^{(1)\rm MF}$ and $\hat{z}_{\Gamma}^{(pa)\rm MF}$ (Iwasaki).}
\label{TAB:hat_z_G_1_hat_z_G_pa_IWASAKI}
\end{center}
\end{table}

\newpage

\begin{table}[H]
\begin{center}
\input{./Tables/table_z_Gamma_ma_IWASAKI.tab}
\end{center}
\begin{center}
\input{./Tables/table_zR_L_1_MF_IWASAKI.tab}
\caption{Numerical values of
$\hat{z}_{\Gamma}^{(ma)\rm MF}$ and $\hat{z}_{L(R)}^{(1)\rm MF}$ (Iwasaki).}
\end{center}
\end{table}

\newpage

\begin{table}[H]
\begin{center}
\input{./Tables/table_z_L_pa_IWASAKI.tab}
\end{center}
\begin{center}
\input{./Tables/table_z_L_ma_IWASAKI.tab}
\caption{Numerical values of
$\hat{z}_L^{(pa)\rm MF}$ and $\hat{z}_L^{(ma)\rm MF}$ (Iwasaki).}
\label{TAB:hat_z_L_pa_hat_z_L_ma_IWASAKI}
\end{center}
\end{table}

\newpage

\begin{table}[H]
\begin{center}
\input{./Tables/table_zR_Gamma_1_MF_DBW2.tab}
\end{center}
\begin{center}
\input{./Tables/table_z_Gamma_pa_DBW2.tab}
\caption{Numerical values of
$\hat{z}_{\Gamma(R)}^{(1)\rm MF}$ and $\hat{z}_{\Gamma}^{(pa)\rm MF}$ (DBW2).}
\end{center}
\end{table}

\newpage

\begin{table}[H]
\begin{center}
\input{./Tables/table_z_Gamma_ma_DBW2.tab}
\end{center}
\begin{center}
\input{./Tables/table_zR_L_1_MF_DBW2.tab}
\caption{Numerical values of
$\hat{z}_{\Gamma}^{(ma)\rm MF}$ and $\hat{z}_{L(R)}^{(1)\rm MF}$ (DBW2).}
\end{center}
\end{table}

\newpage

\begin{table}[H]
\begin{center}
\input{./Tables/table_z_L_pa_DBW2.tab}
\end{center}
\begin{center}
\input{./Tables/table_z_L_ma_DBW2.tab}
\caption{Numerical values of
$\hat{z}_L^{(pa)\rm MF}$ and $\hat{z}_L^{(ma)\rm MF}$ (DBW2).}
\label{TAB:hat_z_L_pa_hat_z_L_ma_DBW2}
\end{center}
\end{table}




\end{document}